\documentclass[a4paper,11pt]{article}
\usepackage{jheppub} %
\usepackage{lineno}

\usepackage[utf8]{inputenc}
\usepackage[table]{xcolor} %
\usepackage{array}         %
\usepackage{multirow}      %
\usepackage{makecell}      %
\usepackage{float}
\usepackage{placeins}     %
\usepackage{booktabs}     %
\usepackage{listings}     %
\usepackage[dvipsnames,svgnames,x11names]{xcolor}
\usepackage{hyperref}

\usepackage[markup=underlined]{changes}

\newcolumntype{?}{!{\vrule width 1.5pt}}
\allowdisplaybreaks[2]

\newcommand{\GeV}{\text{GeV}}
\newcommand{\TeV}{\text{TeV}}
\newcommand{\sss}{\scriptscriptstyle}
\newcommand{\sNN}{\sqrt{s_{\rm NN}}}
\newcommand{\pA}{$pA$}
\newcommand{\pp}{$pp$}
\newcommand{\ppb}{$p\bar{p}$}
\newcommand{\sigexp}{\sigma_{\rm exp}}
\newcommand{\sigsta}{\sigma_{\rm stat}}

\preprint{MS-TP-26-32}
\title{\boldmath NNLO QCD predictions for Drell--Yan production in proton--nucleus collisions}

\author[a,*]{Xuan Chen,}
\emailAdd{xuan.chen@sdu.edu.cn}

\author[b]{Alexander Huss,}
\emailAdd{alexander.huss@cern.ch}

\author[c,*]{Tom\'a\v{s} Je\v{z}o,}
\emailAdd{tomas.jezo@uni-muenster.de}

\author[c]{Michael Klasen,}
\emailAdd{michael.klasen@uni-muenster.de}

\author[a,d,*]{Zhe Li,}
\emailAdd{lizhep@sdu.edu.cn}

\author[a]{Jia-Sheng Liu,}
\emailAdd{jsliu@mail.sdu.edu.cn}

\author[c]{Jost Niehues}
\emailAdd{jniehues@uni-muenster.de}

\affiliation[a]{School of Physics, Shandong University, Jinan, Shandong 250100, China}
\affiliation[b]{CERN, Theoretical Physics Department, CH-1211 Geneva 23, Switzerland}
\affiliation[c]{Institut f{\"u}r Theoretische Physik, Universit{\"a}t M{\"u}nster,
Wilhelm-Klemm-Stra{\ss}e 9, D-48149 M{\"u}nster, Germany}
\affiliation[d]{Zhejiang Institute of Modern Physics, School of Physics, Zhejiang University, Hangzhou 310027, China}

\note[*]{Corresponding authors.}

\abstract{
Drell--Yan production in proton--nucleus collisions is an important probe of the nuclear modifications of parton distributions. We present fully differential NNLO QCD predictions, obtained within a common \texttt{NNLOJET} implementation, for the ATLAS and CMS measurements of $W^{\pm}$ and $Z/\gamma^{*}$ production in proton--lead collisions at $\sNN=5.02$ and $8.16~\TeV$, as well as for the per-nucleon cross-section ratios measured by the fixed-target experiments E772 and E866/NuSea with an $800~\GeV$ proton beam. NNLO corrections are generally at the few-percent level for $W^{\pm}$ production and neutral-current production near the $Z$ peak, but reach $15\%$ in the low-mass dimuon sample. 
In the fixed-target ratios, the higher-order corrections largely cancel except at small target momentum fractions, where the NNLO corrections, reaching $-6\%$, are comparable to the NLO ones. In part of the E866 data, they can therefore not be neglected in NNLO nuclear-PDF analyses. NNLO interpolation grids generated by \texttt{fastNLO} through the \texttt{APPLfast} interface are made publicly available. A comprehensive $\chi^2$ comparison with a selection of the LHC data, based on these grids and on \texttt{nCTEQ26} nuclear PDFs and
employing NNLO evolution of an NLO-fitted input parametrisation without refitting,  shows the largest NNLO improvement in the low-mass window. The resulting framework facilitates the inclusion of collider and fixed-target Drell--Yan data in NNLO nuclear-PDF analyses.
}

\begin{document}
\maketitle
\flushbottom

\section{Introduction}
\label{sec:intro}

Drell--Yan (DY) lepton-pair production~\cite{Drell:1970wh,Christenson:1970um} has long played a central role in precision QCD phenomenology, serving as a benchmark process for both theoretical developments and experimental studies.
Owing to its clean experimental signature and the robustness of perturbative predictions, DY measurements provide essential information on parton dynamics, hadronic structure and electroweak parameters across a wide range of energies
in fixed-target proton--nucleus collisions at Fermilab~\cite{E605:1988dwp,Alde:1990im,NuSea:1998kqi,NuSea:1999egr,E866:2001gdc,FNALE906:2022xdu,FNALE906SeaQuest:2025kjo},
in proton--antiproton collisions at the Tevatron~\cite{D0:2007djv,CDF:2009cjw,CDF:2010vek,CDF:2012gpf,D0:2012kms,D0:2013lql,D0:2014kma,CDF:2016FullRunII,CDF:2018cnj},
and in proton--proton~\cite{CMS:2016qqr,ATLAS:2017TripleDiff,ATLAS:2017rzl,ATLAS:2019fgb,ATLAS:2019zci,CMS:2020cph,LHCb:2021bjt,LHCb:2021FullRunII,CMS:2024FullRunII,ATLAS:2025CCDY}
and proton--nucleus~\cite{LHCb:2014jgh,CMS:2015ehw,ATLAS:2015mwq,TheATLAScollaboration:2015lnm,CMS:2015zlj,ALICE:2016rzo,CMS:2019leu,ALICE:2020jff,CMS:2021ynu,LHCb:2022kph,ALICE:2022cxs} collisions at the Large Hadron Collider (LHC).
In particular, these measurements of DY cross sections and asymmetries are widely used to constrain parton distribution functions (PDFs) in the proton~\cite{Martin:2009iq,Accardi:2016qay,Alekhin:2017kpj,Hou:2019efy,Bailey:2020ooq,NNPDF:2021njg,PDF4LHCWorkingGroup:2022cjn} 
and in nuclei~\cite{Kusina:2016fxy,Kovarik:2015cma,Kusina:2020lyz,Helenius:2021tof,Eskola:2021nhw,AbdulKhalek:2022fyi,PhysRevD.104.034010,Klasen:2023uqj}, 
and also offer a valuable testing ground for perturbative QCD~\cite{Altarelli:1978id,Altarelli:1979ub,Kubar:1980zv,Anastasiou:2003yy,Anastasiou:2003ds,Catani:2009sm,Gehrmann-DeRidder:2016jns}, 
factorisation~\cite{Bodwin:1984hc,Collins:1985ue,Collins:1988ig,Collins:1989gx}, 
and the interplay between hard scattering and hadronic structure~\cite{Gao:2017yyd,Kovarik:2019xvh,Ethier:2020way}.
In proton--nucleus (\pA) collisions in particular, DY data are especially well
suited to disentangling genuine nuclear modifications of parton densities from
final-state effects~\cite{Helenius:2021tof,Kulagin:2014vsa,Arleo:2015qiv}, making them a key ingredient in studies of QCD in nuclear
environments~\cite{Arleo:2025oos}.

A substantial body of DY data in \pA\ collisions has been accumulated over several decades, beginning with fixed-target experiments at moderate centre-of-mass (CM) energies~\cite{E605:1988dwp,Alde:1990im,NuSea:1999egr}.
In particular, the per-nucleon cross-section ratios measured by E772 and E866 constrain the nuclear modifications of the antiquark distributions, complementing charged-lepton deep-inelastic scattering (DIS) in global nuclear-PDF analyses~\cite{Alde:1990im,NuSea:1999egr,Kovarik:2015cma,Eskola:2021nhw,Klasen:2023uqj}.
Measurements on hydrogen and deuterium targets by E866 and E906/SeaQuest also constrain the light-antiquark flavour asymmetry of the proton~\cite{NuSea:1998kqi,E866:2001gdc,SeaQuest:2021zxb,FNALE906:2022xdu,FNALE906SeaQuest:2025kjo,Alekhin:2023uqx}.
Complementing these fixed-target studies, high-precision DY measurements in \pA~collisions have become available at the LHC, extending the kinematic reach to much higher energies and new regions of phase space~\cite{LHCb:2014jgh,CMS:2015ehw,ATLAS:2015mwq,TheATLAScollaboration:2015lnm,CMS:2015zlj,ALICE:2016rzo,CMS:2019leu,ALICE:2020jff,CMS:2021ynu,LHCb:2022kph,ALICE:2022cxs}.
$W^\pm$ and $Z/\gamma^*$ production at $\sqrt{s_{\mathrm{NN}}}=5.02$ and $8.16~\TeV$ probes nuclear quark distributions at typically higher hard scales and, in the direction of the proton, at smaller momentum fractions than the fixed-target measurements~\cite{Paukkunen:2010qg,Klasen:2023uqj}.
Rapidity distributions and forward--backward ratios constrain the $x$ dependence of the nuclear modifications, while lepton-charge asymmetries probe the relative up- and down-quark distributions, with sensitivity to both the free-proton PDFs and flavour-dependent nuclear modifications~\cite{Paukkunen:2010qg,Kusina:2016fxy,Eskola:2022rlm}.

In parallel, fixed-order QCD predictions for DY production have developed considerably since the early next-to-leading-order (NLO) calculations~\cite{Altarelli:1978id,Altarelli:1979ub,Kubar:1980zv}, subsequently extended to the next-to-next-to-leading-order (NNLO) for both inclusive and differential cross sections~\cite{Hamberg:1990np,Harlander:2002wh,Anastasiou:2003yy,Anastasiou:2003ds,Melnikov:2006di,Melnikov:2006kv,Catani:2009sm,Catani:2010en}.
More recently, theoretical predictions for DY production have reached an unprecedented level of accuracy, with inclusive cross sections~\cite{Duhr:2020seh,Duhr:2020sdp,Duhr:2021vwj} as well as differential distributions and fiducial cross sections~\cite{Chen:2021vtu,Chen:2022cgv,Chen:2022lwc,Campbell:2023lcy} now available up to next-to-next-to-next-to-leading order (N$^3$LO) in perturbative QCD.
For proton--proton (\pp) collisions, NNLO predictions have already become standard ingredients in modern phenomenological studies and PDF determinations~\cite{Hou:2019efy,Bailey:2020ooq,NNPDF:2021njg}.
Extending this level of accuracy to asymmetric initial states is conceptually straightforward within collinear factorisation, but its practical realisation in a form suitable for systematic and reusable applications remains far from trivial.
Existing NNLO implementations are largely optimised and validated for symmetric \pp\ and \ppb\ collisions~\cite{Anastasiou:2003ds,Catani:2009sm,Gehrmann-DeRidder:2016jns,Hamberg:1990np,Melnikov:2006kv,Gavin:2010az,Li:2012wna,Gavin:2012sy,Hoche:2014uhw,Boughezal:2016wmq,Caola:2017dug,Grazzini:2017mhc,Campbell:2019dru,Camarda:2019zyx,Alekhin:2021xcu}, and applications to DY production in \pA~collisions typically require analysis-specific adaptations or undocumented extensions~\cite{Kusina:2016fxy,Helenius:2021tof}. 
A general automated framework for asymmetric collisions has only recently become available, and only at NLO~\cite{Flore:2025ync}.
As a consequence, NNLO predictions for DY production in \pA~collisions are not yet routinely available in a form that is robust, reusable, and suitable for systematic phenomenological studies or global analyses.

From a practical perspective, the absence of reusable NNLO infrastructure is particularly limiting for applications that require repeated evaluations of theoretical predictions, most notably global determinations of proton and nuclear parton distribution functions~\cite{Hou:2019efy,Bailey:2020ooq,NNPDF:2021njg,PDF4LHCWorkingGroup:2022cjn,Kovarik:2015cma,Kusina:2016fxy,Kusina:2020lyz,Helenius:2021tof,Eskola:2021nhw,AbdulKhalek:2022fyi}.
In such analyses, fixed-order calculations must be evaluated thousands of times for different PDF parametrisations, making repeated direct NNLO computations prohibitively expensive~\cite{Hou:2019efy,Britzger:2019kkb}.
Interpolation techniques based on precomputed grids provide an effective solution to this problem by factorising the perturbative coefficients from the PDFs and enabling fast and flexible convolutions~\cite{Kluge:2006xs,Carli:2010rw,Britzger:2012bs,Britzger:2019kkb,Carrazza:2020gss,Britzger:2022lbf}.
While such approaches are well established at NNLO for jet production in DIS and $pp$ collisions~\cite{Britzger:2019kkb,Britzger:2022lbf} and have recently become available for DY production in $pp$ and $p\bar{p}$ collisions~\cite{Cruz-Martinez:2025ffa}, their availability for DY production in asymmetric \pA\ collisions at NNLO has so far been limited to the fit-specific grids of the TUJU21 nuclear-PDF analysis~\cite{Helenius:2021tof,Helenius:2022rmf}.\footnote{A closely related step has been taken in this analysis for electroweak boson production in p+Pb collisions, constructed using a modified version of MCFM to handle asymmetric initial states.}

In this work, we present a comprehensive NNLO study of DY production in \pA\ collisions, both in fixed-target experiments at Fermilab and at the LHC, within the \texttt{NNLOJET} framework~\cite{NNLOJET:2025rno}, which employs the antenna subtraction method~\cite{Gehrmann-DeRidder:2005btv,Daleo:2006xa,Currie:2013vh,Chen:2022clm}. 
We compute fully differential NNLO QCD predictions for charged-current and neutral-current DY production over a wide range of kinematic regimes. %
Our analysis encompasses the $W^\pm$ and $Z/\gamma^*$ measurements in $p\mathrm{Pb}$ collisions by the ATLAS and CMS~\cite{ATLAS:2015mwq,CMS:2015ehw,CMS:2015zlj,CMS:2019leu,CMS:2021ynu} at $\sNN=5.02$ and $8.16~\TeV$, together with the per-nucleon nuclear-target ratios measured by the fixed-target experiments E772 and E866/NuSea at Fermilab~\cite{Alde:1990im,NuSea:1999egr}, covering a broad span in Bjorken-$x$, invariant mass, and (pseudo)rapidity.
This unified treatment allows for a consistent comparison of low- and high-energy data within a single perturbative framework and facilitates detailed studies of the sensitivity of DY observables to nuclear effects in quark and antiquark distributions.
Building on these calculations, we construct fast interpolation grids using the APPLfast interface~\cite{Britzger:2019kkb,Britzger:2022lbf} to the fastNLO library~\cite{Kluge:2006xs,Britzger:2012bs}, enabling rapid and flexible evaluation of the NNLO cross sections with arbitrary proton and nuclear PDFs.
The resulting fastNLO grids are made publicly available through the Ploughshare repository~\cite{Ploughshare} and are designed for
direct use in global PDF analyses and related phenomenological applications.

The remainder of this paper is organised as follows.
Section~\ref{sec:framework} introduces the theoretical and computational framework: collinear factorisation and the nuclear PDFs, kinematics and observables in asymmetric and fixed-target collisions, and the \texttt{NNLOJET} implementation together with the interpolation-grid techniques.
Section~\ref{sec:NNLOJET_predictions} presents the NNLO predictions for $W^{\pm}$ and $Z/\gamma^{*}$ production in $p\mathrm{Pb}$ collisions at the LHC and for the fixed-target E772 and E866 nuclear-target ratios, and compares them with the corresponding measurements.
Section~\ref{sec:grid-applications} investigates the impact of the NNLO corrections in the context of nuclear PDF fits by assessing their effect on the $\chi^2$ values of the LHC data using NLO-fitted nPDFs evolved to NNLO.
Section~\ref{sec:Conclusion} contains our conclusions.
The closure test of the interpolation grids and the comparison of matrix-element and full $K$-factors are collected in appendices~\ref{app:closure} and~\ref{app:MEvsFullKfacs}.

\section{Theoretical and computational framework}
\label{sec:framework}

The DY process conventionally refers to the inclusive production of an electroweak gauge boson in hadron--hadron collisions, followed by its leptonic decay, encompassing both the neutral-current channel, in which a $Z/\gamma^{*}$ boson decays into a charged-lepton pair, and the charged-current channel, in which a $W^{\pm}$ boson decays into a charged lepton and the corresponding (anti)neutrino.
Throughout this work, we adopt standard leading-twist collinear factorization to describe DY predictions in both collider and fixed-target \pA\ collisions. This framework is based on factorization theorems for inclusive DY production~\cite{Bodwin:1984hc,Collins:1985ue,Collins:1988ig,Collins:1989gx}, which is assumed to be applicable also to collisions involving nuclei.
Within this framework, the cross section is expressed as a convolution of two distinct ingredients: process-dependent short-distance partonic cross sections, also referred to as coefficient functions, which can be computed order by order in perturbative QCD, and universal non-perturbative parton distribution functions, which encode the longitudinal momentum distributions of partons within each incoming hadron.
In terms of the (nuclear) parton distribution functions of the incoming hadrons $h_1$ and $h_2$ and the partonic hard-scattering cross sections, the cross section is factorized at leading power  as~\cite{Helenius:2021tof,Klasen:2023uqj,Flore:2025ync,Eskola:2022rlm}\footnote{Power-suppressed corrections, including higher-twist contributions, are not shown explicitly in eq.~\eqref{eq:factorisation} and are neglected in the present analysis.}
\begin{multline}
\label{eq:factorisation}
\mathrm{d}\sigma_{\rm\sss DY}
=
\sum_{a,b}\int_0^1\mathrm{d}x_1\int_0^1\mathrm{d}x_2\,
f_a^{h_1}\!\left(x_1,\mu_F^2\right)
f_b^{h_2}\!\left(x_2,\mu_F^2\right)
\mathrm{d}\hat\sigma_{\rm\sss DY}^{ab}\!\left(x_1P_1,x_2P_2;\mu_R^2,\mu_F^2\right),
\end{multline}
where the sum runs over the parton species $a,b$ of the incoming partons, $\mathrm{d}\hat\sigma^{ab}_{\rm\sss DY}$ is the corresponding partonic cross section, $x_{1,2}$ are the partonic momentum fractions of the corresponding nucleon momenta $P_{1,2}$, and $\mu_R$ and $\mu_F$ denote the renormalisation and factorisation scales, respectively.
The collinear factorisation equation~\eqref{eq:factorisation} is assumed to hold for any pair of incoming hadrons $h_1$ and $h_2$,  whose identities specify the collision system.
In $pp$ collisions, both $f^{h_1}$ and $f^{h_2}$ are free-proton PDFs, whereas in collisions involving nuclei either one beam leg or both beam legs are described by per-nucleon nuclear PDFs, whose association with $h_1$ or $h_2$ depends on the chosen beam convention.
Mixed collision systems, most notably $pA$ collisions at the LHC, entail the additional technical complication that the two incoming PDFs are different.
For a nuclear beam, the corresponding momentum fraction can, in principle, extend beyond unity, up to $x=A$.
In the present analysis we restrict ourselves to the conventional region $0<x<1$ and neglect contributions from $x>1$.
Higher-twist effects can become phenomenologically relevant at large $x$ and low hard scales, and have been investigated in global PDF analyses \cite{Accardi:2016qay, Accardi:2023gyr, Accardi:2026jmx} as well as in the nuclear nCTEQ15HIX analysis~\cite{Segarra:2020gtj}.
In DIS, target-mass corrections constitute an additional class of kinematic
power corrections, distinct from genuine higher-twist contributions, and have
been discussed extensively in Refs.~\cite{Schienbein:2007gr,Ruiz:2023ozv}.

Nuclear parton distribution functions (nPDFs) describe the partonic structure of nucleons bound inside nuclei and, like free-proton PDFs, are determined through global analyses of a broad range of hard-scattering data involving nuclear targets.
Compared with free-proton PDFs, they exhibit characteristic dependence on $x$, with shadowing at small $x$, antishadowing at moderate $x$, 
a suppression at larger $x$ known as the EMC effect after its discovery by the European Muon Collaboration~\cite{EuropeanMuon:1983wih},
and Fermi-motion enhancement as $x$ approaches unity~\cite{Klasen:2023uqj}. 
Consequently, they cannot be obtained by simply scaling  an appropriate linear combination of free-nucleon PDFs with the mass number $A$ of the nucleus.
At the parton level, these effects are encoded in flavour-, nucleus- and scale-dependent nuclear PDFs $f_i^A(x,\mu_F^2)$.
Several modern global nPDF analyses are available, differing in their parametrisation choices, free-proton baselines, treatment of experimental uncertainties and datasets included in the fits~\cite{nCTEQ:2026ewr,Helenius:2021tof,Eskola:2021nhw,AbdulKhalek:2022fyi,PhysRevD.104.034010}.

For a nucleus with charge number $Z$ and mass number $A$, the physical PDF per average nucleon can be defined by
\begin{equation}
\label{eq:nPDF}
f_i^{A}\!\left(x,\mu_F^2\right)
=
\frac{Z}{A}\,f_i^{p/A}\!\left(x,\mu_F^2\right)
+\frac{A-Z}{A}\,f_i^{n/A}\!\left(x,\mu_F^2\right),
\end{equation}
where $f_i^{p/A}$ and $f_i^{n/A}$ denote the bound-proton and bound-neutron PDFs, respectively.
Among the analyses employed in this work, the conventional decomposition is adopted, up to the overall normalisation, by the nCTEQ~\cite{nCTEQ:2026ewr}, EPPS~\cite{Eskola:2021nhw} and TUJU~\cite{Helenius:2021tof} analyses.
They parametrise the bound-proton PDFs $f_i^{p/A}$, either directly (nCTEQ, TUJU) or through the nuclear modification factors $R_i^{p/A}$ relative to a free-proton baseline (EPPS), and assemble $f_i^{A}$ from them, rather than parametrising $f_i^{A}$ independently.
Only the bound proton is parametrised independently; the bound-neutron PDFs $f_i^{n/A}\!\left(x,\mu_F^2\right)$ are commonly obtained from those of the bound proton $f_i^{p/A}\!\left(x,\mu_F^2\right)$ by assuming isospin symmetry~\cite{Klasen:2023uqj},
\begin{equation}
\label{eq:boundPDF}
    f_{u,\bar{u}}^{n/A}\left(x, \mu_F^2\right)
    = f_{d,\bar{d}}^{p/A}\left(x, \mu_F^2\right),
    \qquad
    f_{d,\bar{d}}^{n/A}\left(x, \mu_F^2\right)
    = f_{u,\bar{u}}^{p/A}\left(x, \mu_F^2\right).
\end{equation}

\subsection{Kinematics and observables in asymmetric collisions}
\label{sec:kinematics}

Both types of collisions in focus in this work, fixed-target or collider with different beams, are longitudinally asymmetric. 
Hence, the nucleon--nucleon CM frame does not coincide with the laboratory frame: in \pA{} collisions at the LHC the two beams carry in general different energies per nucleon, whereas in fixed-target collisions the target nucleon is at rest.
In this section, we discuss the kinematic differences between symmetric and asymmetric hadronic collisions, covering both collider and fixed-target configurations. Particular attention is paid to the relation between the laboratory and nucleon--nucleon centre-of-mass frames, as well as to the corresponding definitions of observables and partonic momentum fractions. The formulae collected here also provide the basis for the implementation of asymmetric collision systems in \textsc{NNLOjet} discussed in section~\ref{sec:NNLOJET_predictions}.

At the LHC both beams circulate in the two apertures of the same two-in-one magnet system and are therefore constrained to have the same magnetic rigidity $B\rho=p/(Ze)$, with $B$ the bending field, $\rho$ the bending radius, $p$ the ion momentum and $Ze$ its charge~\cite{Salgado:2011wc,AbdulKhalek:2022fyi}.
For a fully stripped ion of mass number $A$, the momentum per nucleon then scales as $Z/A$, so that, in the high-energy limit $E_i^{(N)}\gg m_N$ with $m_N$ the nucleon mass, the per-nucleon beam energies $E_i^{(N)}$ of the two colliding beams obey
\begin{equation}
\frac{E_1^{(N)}}{E_2^{(N)}}=\frac{Z_1A_2}{Z_2A_1} .
\label{eq:rigidity}
\end{equation}
The collision energy is conventionally quoted as the nucleon--nucleon CM energy $\sqrt{s_{\mathrm{NN}}}$, with $s_{\mathrm{NN}}=(P_1+P_2)^2$ built from the per-nucleon four-momenta $P_i$ rather than from those of the full ion.
In the high-energy limit, this reduces to
\begin{equation}
\sqrt{s_{\mathrm{NN}}}\simeq 2\sqrt{E_1^{(N)}E_2^{(N)}} .
\label{eq:sNN}
\end{equation}
For $p\mathrm{Pb}$ collisions, eq.~\eqref{eq:rigidity} gives $E^{(N)}_{\mathrm{Pb}}=(82/208)\,E_p$, the proton beam energies $E_p=4.0$ and $6.5~\TeV$ of the LHC $p\mathrm{Pb}$ Run~1 and Run~2 therefore correspond to $\sqrt{s_{\mathrm{NN}}}\simeq 5.02$ and $8.16~\TeV$, respectively.
A partonic system carrying momentum fractions $x_1$ and $x_2$ then has laboratory rapidity
\begin{equation}
y_q=\frac12\ln\frac{x_1E_1^{(N)}}{x_2E_2^{(N)}}
=\frac12\ln\!\left(\frac{x_1}{x_2}\,\frac{Z_1A_2}{Z_2A_1}\right) .
\label{eq:yq-AP}
\end{equation}
Under the boost from the nucleon–nucleon CM frame to the laboratory frame, this rapidity separates into two terms, $y_q=y^{\ast}+\Delta y$, where $y^{\ast}=\tfrac12\ln(x_1/x_2)$ is the rapidity of the partonic system in the nucleon--nucleon CM frame, and
\begin{equation}
\Delta y=\tfrac12\ln\!\left(Z_1A_2/Z_2A_1\right)
\label{eq:deltay}
\end{equation}
is the rapidity of the nucleon--nucleon CM frame relative to the laboratory frame.
This decomposition makes explicit that the laboratory rapidity of the partonic system contains two conceptually distinct contributions: an event-dependent boost $y^\ast$ arising from the imbalance of the partonic momentum fractions $x_1$ and $x_2$, and a beam-dependent boost $\Delta y$ originating from the unequal per-nucleon beam energies.
For $p\mathrm{Pb}$ collisions, taking the proton as beam~1, $A_1=Z_1=1$, and the lead nucleus as beam~2, $A_2=208$ and $Z_2=82$, gives
\begin{equation}
\Delta y
=
\frac12\ln\!\left(\frac{208}{82}\right)
\simeq 0.465 .
\end{equation}
The shift $\Delta y$ is therefore independent of $x_1$ and $x_2$ and hence common to every event. Throughout, the proton is taken as beam~1 aligned with the proton direction.
\\

In fixed-target nucleon--nucleon collisions, a projectile nucleon of energy $E_b$, momentum $p_b$ and mass $m_1$ along the positive longitudinal direction strikes a target nucleon of mass $m_2$ taken to be at rest in the laboratory. 
For a nuclear projectile, $E_b$ and $p_b$ denote the corresponding energy and momentum per nucleon, while for a nuclear target $m_2$ denotes the mass of a target nucleon rather than that of the full nucleus.
The nucleon--nucleon CM energy is given by $s_{\rm NN}=(P_1+P_2)^2=m_1^2+m_2^2+2m_2E_b$, with $P_1$ and $P_2$ the four-momenta of the projectile proton and the target nucleon, which reduces to $\sqrt{s_{\rm NN}}\simeq\sqrt{2E_bm_2}$ in the high-energy limit $E_b\simeq p_b\gg m_1, m_2$.
A partonic system with momentum fractions $x_1$ and $x_2$ then has laboratory rapidity
\begin{equation}
y_q =
\frac{1}{2}\ln\frac{x_1\left(E_b+p_b\right)+x_2m_2}{x_1\left(E_b-p_b\right)+x_2m_2}
\simeq
\frac{1}{2}\ln\frac{2x_1 E_b}{x_2 m_2}\,.
\label{eq:yq-FT}
\end{equation}
In the high-energy limit, this rapidity separates, in analogy with eq.~\eqref{eq:deltay}, into $y_q=y^{\ast}+\Delta y_{\mathrm{FT}}$, with $y^{\ast}=\tfrac12\ln(x_1/x_2)$ the rapidity of the system in the nucleon--nucleon CM frame and
\begin{equation}
\Delta y_{\mathrm{FT}}
=
\frac{1}{2}\ln\frac{2E_b}{m_2}
\simeq
\frac{1}{2}\ln\frac{s_{\rm NN}}{m_2^{2}}
\label{eq:deltay-FT}
\end{equation}
the rapidity of the nucleon--nucleon CM frame relative to the laboratory frame. 
Here the boost is defined with respect to the nucleon--nucleon CM frame, as appropriate for the hard-scattering kinematics and the definition of $s_{\rm NN}$.%
\footnote{If instead the CM frame of the complete projectile--target nuclear system were considered, the use of the full nuclear masses would introduce an additional, $A$-dependent constant shift. 
Neglecting nuclear binding effects, this is given in the high-energy limit by
$\Delta y_{A_1A_2}\simeq\Delta y_{\mathrm{FT}}+1/2\ln(A_1/A_2)$
where $A_1$ and $A_2$ denote the projectile and target mass numbers, respectively.}
For a proton projectile, $m_1=m_p$, while for a nuclear projectile $m_1$ denotes the corresponding nucleon mass, with $m_2$ analogously denoting the target-nucleon mass.
Neglecting the small proton--neutron mass difference and nuclear binding effects, we take a common nucleon mass $m_N\simeq0.938~\GeV$.
For example,
the $800~\GeV$ proton beam of the E772~\cite{Alde:1990im} and E866/NuSea~\cite{NuSea:1999egr,E866:2001gdc} proton--nucleus experiments gives $\Delta y_{\mathrm{FT}}\simeq3.72$.

In fixed-target DY measurements, it is common to publish data differentially in the reconstructed momentum fractions $x_1$ and $x_2$ or the Feynman variable $x_F$, rather than in the dilepton rapidity. The E772 and E866/NuSea measurements providing prominent examples with their DY cross sections on nuclear targets quoted per nucleon, consistent with the per-nucleon normalisation of the nuclear PDFs $f_i^A$ in eq.~\eqref{eq:nPDF}.
These momentum fractions are reconstructed from the measured dilepton-pair momentum $q$.
Neglecting hadron masses, the reconstructed variables in the nucleon--nucleon CM frame are given by $x_{1,2}=(q^0\pm q^z)/\sNN$; at leading order, they coincide with the momentum fractions of the incoming partons.
A manifestly frame-independent alternative definition is~\cite{FNALE906:2022xdu}
\begin{equation}
  x_1=\frac{P_2\!\cdot q}{P_2\!\cdot P},
  \qquad
  x_2=\frac{P_1\!\cdot q}{P_1\!\cdot P},
  \label{eq:li-reconstructed-x}
\end{equation}
with $P_1$ and $P_2$ the four-momenta of the projectile and target nucleon, respectively, and $P=P_1+P_2$.
The Feynman variable $x_F$ normalises the dilepton's longitudinal momentum to its kinematic maximum~\cite{FNALE906:2022xdu},
\begin{equation}
x_F=\frac{q_z^{\ast}}{q_{z,\max}^{\ast}}
=
\frac{2 q_z^{\ast}}{\sNN\big(1-m_{\ell\ell}^2 / s_{\rm NN}\big)} ,
\label{eq:xF}
\end{equation}
which covers the full range $|x_F|<1$ by construction, with $m_{\ell\ell}=\sqrt{q^2}$ the dilepton invariant mass; here and throughout, an asterisk marks quantities evaluated in the nucleon--nucleon centre-of-mass frame wherever the distinction from the laboratory frame matters, so that $q_z\equiv q_z^{\ast}$.
The alternative definition $x_F=2q_z^{\ast}/\sNN$, also common in the literature, instead spans only $|x_F|<1-m_{\ell\ell}^2/s_{\rm NN}$; the two coincide in the high-energy limit $m_{\ell\ell}^2/s_{\rm NN} \to 0$.
At Born level, the momentum fractions satisfy
\begin{equation}
m_{\ell\ell}^2 = x_1x_2\,s_{\rm NN} ,
\qquad
x_F=\frac{x_1-x_2}{1-x_1x_2},
\label{eq:LOrelations}
\end{equation}
where the latter relation corresponds to the definition in eq.~\eqref{eq:xF} and reduces to the familiar $x_F\simeq x_1-x_2$ in the high-energy limit.
By contrast, eqs.~\eqref{eq:li-reconstructed-x} and~\eqref{eq:xF} define the reconstructed variables $x_1$, $x_2$ and $x_F$ directly in terms of the dilepton momentum $q$ and therefore remain well defined at any perturbative order; these are the definitions used in the fixed-target NNLO QCD predictions presented in section~\ref{sec:FTDY}.

\subsection{NNLOJET implementation and interpolation-grid construction}
\label{sec:nnlojet}

\paragraph{NNLOJET Framework}
All cross-section predictions in this work are obtained within the framework of \texttt{NNLOJET}~\cite{NNLOJET:2025rno}, an open-source parton-level event generator computing NNLO QCD cross sections for a broad class of colour-singlet and jet production processes in lepton--lepton, lepton--hadron and hadron--hadron collisions.
\texttt{NNLOJET} employs the antenna subtraction method~\cite{Gehrmann-DeRidder:2005btv,Daleo:2006xa,Currie:2013vh} to cancel the infrared singularities between the double-real, real--virtual and double-virtual NNLO contributions, 
with the integrated antenna functions computed for the final--final~\cite{Gehrmann-DeRidder:2005btv}, initial--final~\cite{Daleo:2009yj} and initial--initial~\cite{Boughezal:2010mc,Gehrmann:2011wi,Gehrmann-DeRidder:2012too} configurations of the hard radiators. 
Fully differential neutral- and charged-current DY has been studied within the NNLOJET framework through NNLO QCD at low invariant mass~\cite{Gauld:2021pkr}, triple-differentially~\cite{Gehrmann-DeRidder:2023urf}, in the context of proton PDF determinations~\cite{Cruz-Martinez:2025ffa}, as well as predictions that rely both implicitly or explicitly on additional QCD radiation~\cite{Gehrmann-DeRidder:2016jns,Gehrmann-DeRidder:2015wbt,Gehrmann-DeRidder:2016cdi,Gehrmann-DeRidder:2018kng}

For this study, the \texttt{NNLOJET} infrastructure has been extended to support independently specified nucleon or ion beams, each with its own PDF set as well as a new fixed-target mode. 
The beam types determine $Z$ and $A$ internally and set the corresponding collider or fixed-target kinematics described in section~\ref{sec:kinematics}. 
Additional event-level observables have been implemented, including the reconstructed momentum fractions of eq.~\eqref{eq:li-reconstructed-x} and the Feynman variable $x_F$ of eq.~\eqref{eq:xF}.

\paragraph{Interpolation grids}
\label{sec:grids}

NNLO predictions are computationally expensive, while applications such as PDF fits and determinations of $\alpha_s$ require repeated evaluations for different input parameters. 
Repeating the full NNLO calculation for each choice of PDFs, $\alpha_s$ and scales is therefore prohibitively slow for practical use.
Interpolation grids circumvent this bottleneck by storing PDF- and $\alpha_s$-independent perturbative weights on a grid in the momentum fractions and scales, $(x_1,x_2,\mu_R,\mu_F)$. 
The convolution in eq.~\eqref{eq:factorisation} then reduces to a computationally inexpensive weighted sum over interpolation nodes, allowing the PDFs, $\alpha_s$ and the factors multiplying the renormalisation and factorisation scales to be changed a posteriori. 
The costly NNLO calculation is performed only once for a fixed set of observables and histograms to populate the grid with sufficient statistical precision.

Initially used for NLO QCD calculations, this technique is implemented in the publicly available libraries \texttt{fastNLO}~\cite{Kluge:2006xs,Britzger:2012bs}, \texttt{APPLgrid}~\cite{Carli:2010rw}, and \texttt{PineAPPL}~\cite{Carrazza:2020gss,Jezo:2026adf}.
At NNLO, the \texttt{APPLfast} project interfaces \texttt{NNLOJET} to \texttt{fastNLO} and \texttt{APPLgrid}~\cite{Britzger:2019kkb,Britzger:2022lbf}.
More recently, interfaces to \texttt{PineAPPL} have been developed for \texttt{NNLOJET}, providing DY grids for $pp$ and $p\bar p$ collisions~\cite{Cruz-Martinez:2025ffa}, and for \texttt{MATRIX} through the \texttt{MATRIX HAWAII} project~\cite{Devoto:2025cuf}.
Grids produced with \texttt{NNLOJET} have found broad phenomenological applications, including determinations of $\alpha_s$ and its running~\cite{Ahmadova:2024emn,CMS:2024trs} and global proton-PDF analyses~\cite{Ablat:2024uvg}.

For collisions involving nuclei, the TUJU21 analysis used NNLO grids generated with a modified version of \texttt{MCFM} to include electroweak-boson production in asymmetric $p\mathrm{Pb}$ collisions~\cite{Helenius:2021tof}. However, neither the modified MCFM implementation nor the corresponding interpolation grids have been made publicly available.
For fixed-target DY production, a modified version of \texttt{Vrap} interfaced to \texttt{PineAPPL} has been used within the \texttt{Pineline} framework to provide NNLO predictions for proton-PDF fits~\cite{Barontini:2023vmr}. This implementation provides an isoscalar-target option restricted to dilepton invariant-mass and rapidity distributions. %

In this work, we use \texttt{NNLOJET} interfaced to \texttt{fastNLO} through \texttt{APPLfast} to produce NNLO grids for both collider and fixed-target proton--nucleus DY production, with independently specified proton and nuclear PDFs.
All grids produced in this study are made publicly available through the Ploughshare repository~\cite{Ploughshare}.

\section{NNLO QCD predictions in proton--nucleus collisions}
\label{sec:NNLOJET_predictions}
The NNLO QCD predictions presented in this section cover the observables summarised in table~\ref{tab:AP_obs} for charged-current $W^{\pm}$ and neutral-current $Z/\gamma^{*}$ production in $p\mathrm{Pb}$ collisions at $\sNN=5.02$ and $8.16~\TeV$, and those in table~\ref{tab:FT_obs} for neutral-current DY production in fixed-target \pA\ collisions with an $800~\GeV$ proton beam.
The predictions are compared with measurements from ATLAS Run~1~\cite{ATLAS:2015mwq} and CMS Runs~1 and~2~\cite{CMS:2015ehw,CMS:2015zlj,CMS:2019leu,CMS:2021ynu} at the LHC, as well as with the Fermilab E772~\cite{Alde:1990im} and E866/NuSea~\cite{NuSea:1999egr} measurements.
\begin{table}[htp]
\centering
 \small
 \renewcommand{\arraystretch}{1.5}
\begin{tabular}{p{4.2cm}<{\centering} p{5.6cm}<{\centering} p{4.2cm}<{\centering}}
\toprule[1.2pt]
 Experiment &  Fiducial selections & Observables\\
 \midrule
 \midrule
\multirow{3}{*}{\makecell{$W$ production in CMS \\  at $\sNN=5.02~\TeV$~\cite{CMS:2015ehw}\\
\href{https://www.hepdata.net/record/ins1353541}{HEPData Link}}
}
 & \multirow{3}{*}{\makecell{$\lvert \eta_{\rm lab}^{\ell^{\pm}} \rvert < 2.4$, $ p_{T}^{\ell^{\pm}} > 25~\GeV$} }
 & \makecell{$d\sigma(W^+ \to \ell^+ \nu) / d\eta^{\ell^+}_{\rm lab}$ }
 \\ \cmidrule{3-3}
 &
 & \makecell{$d\sigma(W^- \to \ell^-\bar{\nu}) / d\eta^{\ell^-}_{\rm lab}$ }
  \\ \cmidrule{3-3}
 &
 & \makecell{$R_{\mathrm{FB}}^{W^+},\, R_{\mathrm{FB}}^{W^-}, \, \mathcal{A}_{\mathrm{pPb}}$ }
 \\ \midrule\midrule
\multirow{3}{*}{ \makecell{$W$ production in CMS \\ at $\sNN=8.16~\TeV$~\cite{CMS:2019leu}  \\
\href{https://www.hepdata.net/record/ins1733223}{HEPData Link}} }
 & \multirow{3}{*}{ \makecell{$\lvert \eta_{\rm lab}^{\mu} \rvert < 2.4$, $ p_{T}^{\mu} > 25~\GeV$}}
& \makecell{$d\sigma(W^+ \to \mu^+ \nu_\mu) / d\eta^{\mu^+}_{\rm CM}$ }
 \\ \cmidrule{3-3}
 &
 & \makecell{$d\sigma(W^- \to \mu^-\bar{\nu}_{\mu}) / d\eta^{\mu^-}_{\rm CM}$ }
  \\ \cmidrule{3-3}
 &
 & \makecell{$R_{\mathrm{FB}}^{W^+},\, R_{\mathrm{FB}}^{W^-}, \, \mathcal{A}_{\mathrm{pPb}}$ }
 \\ \midrule \midrule
\makecell{$Z$ production in ATLAS \\ at $\sNN=5.02~\TeV$~\cite{ATLAS:2015mwq}\\
\href{https://www.hepdata.net/record/ins1384272}{HEPData Link}}
 & \makecell{$\lvert y^{Z}_{\rm CM}\rvert < 3.5$,\\ $66~\GeV < m_{\ell^+ \ell^-} < 116~\GeV$}
 & \makecell{$d\sigma(Z/\gamma^{\ast} \rightarrow \ell^+ \ell^-) / dy^{Z}_{\rm CM}$  }
 \\ \midrule \midrule
\multirow{2}{*}{ \makecell{$Z$ production in CMS \\ at $\sNN = 5.02~\TeV$~\cite{CMS:2015zlj}\\
\href{https://www.hepdata.net/record/ins1410832}{HEPData Link}}}
 & \multirow{2}{*}{\makecell{$|\eta^{\ell^{\pm}}_{\rm lab}| < 2.4$, $ p_{T}^{\ell^+ (\ell^-) }  > 20~\GeV$, \\$60~\GeV < m_{\ell^+ \ell^- } < 120~\GeV$}}
 & \makecell{$d\sigma(Z/\gamma^{\ast} \to \ell^+ \ell^-) / dy^{Z}_{\rm CM}$  }
 \\ \cmidrule{3-3}
 &
 & \makecell{$R_{\mathrm{FB}}^{Z}$ }
 \\ \midrule \midrule
\multirow{3}{*}{  \makecell{$Z$ production in CMS \\ at $\sNN = 8.16~\TeV$~\cite{CMS:2021ynu}\\
\href{https://www.hepdata.net/record/ins1849180}{HEPData Link}}}
 &  \makecell{$|\eta^{\mu^{\pm}}_{\rm lab}| < 2.4$, $ p_{T,1(2)}^{\mu}  > 15(10)~\GeV$, \\$15~\GeV < m_{\mu^+ \mu^- } < 60~\GeV$} 
 &  \makecell{$d\sigma(Z/\gamma^{\ast} \to \mu^+ \mu^-) / dy^{Z}_{\rm CM}$  }
 \\ \cmidrule{2-3}
 &  \makecell{$|\eta^{\mu^{\pm}}_{\rm lab}| < 2.4$, $ p_{T,1(2)}^{\mu}  > 15(10)~\GeV$, \\$60~\GeV < m_{\mu^+ \mu^- } < 120~\GeV$}
 &  \makecell{$d\sigma(Z/\gamma^{\ast} \to \mu^+ \mu^-) / dy^{Z}_{\rm CM}$  }
 \\
 \bottomrule[1.2pt]
\end{tabular}
\caption{Summary of the fiducial selections and observables used in the comparisons with measurements of $W^{\pm}$ and $Z/\gamma^{*}$ production in $p\mathrm{Pb}$ collisions.
The datasets comprise the CMS measurements of $W^{\pm}$ production at $\sNN=5.02$ and $8.16~\TeV$~\cite{CMS:2015ehw,CMS:2019leu}, the ATLAS measurement of $Z/\gamma^{*}$ production at $\sNN=5.02~\TeV$~\cite{ATLAS:2015mwq}, and the CMS measurements of $Z/\gamma^{*}$ production at $\sNN=5.02$ and $8.16~\TeV$~\cite{CMS:2015zlj,CMS:2021ynu}.
For each dataset, a link to the corresponding HEPData record is provided.
All quantities are quoted in the frames used by the respective publications: $\eta_{\rm lab}$ denotes the laboratory-frame pseudorapidity, while $\eta_{\rm CM}$ and $y^{Z}_{\rm CM}$ denote the pseudorapidity and rapidity in the nucleon--nucleon CM frame, respectively.}
\label{tab:AP_obs}
\end{table}
\begin{table}[htp]
\centering
 \small
 \renewcommand{\arraystretch}{1.5}
\begin{tabular}{p{4.5cm}<{\centering} p{6.0cm}<{\centering} p{4.5cm}<{\centering}}
\toprule[1.2pt]
 Experiment & Fiducial selections & Observables\\
 \midrule
 \midrule
\makecell{E772 at Fermilab, \\ with $E_{\rm beam}=800~\GeV$~\cite{Alde:1990im}\\
\href{https://doi.org/10.17182/hepdata.19997.v1}{HEPData Link}}
 & \makecell{$4<m_{\mu\mu}<9~\GeV$ or $m_{\mu\mu}>11~\GeV$,\\ $x_F>0$}
 & \makecell{$\sigma_{p\mathrm{A}}/\sigma_{p\mathrm{D}}$ versus $x_2$, \\ $\mathrm{A}=\mathrm{C},\,\mathrm{Ca},\,\mathrm{Fe},\,\mathrm{W}$}
 \\ \midrule \midrule
\makecell{E866/NuSea at Fermilab, \\ with $E_{\rm beam}=800~\GeV$~\cite{NuSea:1999egr}\\
\href{https://doi.org/10.17182/hepdata.19445.v1}{HEPData Link}}
 & \makecell{$4<m_{\mu\mu}<5~\GeV$,\\ $5<m_{\mu\mu}<6~\GeV$,\\ $6<m_{\mu\mu}<7~\GeV$,\\ $7<m_{\mu\mu}<8~\GeV$}
 & \makecell{$\sigma_{p\mathrm{Fe}}/\sigma_{p\mathrm{Be}}$ and \\ $\sigma_{p\mathrm{W}}/\sigma_{p\mathrm{Be}}$ versus $x_1$}
 \\
 \bottomrule[1.2pt]
\end{tabular}
\caption{Summary of the fiducial selections and observables used in the comparisons with the E772 and E866 fixed-target DY measurements.
The datasets comprise the E772 measurement of per-nucleon cross-section ratios of carbon, calcium, iron and tungsten targets to deuterium~\cite{Alde:1990im} and the E866/NuSea measurement of the per-nucleon ratios of iron and tungsten to beryllium~\cite{NuSea:1999egr}, both taken with the $800~\GeV$ proton beam at Fermilab.
For each dataset, a link to the corresponding HEPData record is also provided.
The momentum fractions $x_1$ and $x_2$ refer to the beam proton and the target nucleon, respectively, and $x_F$ denotes the Feynman variable, as defined in section~\ref{sec:kinematics}.}
\label{tab:FT_obs}
\end{table}

Two ratio observables built from the pseudorapidity and rapidity distributions are used throughout section~\ref{sec:AP}.
The forward-to-backward ratio is defined as
\begin{equation}
\label{eq:RFB}
R_{\mathrm{FB}}\!\left(\xi\right)
=\frac{\mathrm{d} \sigma_{\mathrm{pPb}} /\left.\mathrm{d} \xi\right|_{+\xi}}{\mathrm{d} \sigma_{\mathrm{pPb}} /\left.\mathrm{d} \xi\right|_{-\xi}} ,
\qquad
\xi=\eta_{\ell}\;(W^{\pm}),
\quad 
\xi=y^Z_{\rm CM}\;(Z),
\end{equation}
and the $W$ charge asymmetry is defined as
\begin{equation}
\label{eq:A_pPb}
    \mathcal{A}_{\mathrm{pPb}}\!\left(\eta_\ell\right)=\frac{\mathrm{d} \sigma_{\mathrm{pPb}}^{W^{+}} / \mathrm{d} \eta_\ell-\mathrm{d} \sigma_{\mathrm{pPb}}^{W^{-}} / \mathrm{d} \eta_\ell}{\mathrm{d} \sigma_{\mathrm{pPb}}^{W^{+}} / \mathrm{d} \eta_\ell+\mathrm{d} \sigma_{\mathrm{pPb}}^{W^{-}} / \mathrm{d} \eta_\ell} ,
\end{equation}
both of which benefit from the cancellation of the luminosity uncertainty and a partial cancellation of other experimental and theoretical uncertainties~\cite{CMS:2015ehw,CMS:2015zlj,CMS:2019leu}.
The forward-to-backward ratio is expected to be sensitive to the $x$ dependence of nuclear modifications by comparing proton-going and lead-going (forward and backward) production, which sample the lead nucleus at typically smaller and larger momentum fractions, respectively~\cite{CMS:2015ehw,CMS:2015zlj,CMS:2019leu}.
The lepton charge asymmetry provides complementary sensitivity to the relative up- and down-quark distributions; in particular, measurements in the lead-going region probe the down-to-up quark ratio in a nucleon bound inside the lead nucleus~\cite{CMS:2015ehw,CMS:2019leu}.
The nuclear modifications of the PDFs mostly cancel in this quantity~\cite{Paukkunen:2010qg,CMS:2019leu}, which nevertheless retains sensitivity to the free-proton PDFs, so that its interpretation involves both the proton flavour structure and residual flavour-dependent nuclear modifications~\cite{Eskola:2022rlm}.
We compute both observables at NNLO in the frames and fiducial regions of the respective measurements documented in table~\ref{tab:AP_obs}, with $R_{\mathrm{FB}}$ evaluated separately for each $W$ charge.

The central renormalisation and factorisation scales are set to the invariant mass of the electroweak final state, $\mu_0=m_{\ell\nu}$ for charged-current production and $\mu_0=m_{\ell\ell}$ for neutral-current production.
For absolute differential cross sections, the perturbative uncertainty is estimated with the standard seven-point variation
\begin{equation}
\label{eq:scale7}
\mathcal{V}_7=
\left\{(1,1),\left(\tfrac12,\tfrac12\right),(2,2),
\left(\tfrac12,1\right),\left(1,\tfrac12\right),(1,2),(2,1)\right\},
\qquad
(\mu_R,\mu_F)=(\kappa_R,\kappa_F)\mu_0 .
\end{equation}
For a fixed-target ratio of numerator $N$ over denominator $D$, $N/D$, where $N$ and $D$ are the per-nucleon cross sections for the target and reference nucleus, respectively, a fully correlated variation would cancel much of the scale dependence common to the numerator and denominator.
We therefore use the partially correlated 15-point prescription: the renormalisation-scale multiplier is common, while the two factorisation-scale multipliers are varied separately subject to the constraints below
\begin{equation}
\label{eq:scale15}
\mathcal{V}_{15}=
\left\{(\kappa_R,\kappa_F^N,\kappa_F^D):
(\kappa_R,\kappa_F^N),(\kappa_R,\kappa_F^D)\in\mathcal{V}_7,\;
\frac{1}{2}\leq\frac{\kappa_F^N}{\kappa_F^D}\leq2\right\}.
\end{equation}
The constraint removes combinations containing scale multipliers that differ by a factor of four and leaves 15 tuples.
The ratio is evaluated for each complete tuple before the envelope is formed.
For the forward--backward ratio in eq.~\eqref{eq:RFB}, the two factorisation-scale multipliers apply to the forward and backward cross sections, respectively.
For the charge asymmetry in eq.~\eqref{eq:A_pPb}, they apply instead to the $W^+$ and $W^-$ cross sections, respectively; for each scale combination, the two varied cross sections enter both the difference in the numerator and the sum in the denominator.
A correlated variation of $\mu_R$ combined with an independent variation of the two factorisation scales subject to the same constraint has been used for ratios of $W$- and $Z$-boson transverse-momentum distributions in refs.~\cite{Bizon:2019zgf,Buonocore:2024xmy}.

\subsection{Drell--Yan production in proton--lead (pPb) collisions at the LHC}
\label{sec:AP}

All predictions presented in this section are computed directly within \texttt{NNLOJET}, using a common numerical setup.
The electroweak input parameters are specified in the $G_\mu$ scheme, with $M_Z=91.1876~\GeV$, $M_W=80.379~\GeV$, $\Gamma_Z=2.4952~\GeV$, $\Gamma_W=2.085~\GeV$ and $G_F=1.1663787\times10^{-5}~\GeV^{-2}$, together with a diagonal CKM matrix.
The proton beam is described by the \texttt{CT18NNLO\_as\_0118 } PDF set~\cite{Hou:2019efy} and the lead beam by the \texttt{EPPS21nlo\_CT18Anlo\_Pb208} nuclear PDF set~\cite{Eskola:2021nhw} with $\alpha_s(M_Z)=0.118$. 
Here, the \texttt{EPPS21} nuclear PDF set provides an established reference for validating the collider implementation against published predictions: the EPPS nuclear PDFs have been compared with the LHC $p\mathrm{Pb}$ collision data from Run~1 and Run~2 in the form of forward-to-backward and nuclear-modification ratios in the EPPS16 and EPPS21 analyses~\cite{Eskola:2016oht,Eskola:2021nhw} and in the form of absolute lepton-pseudorapidity distributions in refs.~\cite{Eskola:2022rlm,Flore:2025ync,Klasen:2023uqj}.
We have cross-checked our NLO predictions against published results, finding sub-percent agreement for the fiducial setup based on CMS Run 2 $W^\pm$ measurements.
The same PDF sets and strong-coupling input are used at LO, NLO and NNLO, so that the order-by-order comparison isolates the successive corrections to the partonic coefficient functions.
A dedicated comparison with the LHC data using the recent \texttt{nCTEQ26} nuclear PDFs~\cite{nCTEQ:2026ewr}, including a $\chi^2$ assessment of the impact of NNLO corrections and their interplay with the perturbative order of the PDFs, is presented in section~\ref{sec:grid-applications}.

\subsubsection{Charged-current Drell-Yan production}
\label{sec:AP-W}

\begin{figure}[tbp]
    \centering
    \includegraphics[width=0.49\textwidth]{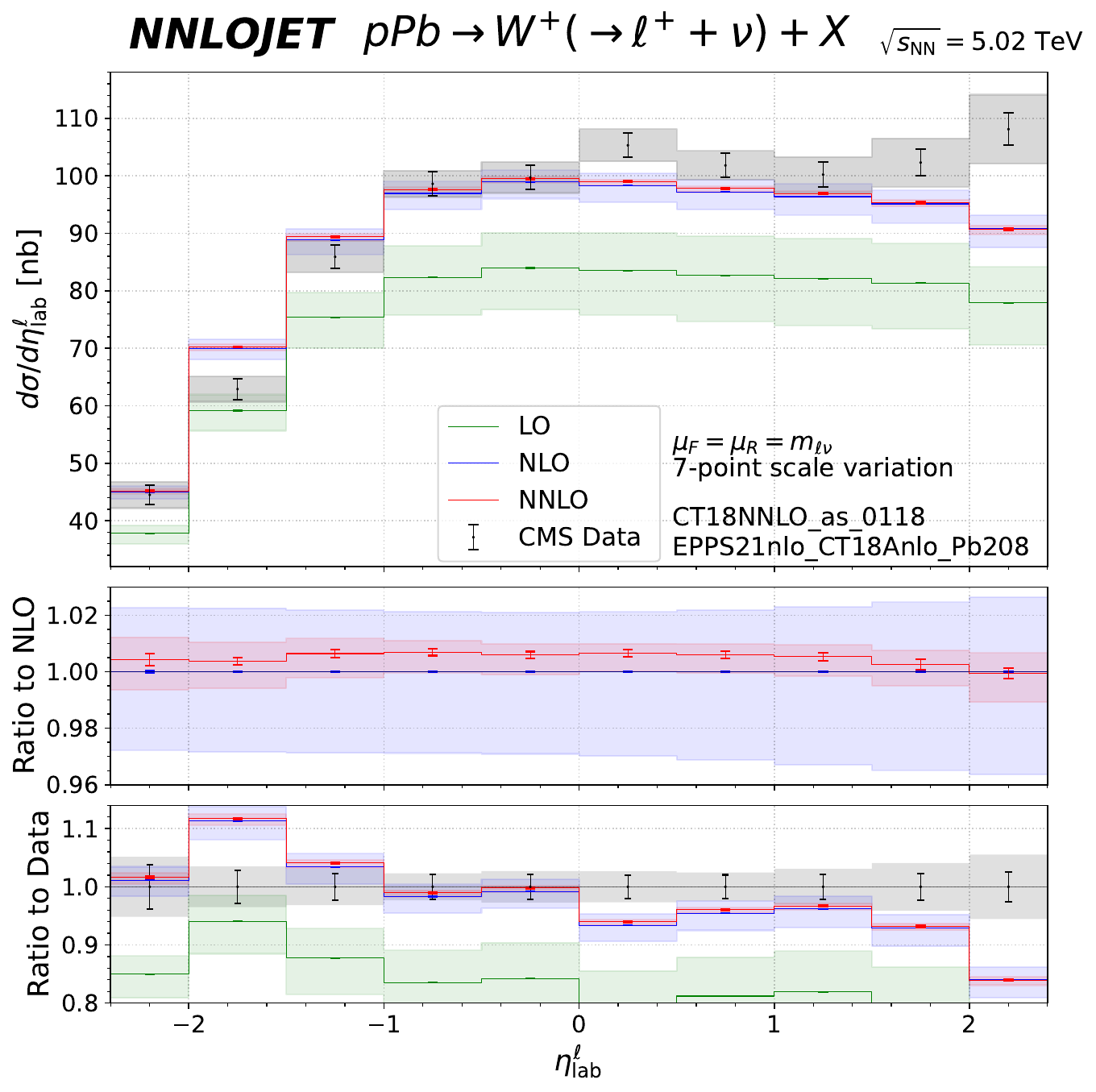}
    \includegraphics[width=0.49\textwidth]{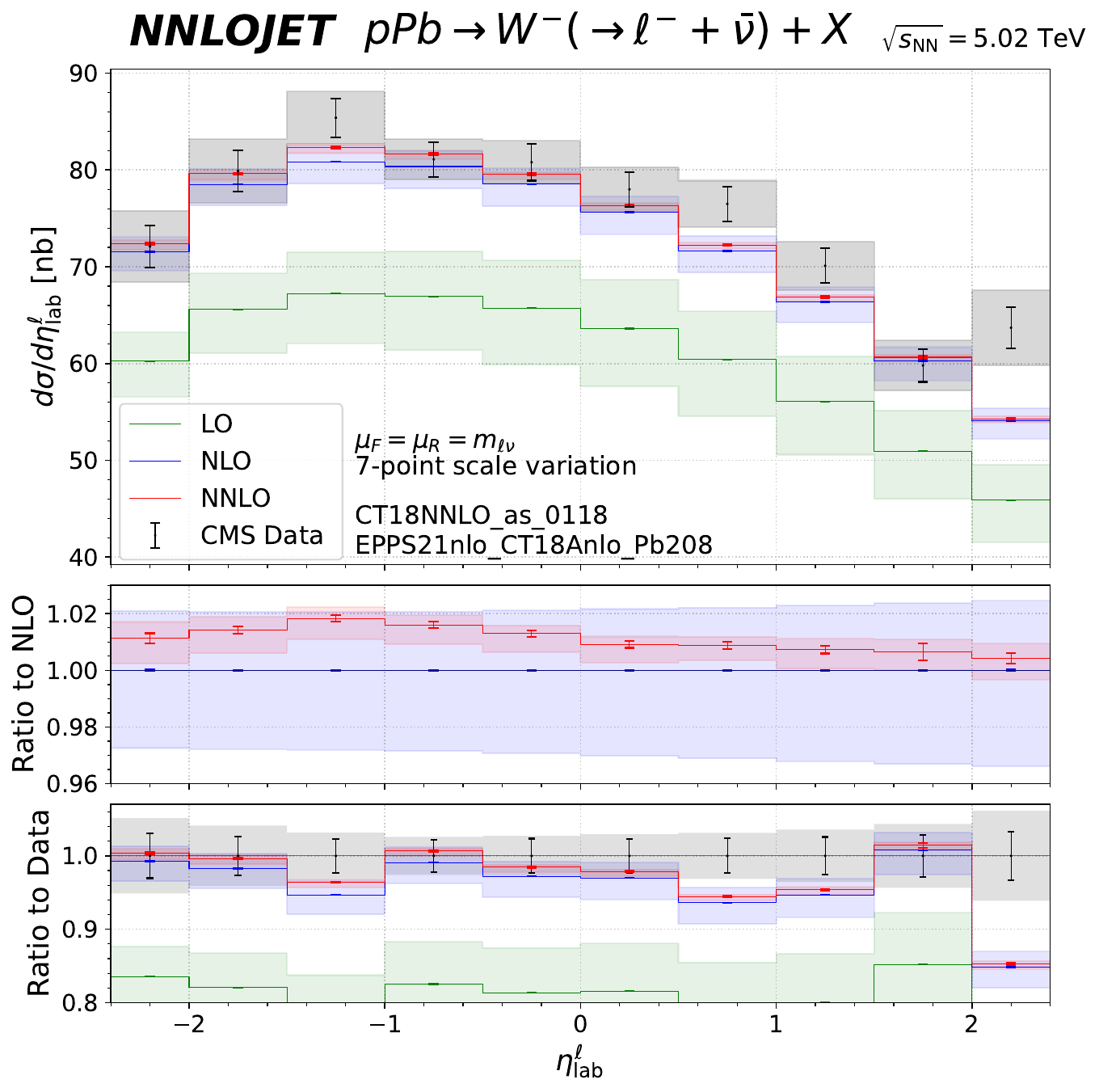}
    \includegraphics[width=0.49\textwidth]{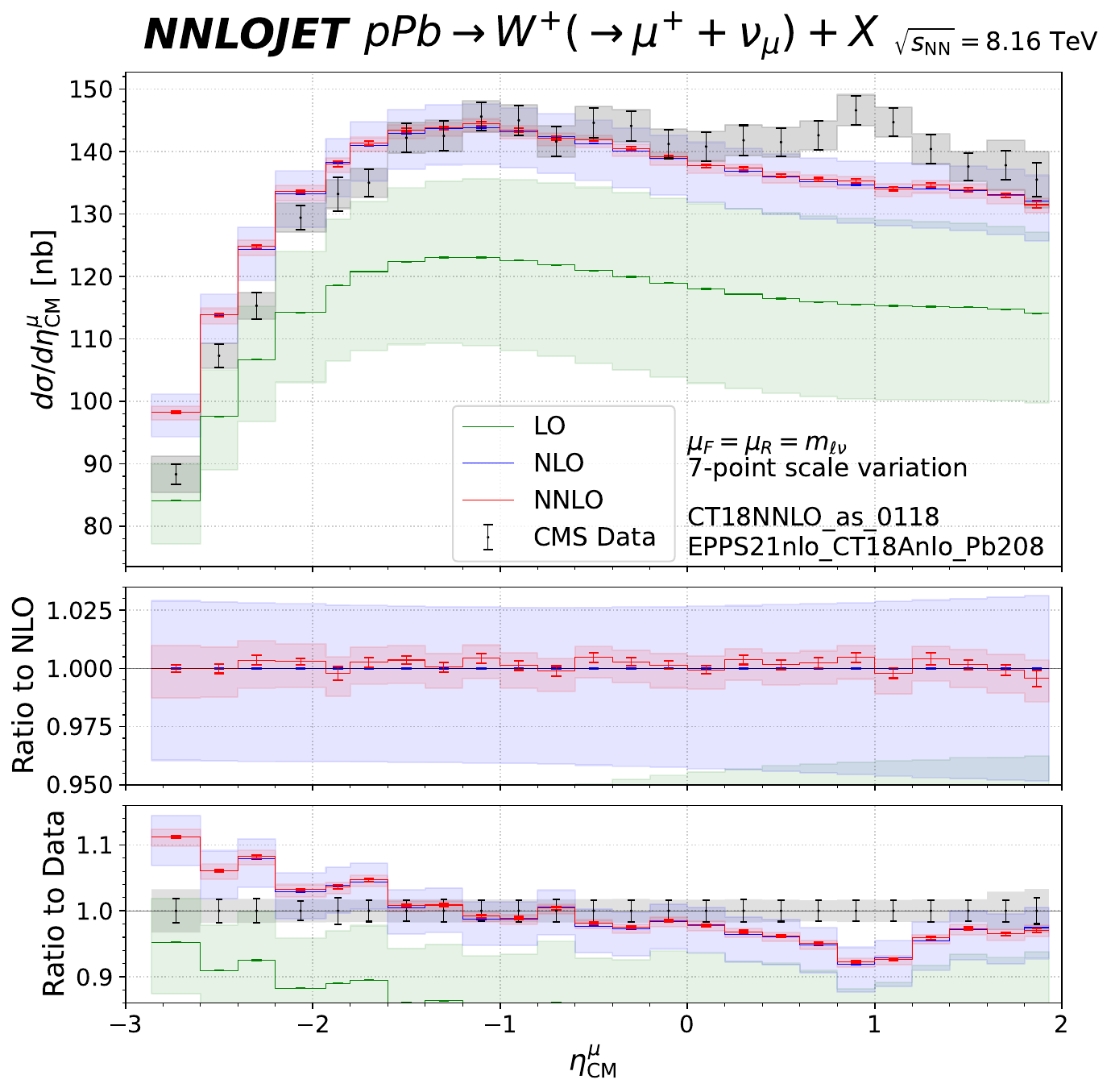}
    \includegraphics[width=0.49\textwidth]{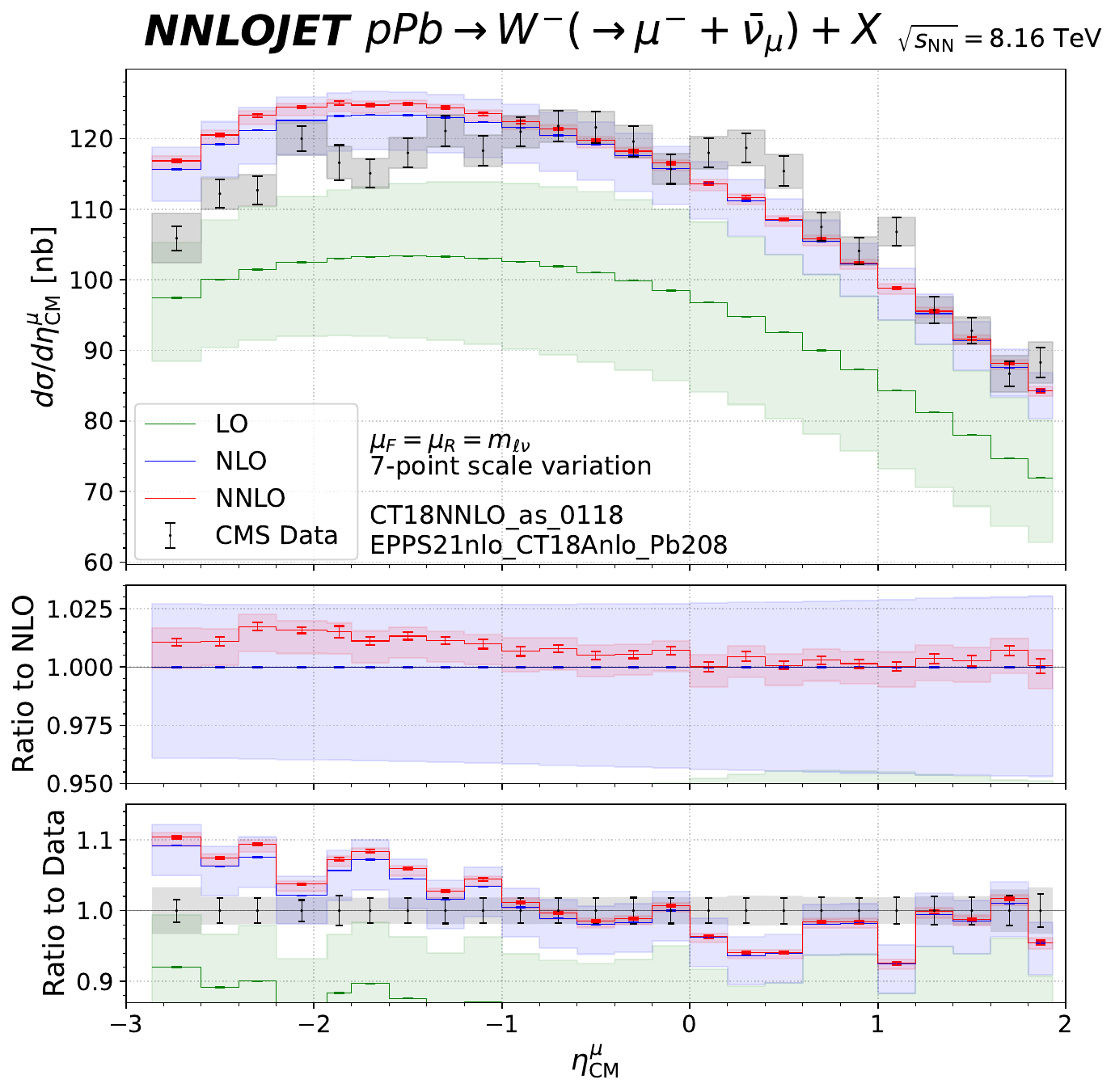}
\caption{Differential cross sections for $p\mathrm{Pb}\to W^{\pm}+X\to\ell\nu+X$ as functions of the charged-lepton pseudorapidity, compared with the \href{https://www.hepdata.net/record/ins1353541}{CMS Run~1} measurements at $\sqrt{s_{\mathrm{NN}}}=5.02~\TeV$~\cite{CMS:2015ehw} (upper row) and the \href{https://www.hepdata.net/record/ins1733223}{CMS Run~2} measurements at $\sqrt{s_{\mathrm{NN}}}=8.16~\TeV$~\cite{CMS:2019leu} (lower row), for $W^{+}$ (left) and $W^{-}$ (right).
The predictions are evaluated up to NNLO QCD using the \texttt{CT18NNLO\_as\_0118} and \texttt{EPPS21nlo\_CT18Anlo\_Pb208} PDF sets at $\mu_F=\mu_R=m_{\ell\nu}$. 
The coloured bands show the seven-point scale-variation uncertainty of eq.~\eqref{eq:scale7} and the error bars on the predictions the numerical integration uncertainties, while the grey bands and error bars on the data denote systematic and statistical uncertainties, respectively.
The middle and lower panels show the predictions normalised to the central NLO result and to the CMS data, respectively,
}
\label{fig:AP2W_eta-dist}
\end{figure}

\begin{figure}[tbp]
    \centering
    \includegraphics[width=0.49\textwidth]{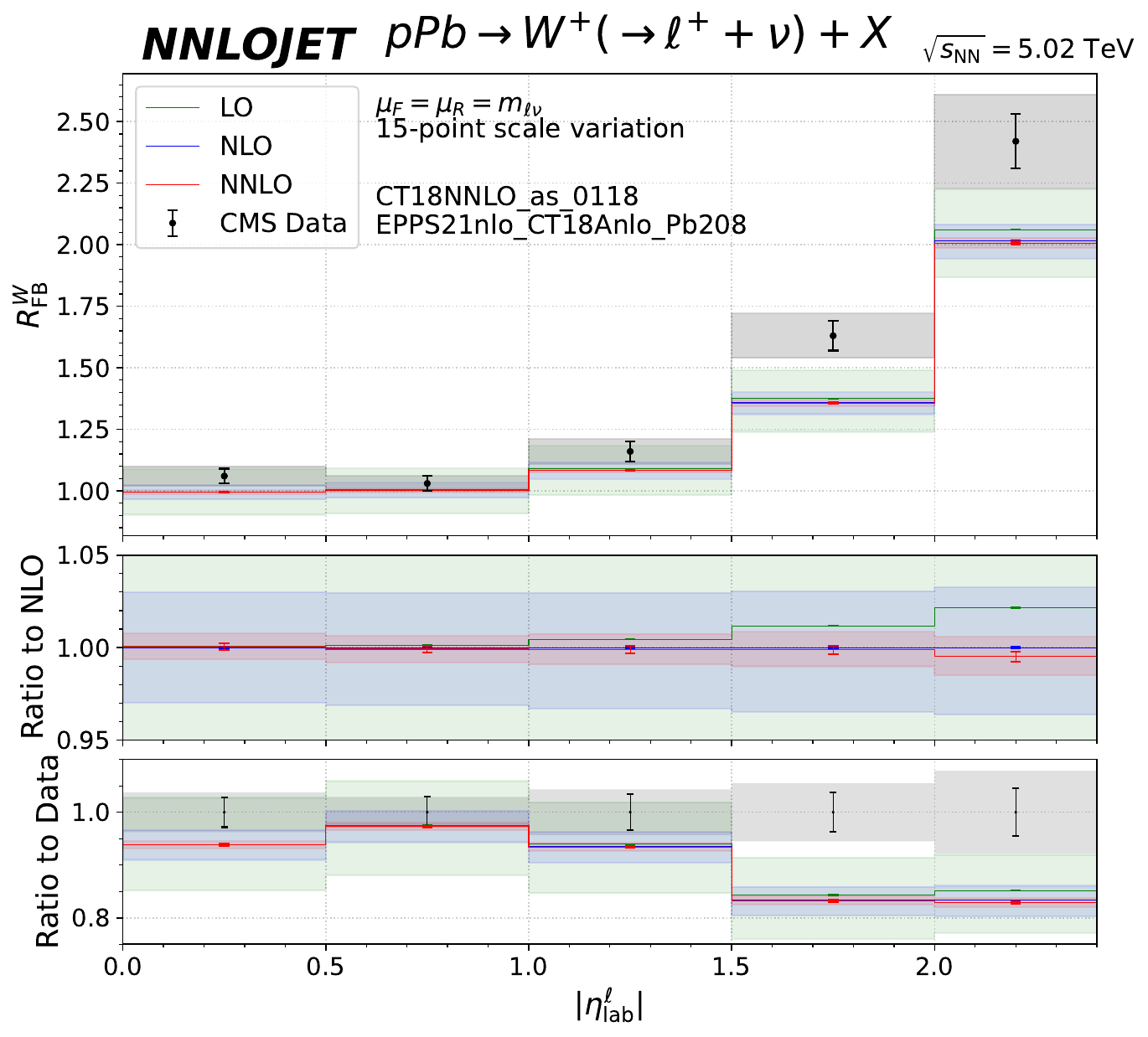}
    \includegraphics[width=0.49\textwidth]{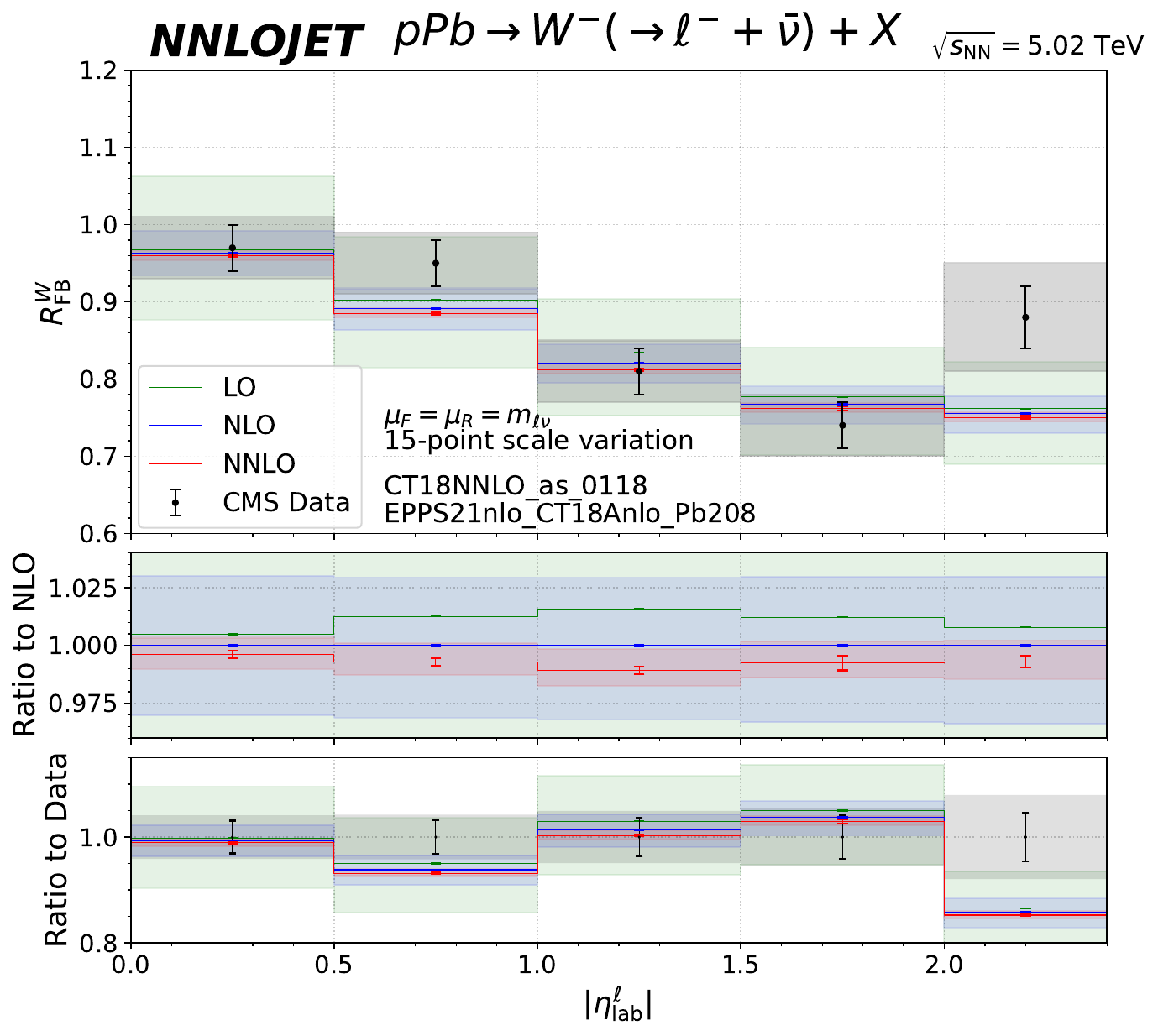}
    \includegraphics[width=0.49\textwidth]{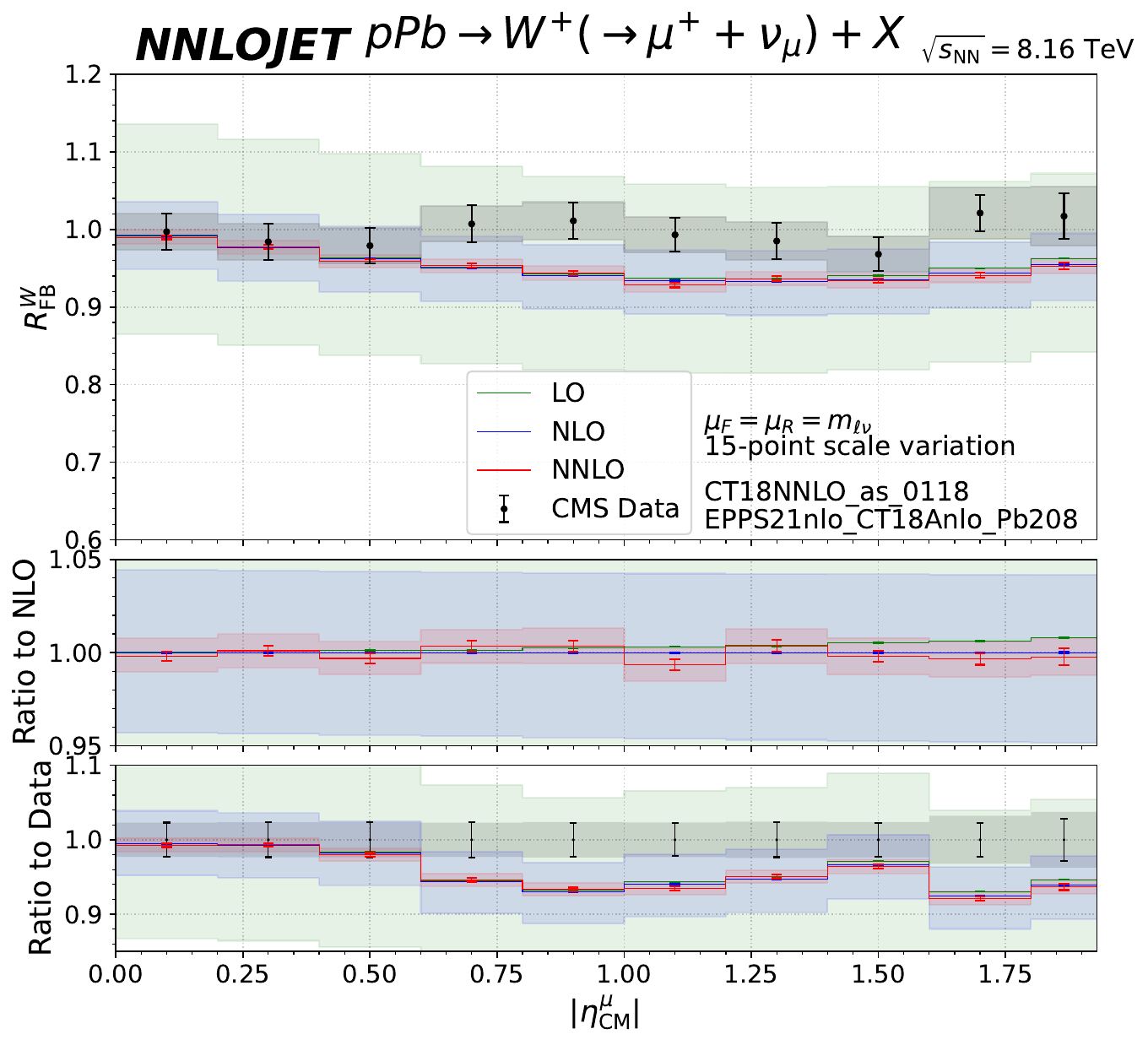}
    \includegraphics[width=0.49\textwidth]{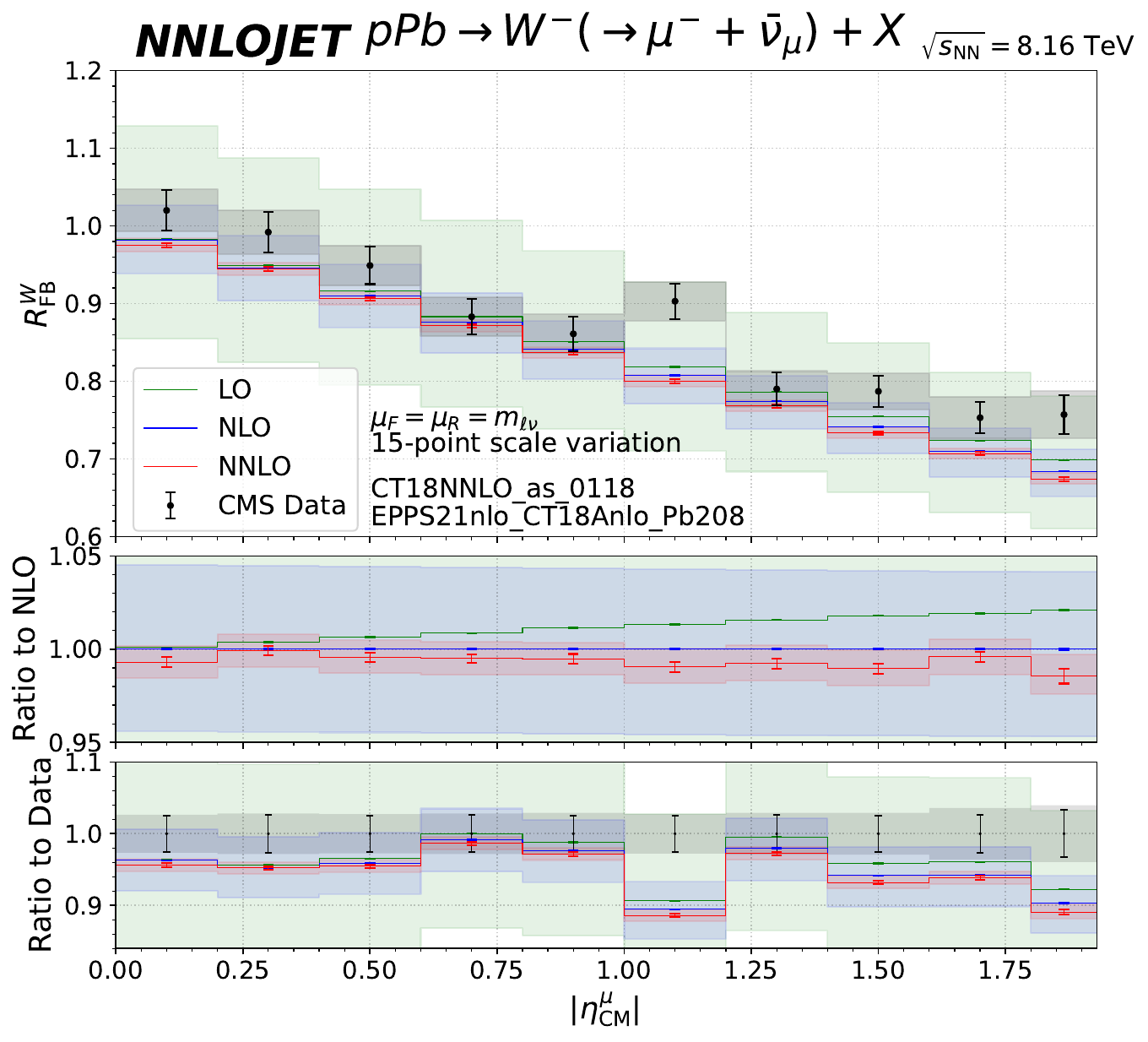}
\caption{Same as figure~\ref{fig:AP2W_eta-dist}, but for the forward--backward ratios $R_{\mathrm{FB}}^{W^{\pm}}$, defined in eq.~\eqref{eq:RFB}, as functions of the absolute charged-lepton pseudorapidity.
The coloured bands show the partially correlated 15-point scale-variation uncertainty of eq.~\eqref{eq:scale15}.
}
\label{fig:AP2W_RFB}
\end{figure}

\begin{figure}[tbp]
    \centering
    \includegraphics[width=0.49\textwidth]{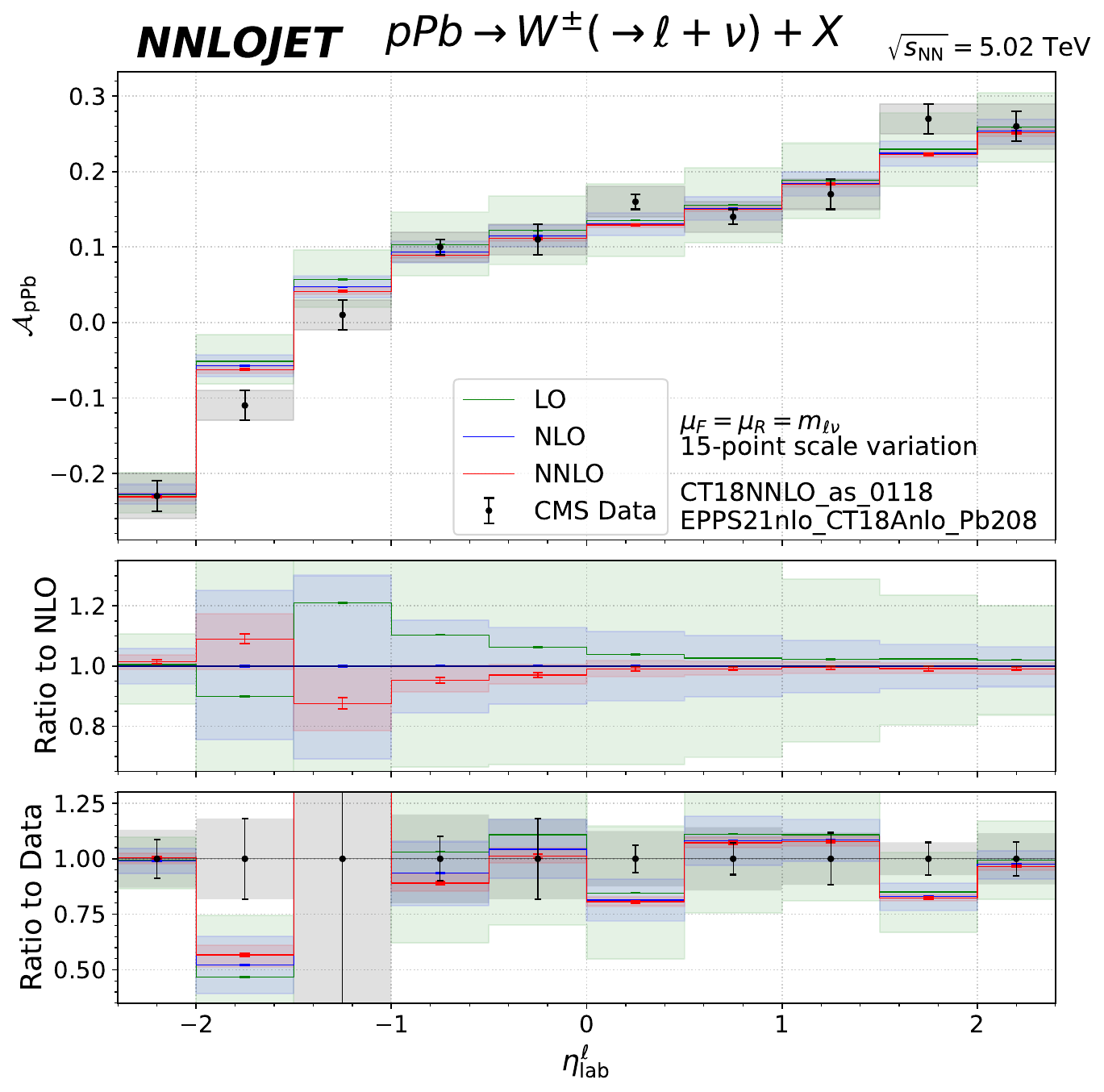}
    \includegraphics[width=0.49\textwidth]{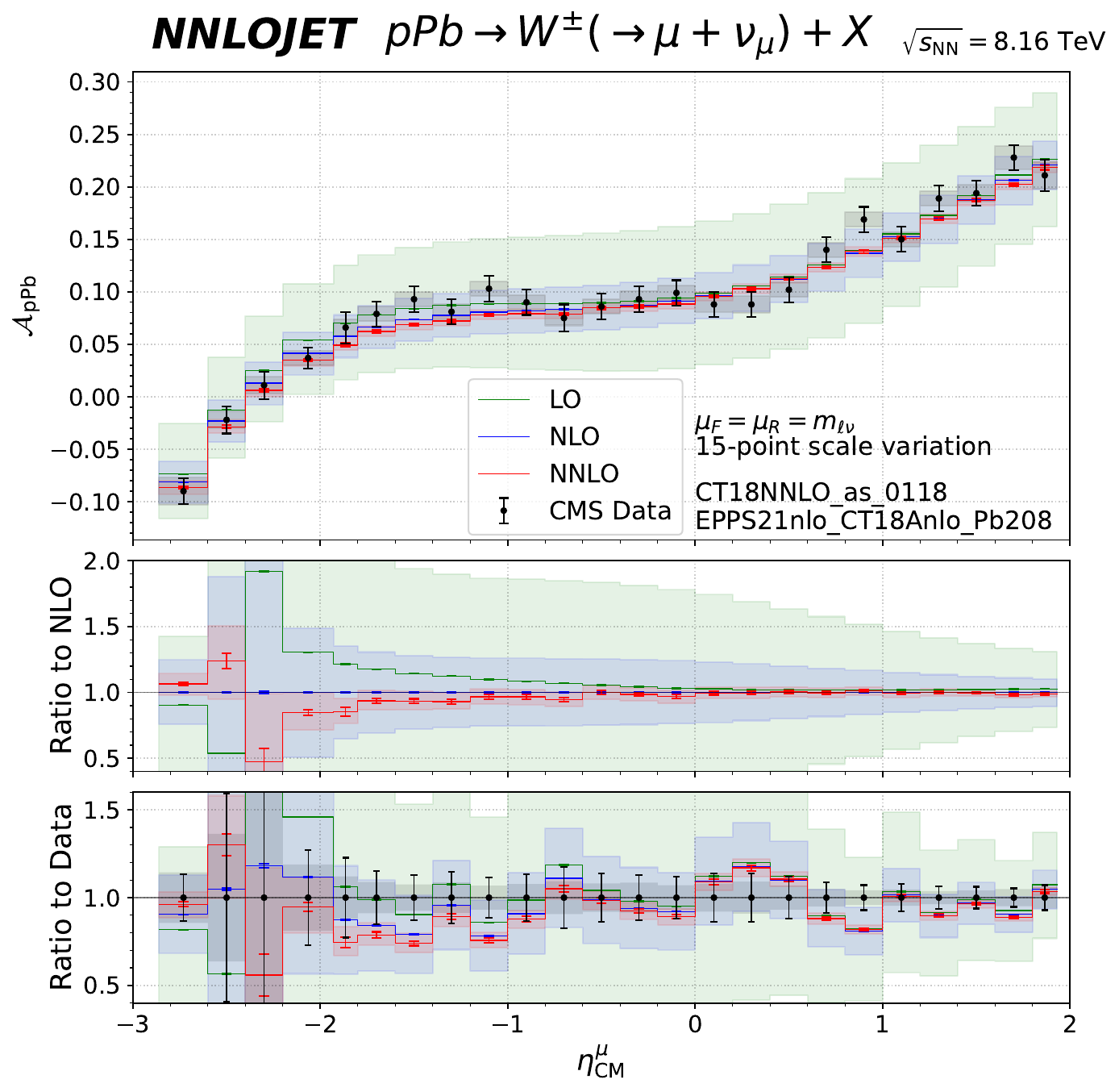}
\caption{Same as figure~\ref{fig:AP2W_RFB}, but for the lepton charge asymmetry $\mathcal{A}_{\mathrm{pPb}}$, defined in eq.~\eqref{eq:A_pPb}, as a function of the charged-lepton pseudorapidity.
The left column corresponds to CMS Run~1, with the pseudorapidity $\eta^{\ell}_{\rm lab}$ defined in the laboratory frame, and the right column to CMS Run~2, with the pseudorapidity $\eta^{\mu}_{\rm CM}$ defined in the nucleon--nucleon CM frame. 
}
\label{fig:AP2W_charge_Asy}
\end{figure}

Figure~\ref{fig:AP2W_eta-dist} shows the differential cross sections for $W^{+}$ and $W^{-}$ production as functions of the charged-lepton pseudorapidity up to NNLO QCD, compared with CMS Run~1 measurements at $\sNN=5.02~\TeV$~\cite{CMS:2015ehw} and the CMS Run~2 measurements at $\sNN=8.16~\TeV$~\cite{CMS:2019leu}.
Fixed-order predictions obtained with the \texttt{CT18NNLO\_as\_0118}~\cite{Hou:2019efy} and \texttt{EPPS21nlo\_CT18Anlo\_Pb208}~\cite{Eskola:2021nhw} PDF sets in the laboratory frame for Run~1 and the nucleon--nucleon CM frame for Run~2, at the dynamical central scale $\mu_F=\mu_R=m_{\ell\nu}$ are shown, with the middle panels showing the predictions normalised to the central NLO result and the lower panels their ratios to the CMS experimental data.
The coloured bands represent the seven-point scale variation of eq.~\eqref{eq:scale7} and the error bars on the fixed-order predictions indicate the numerical integration uncertainties; the grey bands on the CMS data denote the systematic uncertainty and the error bars the statistical uncertainty.
In this figure, the vertical ranges of the middle and lower panels are restricted so as to resolve the NLO and NNLO predictions; the LO predictions, shown in full in the upper panels, may therefore lie partly or entirely outside these ranges, and bands and error bars extending beyond them are truncated.
The same treatment applies to figures~\ref{fig:AP2W_RFB}--\ref{fig:FTDY_E866_WoBE}.
In the numerical discussion below, the quoted scale uncertainty is defined by assigning the maximum variation among the upward and downward deviations symmetrically from the central value, whereas the plotted scale bands retain the original, generally asymmetric envelopes; the NLO and NNLO corrections are quoted relative to the central LO and NLO predictions, respectively.

Both charges exhibit broad maxima, with the $W^{-}$ maximum at more backward pseudorapidities and a steeper decline towards the forward edge than for $W^{+}$.
The charge-dependent shapes reflect the different quark-flavour combinations and leptonic decay kinematics, together with the frame and fiducial selections~\cite{CMS:2015ehw,CMS:2019leu}.
The four distributions exhibit a similar perturbative behaviour.
The NLO corrections enhance the LO predictions by $16$--$20\%$ for $W^{\pm}$ across the two collision energies, well outside the LO scale bands, as seen in the upper panels.
The subsequent NNLO corrections are considerably smaller: they remain within $0.7\%$ of NLO for $W^{+}$ at both energies, whereas for $W^{-}$ they are positive throughout the measured regions and reach $1.8\%$ at $5.02~\TeV$ and $1.7\%$ at $8.16~\TeV$ in the backward hemisphere.
The bin-to-bin fluctuations of the NNLO-to-NLO ratio visible at $8.16~\TeV$ are compatible with the numerical integration uncertainties of $0.1$--$0.4\%$.
The NNLO central predictions lie within the NLO scale bands over the entire regions, consistent with improved perturbative stability.
At $5.02~\TeV$ ($8.16~\TeV$), the scale uncertainty decreases from $5$--$10\%$ ($8$--$13\%$) at LO to $2.7$--$3.6\%$ ($3.9$--$4.8\%$) at NLO and $0.5$--$1.1\%$ ($0.8$--$1.3\%$) at NNLO, which clearly demonstrate that the scale dependence decreases substantially with increasing perturbative order, as visible from the widths of the bands in the upper and middle panels of figure~\ref{fig:AP2W_eta-dist}.

The NNLO predictions reproduce the broad shapes of the measured distributions, but rapidity-dependent differences between the data and the central predictions remain, as shown in the lower panels of figure~\ref{fig:AP2W_eta-dist}.
These differences are quoted below in units of the experimental uncertainty $\sigma_{\rm exp}$, the quadrature sum of the statistical and systematic uncertainties of each data point, which amounts to $3$--$7\%$ at $5.02~\TeV$ and to $2$--$4\%$ at $8.16~\TeV$; the luminosity uncertainty of $3.5\%$~\cite{CMS:2015ehw,CMS:2019leu}, common to all bins of a measurement, is not included.
At $5.02~\TeV$, the $W^{+}$ prediction exceeds the data by $2.6\sigexp$ in $-2<\eta^{\ell}_{\rm lab}<-1.5$ and agrees with them within about $1\sigexp$ elsewhere in the backward hemisphere, while the largest deficits, of $2.6\sigexp$ for $W^{+}$ and $2.1\sigexp$ for $W^{-}$, occur in the extreme forward region; in all other bins the predictions differ from the data by less than $2\sigexp$.
At $8.16~\TeV$, the predictions for both charges lie above the data towards the backward edge of the acceptance, with the largest excesses reaching about $3.2\sigma_{\rm exp}$ for $W^{+}$ and $3.6\sigma_{\rm exp}$ for $W^{-}$.
The largest forward deficits occur around $\eta^\mu_{\rm CM}\simeq1$, where the data exhibit local enhancements: they reach about $3.2$--$3.3\sigma_{\rm exp}$ in $0.8<\eta^{\mu}_{\rm CM}<1.2$ for $W^{+}$ and $2.8\sigma_{\rm exp}$ in $1.0<\eta^{\mu}_{\rm CM}<1.2$ for $W^{-}$. 
Smaller deficits of about $2.2\sigma_{\rm exp}$ occur in $0.6<\eta^\mu_{\rm CM}<0.8$ for $W^+$ and $0.2<\eta^\mu_{\rm CM}<0.6$ for $W^-$; elsewhere the forward differences remain below $2\sigma_{\rm exp}$.

Backward and forward production probe, on average, larger and smaller momentum fractions in the lead nucleus, respectively, making these distributions sensitive to the $x$ dependence of the nuclear PDFs~\cite{CMS:2019leu}.
The residual differences cannot, however, be assigned uniquely to nuclear modifications:
the displayed theory bands include scale uncertainties but no proton- or nuclear-PDF uncertainties, 
and an assessment of their significance must also account for the experimental correlations~\cite{Eskola:2022rlm}, which the bin-by-bin comparison above neglects.
\\

The forward--backward ratios $R_{\mathrm{FB}}^{W^{\pm}}$ of eq.~\eqref{eq:RFB} are shown in figure~\ref{fig:AP2W_RFB}, with the middle and lower panels displaying the ratios to the central NLO prediction and to the CMS data, respectively.
The shapes differ markedly between the two charges and collision energies.
In Run~1, where the ratio compares opposite laboratory pseudorapidities, the NNLO $W^{+}$ prediction rises from $0.995$ in $0<|\eta^{\ell}_{\rm lab}|<0.5$ to $2.01$ in $2<|\eta^{\ell}_{\rm lab}|<2.4$, reflecting a backward cross section that is only half the forward one at the edge of the laboratory acceptance.
In Run~2, where the ratio is instead formed in the nucleon--nucleon CM frame, the $W^+$ prediction remains below unity, decreasing from $0.99$ to a minimum of $0.93$ in $1.0<|\eta^\mu_{\rm CM}|<1.2$ before increasing overall to $0.95$ in the last bin.
The $W^{-}$ ratios decrease monotonically across both measured ranges, from $0.96$ to $0.75$ in Run~1 and from $0.98$ to $0.67$ in Run~2.
The large Run~1 $W^{+}$ ratio reflects the asymmetric CM-frame coverage of opposite laboratory bins, whereas the decrease of $R_{\mathrm{FB}}^{W^{-}}$ in both runs is consistent with enhanced backward production associated with the neutron excess of lead, with small-$x$ shadowing in the numerator and larger-$x$ antishadowing in the denominator potentially further reducing the ratios for both charges~\cite{CMS:2015ehw,Eskola:2022rlm}.

As can be seen from the middle panels of figure~\ref{fig:AP2W_RFB}, the perturbative corrections largely cancel between the forward and backward cross sections: the NLO corrections, which enhance the individual cross sections by $16$--$20\%$, reduce the ratios by at most $2.1\%$.
The NNLO corrections relative to NLO remain within $0.7\%$ for $W^{+}$ at both energies, while for $W^{-}$ they are negative throughout and reach $-1.1\%$ at $5.02~\TeV$ and $-1.4\%$ at $8.16~\TeV$, in the same direction as the NLO corrections.
The scale bands of successive orders are nested, each lying within the band of the preceding order in every bin.
With the partially correlated 15-point prescription of eq.~\eqref{eq:scale15}, the scale uncertainty decreases from approximately $9$--$10\%$ ($13$--$15\%$) at LO to $3.0$--$3.6\%$ ($4.4$--$4.8\%$) at NLO and $0.7$--$1.1\%$ ($0.9$--$1.2\%$) at NNLO for Run~1 (Run~2).

The lower panels of figure~\ref{fig:AP2W_RFB} compare the predictions with the CMS data from the two runs, which differ in collision energy and pseudorapidity-frame conventions.
The experimental uncertainty $\sigma_{\rm exp}$ amounts to $4$--$9\%$ in Run~1 and to $3$--$5\%$ in Run~2.
In Run~1, the largest departure occurs for $R_{\mathrm{FB}}^{W^+}$ in $1.5<|\eta^\ell_{\rm lab}|<2$, where the NNLO prediction lies about $2.5\sigma_{\rm exp}$ below the data; all other bins for both charges differ by less than $2\sigma_{\rm exp}$.
In Run~2, the predictions lie below the data throughout the measured range. For $W^+$ the deficits remain within about $2\sigma_{\rm exp}$, while for $W^-$ they reach $3.0\sigma_{\rm exp}$ in $1.0<|\eta^\mu_{\rm CM}|<1.2$ and $2.1\sigma_{\rm exp}$ in the last bin; the remaining $W^-$ differences are below $1.8\sigma_{\rm exp}$.
The larger data--theory differences exceed the NNLO corrections and are therefore not resolved by the higher-order corrections.
The cancellation of normalisation uncertainties is expected to enhance the sensitivity to the relative nuclear modifications probed in the two hemispheres, but a quantitative interpretation still requires the proton- and nuclear-PDF uncertainties and the experimental correlations~\cite{CMS:2019leu,Eskola:2022rlm}.
\\

The lepton charge asymmetry $\mathcal{A}_{\mathrm{pPb}}$ of eq.~\eqref{eq:A_pPb} is shown in figure~\ref{fig:AP2W_charge_Asy}, with the middle and lower panels displaying the ratios to the central NLO prediction and to the CMS data, respectively.
The NNLO predictions reproduce the overall increase of the asymmetry from negative values in the backward region to positive values in the forward region, rising from $-0.231$ to $0.251$ across the Run~1 acceptance and from $-0.087$ to $0.219$ in Run~2.
In the forward region, the NLO corrections lower the asymmetry by $2$--$4\%$ in Run~1 and by $1.7$--$2.6\%$ in Run~2, and the NNLO corrections relative to NLO remain within $1\%$ and $2\%$, respectively.
In the backward region, where the asymmetry changes sign, the relative corrections grow accordingly: away from the sign change, the NLO corrections reach $9\%$ ($23\%$) and the NNLO corrections reach $5\%$ ($15\%$), while in the two bins on either side of the sign change, the NNLO-to-NLO ratio is $1.090$ ($1.240$) and $0.876$ ($0.473$) in Run~1 (Run~2).
These relative changes are amplified by the small asymmetry in the denominator and do not imply comparably large corrections to the individual $W^{\pm}$ cross sections.

The scale dependence decreases substantially with increasing perturbative order.
The scale uncertainty, decreases in the forward region from $18$--$36\%$ ($28$--$70\%$) at LO to $6.9$--$11.5\%$ ($10.7$--$23.1\%$) at NLO and $1.8$--$2.8\%$ ($2.2$--$4.5\%$) at NNLO in Run~1 (Run~2); the scale bands of successive orders are nested in every bin.
In the backward region, the relative uncertainties are larger: away from the sign change, the NNLO uncertainty reaches $5\%$ ($18\%$), and in the two bins adjacent to it $9\%$ ($24\%$) and $13\%$ ($110\%$) in Run~1 (Run~2).
The relative uncertainties are likewise enhanced near the sign change for the same reason.

The NNLO predictions capture the overall behaviour of the measured asymmetries, but pseudorapidity-dependent differences remain.
In Run~1, all bins agree within $1.7\sigma_{\rm exp}$.
In Run~2, the largest difference is a deficit of $2.2\sigma_{\rm exp}$ in $0.8<\eta^\mu_{\rm CM}<1.0$, while all other bins differ by less than $1.9\sigma_{\rm exp}$.
Large ratios near the sign change do not imply correspondingly large discrepancies: the Run~1 value $\mathcal{A}_{\rm NNLO}/\mathcal{A}_{\rm data}\simeq4.1$ in $-1.5<\eta^\ell_{\rm lab}<-1$, outside the displayed lower-panel range, corresponds to only $1.1\sigma_{\rm exp}$ because the measured asymmetry is compatible with zero.

Nuclear modifications largely cancel in the charge asymmetry~\cite{CMS:2019leu,Paukkunen:2010qg}, while sensitivity to the free-proton flavour structure remains, particularly to the valence-quark difference $u_{\rm V}-d_{\rm V}$ at large positive rapidities~\cite{Kusina:2016fxy,Eskola:2022rlm}.
The residual differences therefore cannot be attributed uniquely to nuclear modifications; interpreting them requires both proton- and nuclear-PDF uncertainties alongside experimental correlations~\cite{CMS:2019leu,Eskola:2022rlm}.

\subsubsection{Neutral-current Drell-Yan production}
\label{sec:AP-Z}

\begin{figure}[tbp]
    \centering
    \includegraphics[width=0.49\textwidth]{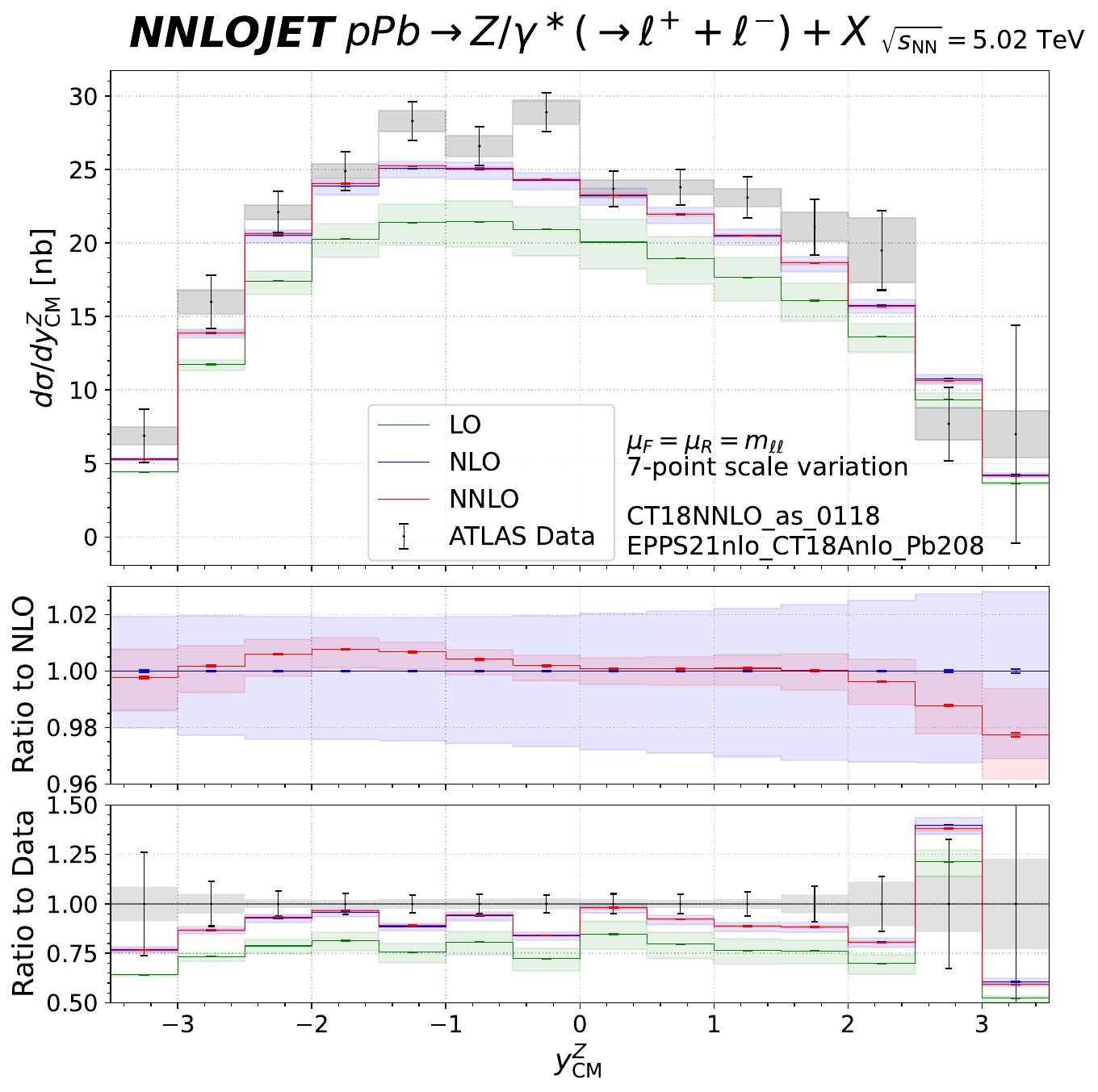}
    \includegraphics[width=0.49\textwidth]{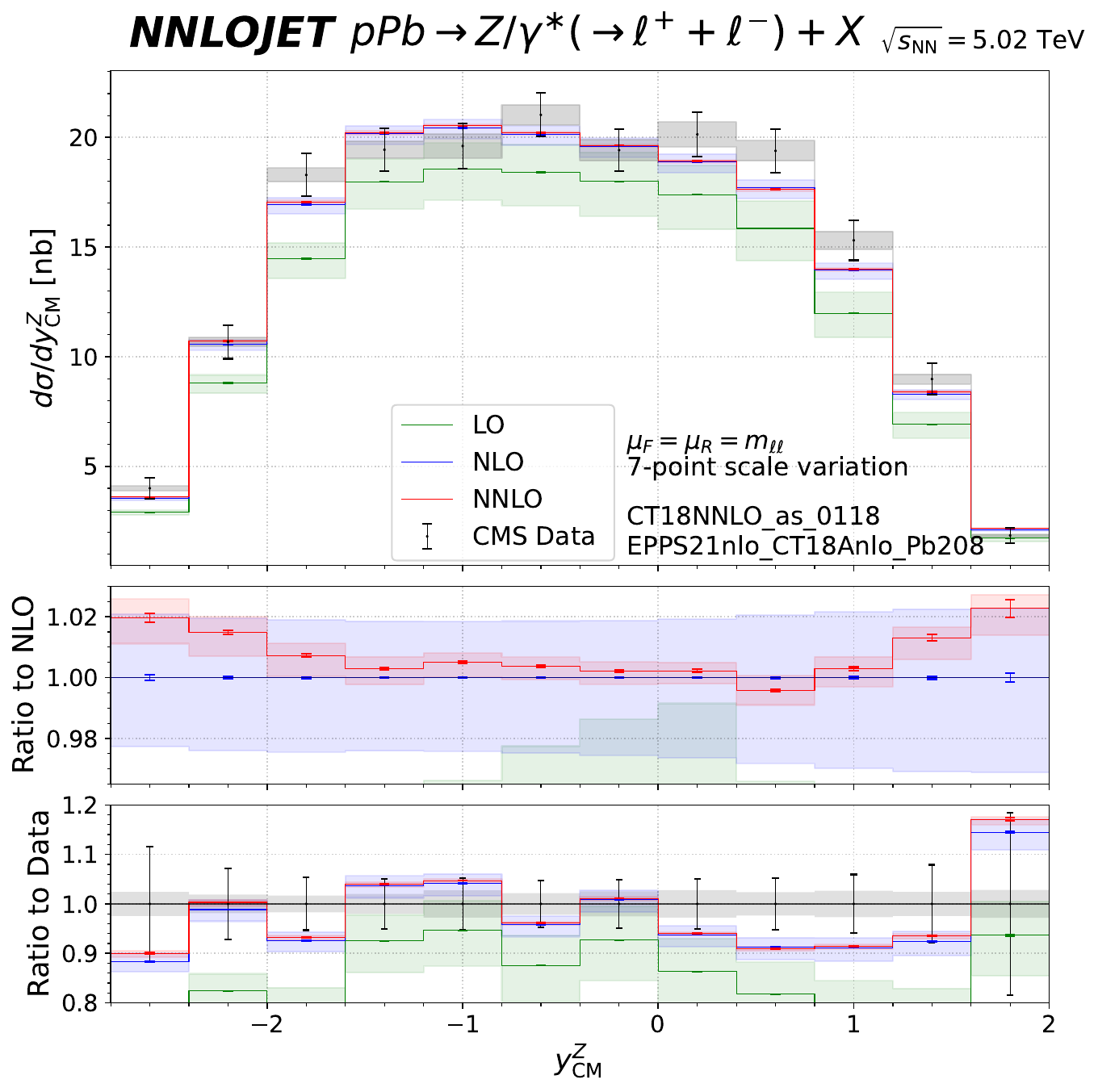}
    \includegraphics[width=0.49\textwidth]{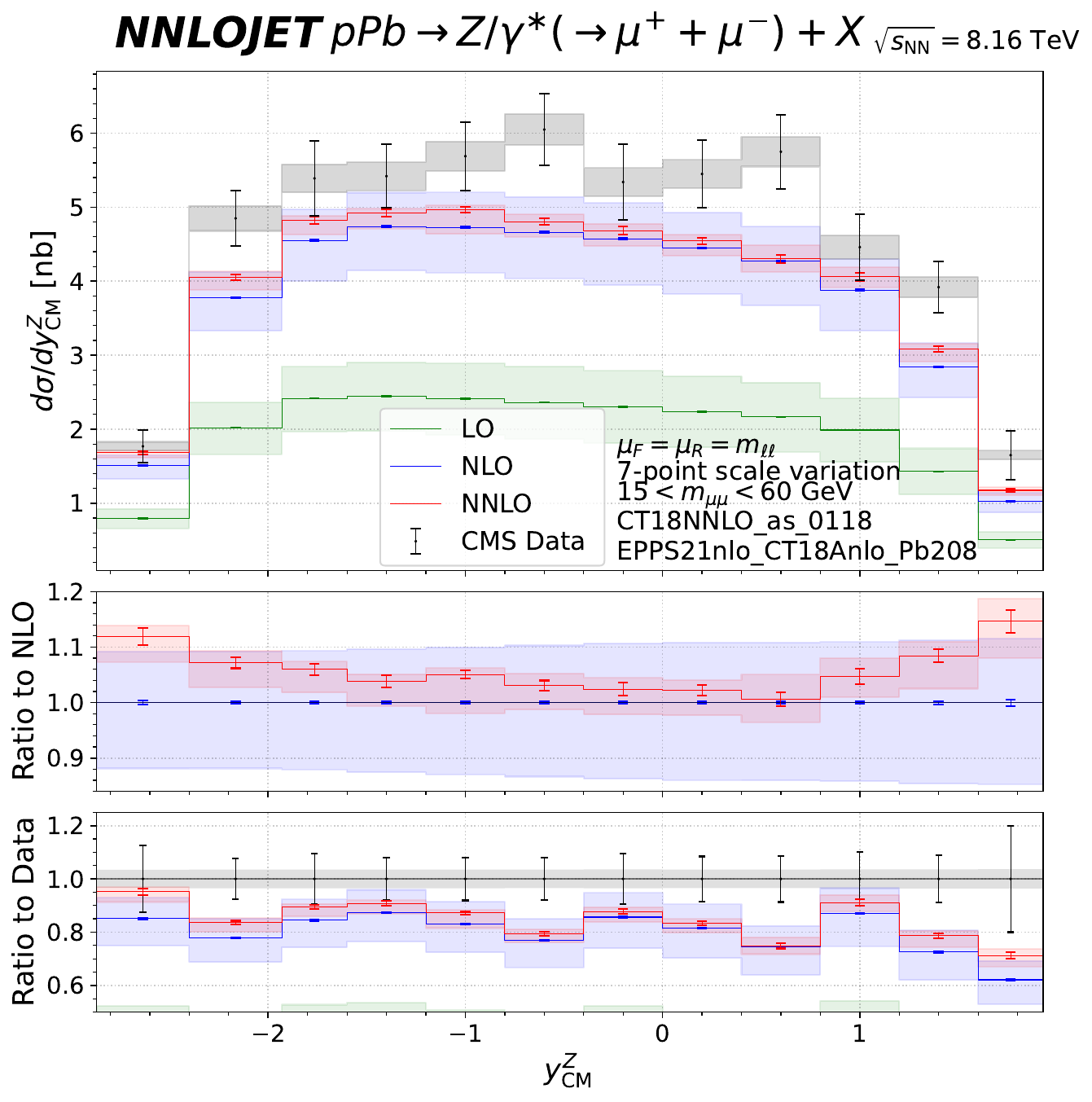}
    \includegraphics[width=0.49\textwidth]{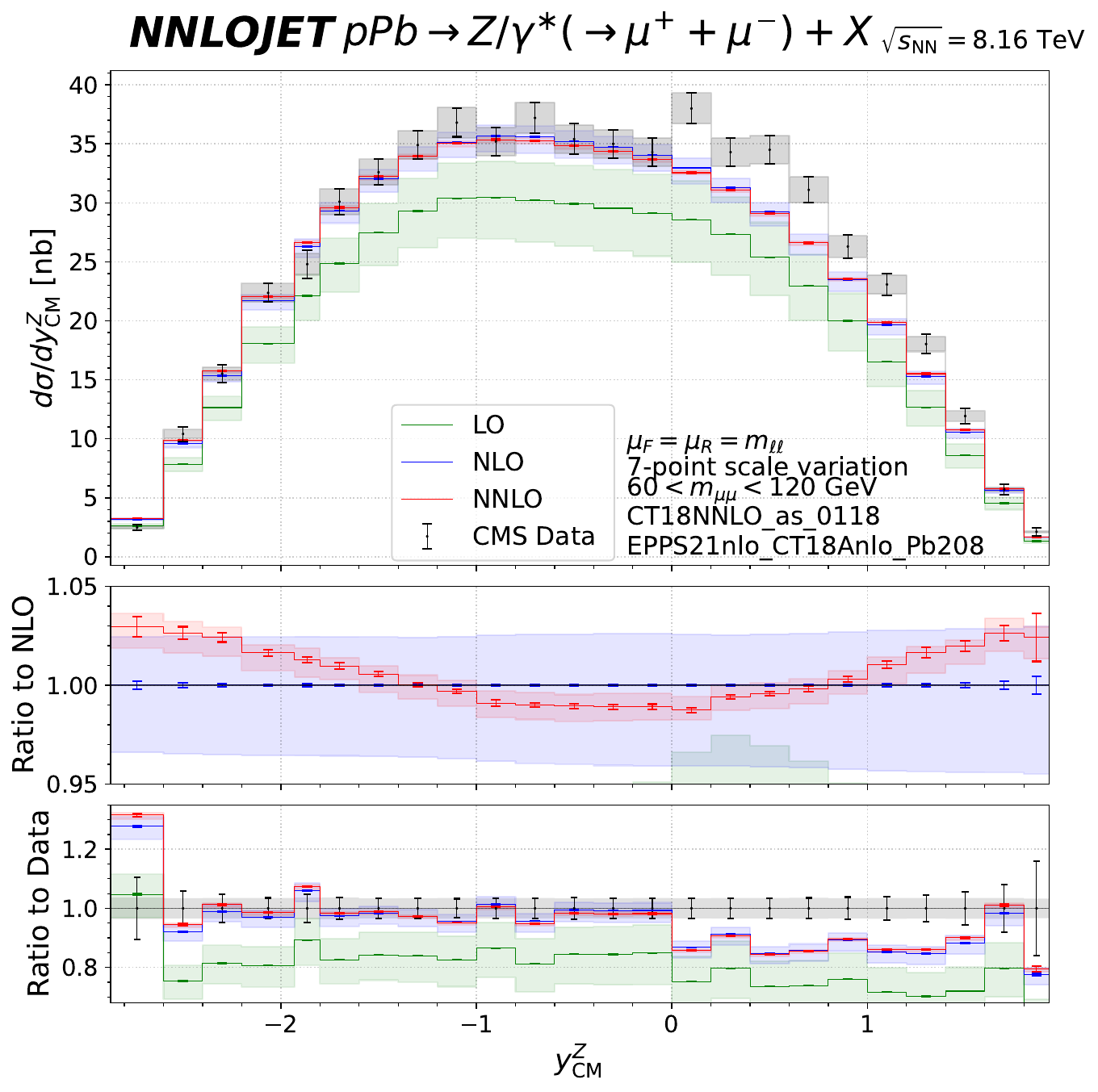}
\caption{Same as figure~\ref{fig:AP2W_eta-dist}, but for $p\mathrm{Pb}\to Z/\gamma^{*}+X\to\ell^{+}\ell^{-}+X$ as a function of the dilepton rapidity $y^{Z}_{\rm CM}$ in the nucleon--nucleon CM frame, evaluated at the central scale $\mu_F=\mu_R=m_{\ell\ell}$ and compared with the \href{https://www.hepdata.net/record/ins1384272}{ATLAS Run~1} and \href{https://www.hepdata.net/record/ins1410832}{CMS Run~1} measurements at $\sqrt{s_{\mathrm{NN}}}=5.02~\TeV$~\cite{ATLAS:2015mwq,CMS:2015zlj} and the \href{https://www.hepdata.net/record/ins1849180}{CMS Run~2} measurements at $\sqrt{s_{\mathrm{NN}}}=8.16~\TeV$~\cite{CMS:2021ynu}.
For the CMS Run~2 data, the error bars denote the statistical and systematic uncertainties added in quadrature and the grey bands the luminosity uncertainty of $3.5\%$.
The upper row corresponds to ATLAS Run~1 (left) and CMS Run~1 (right); the lower row corresponds to the low-mass window $15<m_{\mu\mu}<60~\GeV$ (left) and high-mass window $60<m_{\mu\mu}<120~\GeV$ (right) at CMS Run~2.
}
\label{fig:AP2Z_yz-dist}
\end{figure}

Figure~\ref{fig:AP2Z_yz-dist} shows the differential cross sections for neutral-current DY production as functions of the dilepton rapidity $y^{Z}_{\rm CM}$ up to NNLO QCD, compared with the ATLAS and CMS Run~1 measurements at $\sNN=5.02~\TeV$~\cite{ATLAS:2015mwq,CMS:2015zlj} and with the CMS Run~2 measurements at $\sNN=8.16~\TeV$~\cite{CMS:2021ynu}.
The fixed-order predictions are again obtained with the \texttt{CT18NNLO\_as\_0118}~\cite{Hou:2019efy} and \texttt{EPPS21nlo\_CT18Anlo\_Pb208}~\cite{Eskola:2021nhw} PDF sets at the dynamical central scale $\mu_F=\mu_R=m_{\ell\ell}$, here in the nucleon--nucleon CM frame for both runs, 
with the middle and lower panels showing the predictions normalised to the central NLO result and to the corresponding ATLAS or CMS data, respectively.
All four NNLO distributions have broad maxima at negative CM rapidities, around $y^{Z}_{\rm CM}\simeq-1$, and fall towards the edges of the measured ranges.
Forward and backward rapidities probe, on average, smaller and larger momentum fractions in the lead nucleus, respectively, so that the asymmetry of the distributions about $y^{Z}_{\rm CM}=0$ reflects both the nuclear modifications of the parton densities and the experimental acceptance~\cite{ATLAS:2015mwq,CMS:2021ynu}.

As can be seen from the upper and middle panels of the figure~\ref{fig:AP2Z_yz-dist}, in the $Z$-mass region, the NLO corrections enhance the LO cross sections by $15$--$20\%$ for ATLAS Run~1, $9$--$22\%$ for CMS Run~1 and $14$--$24\%$ for CMS Run~2, 
where this enhancement is smallest near central rapidities and grows towards the acceptance boundaries for both CMS datasets.
The subsequent NNLO corrections are considerably smaller, ranging from $-2.3\%$ to $+0.8\%$ for ATLAS Run~1, from $-0.4\%$ to $+2.3\%$ for CMS Run~1 and from $-1.3\%$ to $+3.0\%$ for the CMS Run~2 in the $Z$-mass region.
For the CMS fiducial distributions, the corrections are small around central rapidities and become positive and larger near both acceptance boundaries, whereas the largest ATLAS corrections are negative and occur in the most forward bins.
The NNLO central predictions lie within the plotted NLO scale bands throughout the $Z$-peak distributions except the two most backward bins at $8.16~\TeV$, where the corrections of $+3.0$ and $+2.6\%$ exceed the upper edge of the NLO band, about $+2.5\%$, by a small margin.
The scale uncertainty decreases overall from $1$--$9\%$ at LO to $2.0$--$3.2\%$ at NLO and $0.5$--$1.7\%$ at NNLO for ATLAS, from $4$--$9\%$ to $2.3$--$3.1\%$ and $0.4$--$0.9\%$ for CMS at $5.02~\TeV$, and from $8$--$13\%$ to $3.4$--$4.5\%$ and $0.7$--$1.1\%$ at $8.16~\TeV$; the NNLO uncertainty reaches $1.7\%$ only in the most forward ATLAS bin and stays below $1.2\%$ elsewhere.

The low-mass window, $15<m_{\mu\mu}<60~\GeV$, for CMS at $8.16~\TeV$ exhibits a markedly different perturbative behaviour.
The NLO corrections enhance the LO prediction by approximately $87$--$102\%$, nearly doubling the cross section, while the additional NNLO corrections are generally a few percent and reach $12\%$ and $15\%$ in the most backward and forward regions, respectively.
The large corrections are enhanced by the fiducial cuts: at Born level the two muons have equal transverse momenta, bounded by $m_{\mu\mu}/2$, so that the leading-muon requirement $p_T^{\mu}>15~\GeV$ excludes masses below $30~\GeV$.
Below this boundary, accepted configurations arise only through the transverse recoil of the dilepton pair against QCD radiation, which first appears at NLO and contributes to the large corrections in this sample.
For this part of the mass interval the NNLO prediction therefore supplies only the first correction to the leading non-vanishing contribution~\cite{Gehrmann-DeRidder:2023urf}.
The scale uncertainty decreases from $17$--$22\%$ at LO to $12$--$15\%$ at NLO and $3.5$--$6.6\%$ at NNLO, remaining substantially larger than in the $Z$-mass region.
In both edge bins the NNLO central prediction lies above the NLO scale band, although the two bands still overlap.

The NNLO predictions reproduce the broad shapes of the measured distributions, but rapidity-dependent differences remain, as visible from the lower panels of figure~\ref{fig:AP2Z_yz-dist}.
They are again quoted in units of $\sigma_{\rm exp}$, which excludes the luminosity uncertainties of $2.7\%$ for ATLAS~\cite{ATLAS:2015mwq} and $3.5\%$ for CMS~\cite{CMS:2015zlj,CMS:2021ynu}.
For ATLAS Run~1, the central prediction is below the data in most bins, with the largest deficits of about $3.0\sigma_{\rm exp}$ in $-0.5<y^Z_{\rm CM}<0$ and $2.1\sigma_{\rm exp}$ in $-1.5<y^Z_{\rm CM}<-1$; all other bins differ by less than $1.8\sigma_{\rm exp}$.
Larger relative deviations occur in the outermost forward bins, remaining within about $1.1\sigexp$, where the statistical uncertainties of the data are also substantially larger.
For CMS Run~1, the prediction lies within one experimental standard deviation in most bins, with no deviation exceeding $1.6\sigma_{\rm exp}$.
For CMS Run~2 in the $Z$-mass region, the NNLO prediction agrees with the data within $1\sigma_{\rm exp}$ in most bins over $-2.6<y^Z_{\rm CM}<0$, with the remaining differences below $1.6\sigma_{\rm exp}$.
In contrast, it lies $1.8$--$4.5\sigma_{\rm exp}$ below the data throughout $0<y^Z_{\rm CM}<1.6$.
The outermost backward bin, $-2.87<y^Z_{\rm CM}<-2.6$, shows an excess of about $3.0\sigma_{\rm exp}$, whereas the outermost forward bin, $1.8<y^Z_{\rm CM}<1.93$, shows a deficit of only $1.3\sigma_{\rm exp}$.
In the low-mass window of CMS Run~2, the NNLO prediction lies below the data throughout, with the largest deficits of $2.6\sigma_{\rm exp}$ in $-0.8<y^Z_{\rm CM}<-0.4$ and $2.9\sigma_{\rm exp}$ in $0.4<y^Z_{\rm CM}<0.8$. The most backward bin and $0.8<y^Z_{\rm CM}<1.2$ agree within one experimental standard deviation; elsewhere the deficits are between one and three. The positive NNLO corrections improve the description, as also observed in ref.~\cite{Helenius:2021tof}, but the data central values remain above the NNLO scale band.

The different rapidity and mass intervals provide complementary sensitivity to the momentum fractions probed in the lead nucleus~\cite{CMS:2021ynu}.
As for the $W^{\pm}$ distributions, the residual differences cannot be attributed uniquely to nuclear modifications: the low-mass window remains sensitive to higher-order corrections, and a quantitative interpretation requires the proton- and nuclear-PDF uncertainties as well as the experimental correlations~\cite{CMS:2019leu,Eskola:2022rlm}.
\\

\begin{figure}[tbp]
    \centering
      \includegraphics[width=0.49\textwidth]{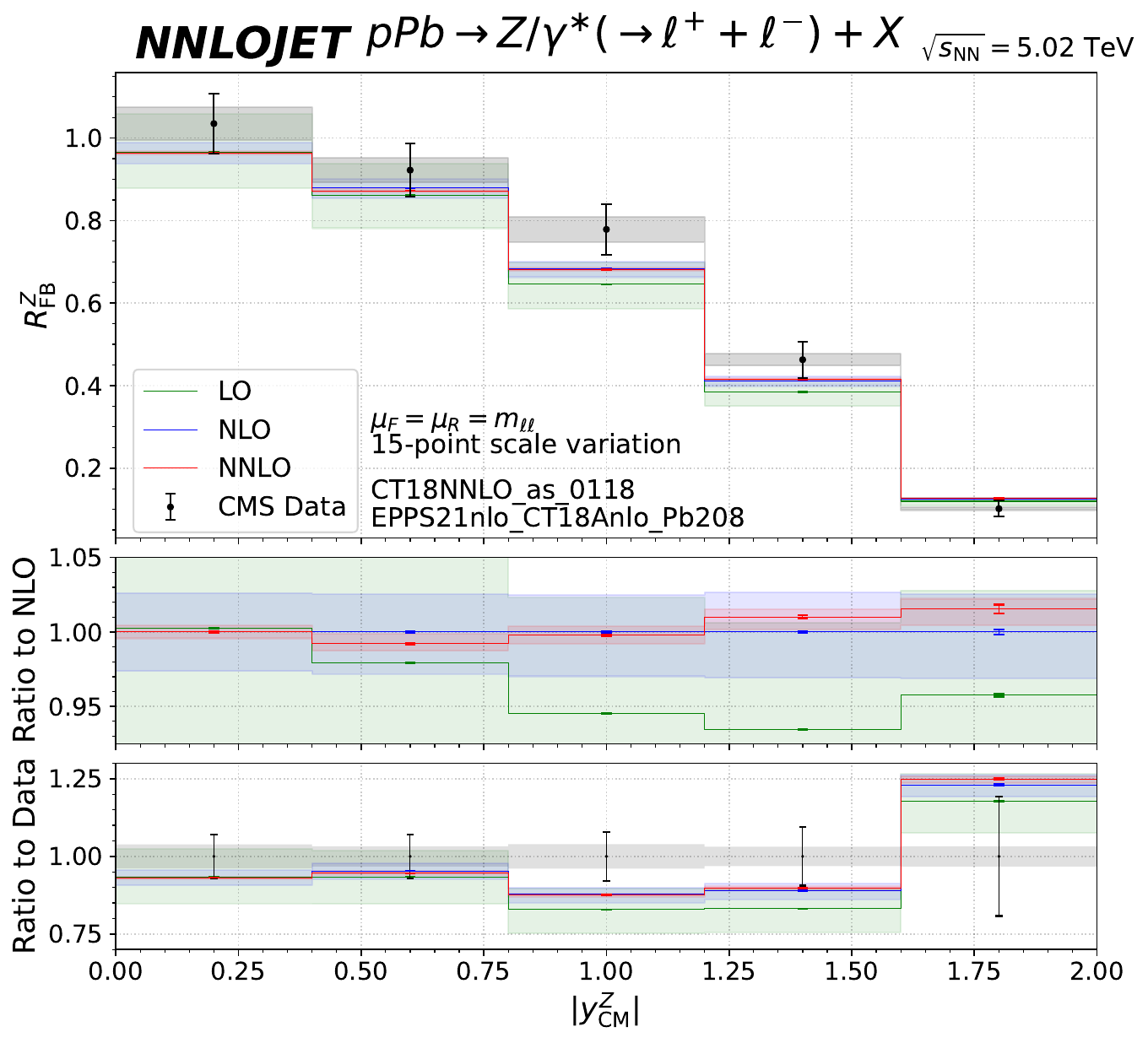}
\caption{Same as figure~\ref{fig:AP2Z_yz-dist}, but for the forward--backward ratio $R_{\mathrm{FB}}^{Z}$, defined in eq.~\eqref{eq:RFB}, as a function of $|y^{Z}_{\rm CM}|$, compared with the \href{https://www.hepdata.net/record/ins1410832}{CMS Run~1} measurements at $\sqrt{s_{\mathrm{NN}}}=5.02~\TeV$~\cite{CMS:2015zlj}.
The coloured bands show the partially correlated 15-point scale-variation uncertainty of eq.~\eqref{eq:scale15}.
}
\label{fig:AP2Z_RFB}
\end{figure}

Figure~\ref{fig:AP2Z_RFB} shows the comparison of the forward-to-backward ratio $R^{Z}_{\mathrm{FB}}$ of eq.~\eqref{eq:RFB} with the CMS measurement at $5.02~\TeV$~\cite{CMS:2015zlj}, with the middle and lower panels displaying the ratios to the central NLO prediction and to the CMS data, respectively.
The NNLO prediction decreases monotonically from $0.96$ in the $0<|y^Z_{\rm CM}|<0.4$ bin to $0.13$ in $1.6<|y^Z_{\rm CM}|<2.0$.
This pronounced decrease is driven mainly by the fiducial acceptance: the lepton pseudorapidity selection is imposed in the laboratory frame, which is shifted by ${\Delta y=0.465}$ with respect to the CM frame, eq.~\eqref{eq:deltay}, so that the forward bins are depleted far more strongly than their backward counterparts; the departure from unity therefore cannot be identified with the size of a nuclear modification~\cite{CMS:2015zlj}.
The perturbative corrections cancel less completely than for the $W^{\pm}$ ratios: the NLO corrections change the ratio by up to $+7.0\%$, in $1.2<|y^{Z}_{\rm CM}|<1.6$, while the NNLO corrections remain within $-0.8$ to $+1.5\%$ and the NNLO central prediction lies within the NLO band throughout the measured rapidity regions.
The scale uncertainty decreases from approximately $9$--$10\%$ at LO to $2.6$--$3.1\%$ at NLO and $0.4$--$1.1\%$ at NNLO.
The NNLO prediction agrees with the data within one experimental standard deviation for $|y^Z_{\rm CM}|<0.8$, with differences of only about $1.0$--$1.4\sigma_{\rm exp}$ in the remaining bins. Thus, despite the sizeable relative deviation in the outermost bin, no large binwise discrepancy is observed.
The normalisation uncertainties cancel in this ratio, which may enhance the sensitivity to the nuclear modifications~\cite{CMS:2015zlj}, but a quantitative interpretation requires, as for the charged-current ratios, the proton- and nuclear-PDF uncertainties as well as the experimental correlations.

In summary, we have presented a systematic study of NNLO QCD corrections to charged- and neutral-current Drell--Yan production in $p$Pb collisions at the LHC, using the state-of-the-art EPPS21 nuclear PDFs. This work provides the first publicly available implementation of NNLO QCD calculations for Drell--Yan production in proton--nucleus collisions delivering fast convolution grids, within the NNLOJET framework. In the $W^\pm$ and $Z$ mass regions, the NNLO corrections are generally at the percent level, with a substantial reduction in scale dependence, while the corrections to forward-to-backward ratios are further suppressed by cancellations. In contrast, the low-mass Drell--Yan distributions receive considerably larger corrections, reaching 15\% in some kinematic regions. The NNLO predictions broadly reproduce the measured distributions, although rapidity-dependent differences remain that cannot be resolved by higher-order corrections alone. These results provide a basis for assessing the impact of NNLO corrections on nuclear-PDF determinations.

\subsection{Drell--Yan production in fixed-target proton--nucleus collisions}
\label{sec:FTDY}

The fixed-target measurements of E772~\cite{Alde:1990im} and E866/NuSea~\cite{NuSea:1999egr} summarised in table~\ref{tab:FT_obs}, both taken with the $800~\GeV$ projectile proton beam incident on stationary nuclear targets at Fermilab, probe the DY process at lower dilepton invariant masses, in a kinematic regime complementary to that of the LHC measurements discussed in section~\ref{sec:AP}.
The dilepton masses lie between the $J/\psi$ and $\Upsilon$ resonances or, for part of the E772 sample, above the $\Upsilon$, and the nuclear-target ratios are measured differentially in the target momentum fraction $x_2$ in E772 and the beam momentum fraction $x_1$ in E866, with E772 probing target momentum fractions up to $x_2\simeq0.3$~\cite{Alde:1990im} and E866 extending the coverage towards larger beam and smaller target momentum fractions, $0.21<x_1<0.95$ and $0.01<x_2<0.12$~\cite{NuSea:1999egr}.
In this regime, the leading-order cross section is dominated by the annihilation of a beam valence quark with a target antiquark, providing sensitivity to the target antiquark distributions complementary to that of charged-lepton DIS~\cite{NuSea:1999egr,Klasen:2023uqj}.
The per-nucleon cross-section ratios between different nuclear targets therefore constrain the nuclear modification of the antiquark distributions at small and moderate $x_2$~\cite{NuSea:1999egr,Eskola:2021nhw}.
The E772 and E866 ratios are included in the NLO nuclear-PDF analyses nCTEQ26~\cite{nCTEQ:2026ewr}, EPPS21~\cite{Eskola:2021nhw} and partly in nNNPDF3.0~\cite{AbdulKhalek:2022fyi}, where, together with DIS data, they help disentangle the nuclear modifications of valence- and sea-quark distributions~\cite{Klasen:2023uqj}.
By contrast, TUJU21 did not include these measurements, prioritising datasets expected to provide the strongest nuclear-PDF constraints in view of the computational cost of NNLO grid generation~\cite{Helenius:2021tof}.
The NNLO interpolation grids presented here facilitate the inclusion of these fixed-target data in NNLO nuclear-PDF analyses.

All predictions presented in this section are computed directly within \texttt{NNLOJET}, retaining only virtual-photon exchange, which dominates DY production in the low-mass region considered here~\cite{Klasen:2023uqj}.
The electromagnetic coupling is fixed in the $\alpha(0)$ scheme with $\alpha = 1/137.036$ and cancels exactly in the nuclear-target ratios at the QCD orders considered.
The projectile proton and the target nuclei are described by the \texttt{nCTEQ26} PDF sets~\cite{nCTEQ:2026ewr}, with the proton baseline for the projectile and the full nuclear sets for deuterium, beryllium, carbon, calcium, iron and tungsten; the names of the individual sets are indicated in the figures.
For the fixed-target ratios, we adopt the \texttt{nCTEQ26} sets to describe the projectile proton and all target and reference nuclei within a single PDF analysis. %
This choice provides a common nuclear-PDF framework for studying higher-order QCD corrections in the fixed-target predictions and in the interpolation-grid $\chi^2$ comparison with the LHC data in section~\ref{sec:grid-applications}.
The momentum fractions $x_1$ and $x_2$ and Feynman variables $x_F$ are reconstructed event by event from the dilepton momentum according to eqs.~\eqref{eq:li-reconstructed-x} and~\eqref{eq:xF}, the selections of table~\ref{tab:FT_obs} are applied, and the ratios are binned as published, in $x_2$ for E772 and in $x_1$ for each E866 mass window.
The central scale is chosen as the dynamical scale $\mu_F=\mu_R=m_{\mu\mu}$, and the scale uncertainty of the ratios is obtained from the partially correlated 15-point prescription of eq.~\eqref{eq:scale15}, quoted as in section~\ref{sec:AP} as the larger of the upward and downward deviations from the central prediction.
As in section~\ref{sec:AP}, the same PDFs are used at LO, NLO and NNLO, so that the order-by-order comparison isolates the corrections to the partonic coefficient functions.
The E772 and E866 ratios are included in the \texttt{nCTEQ26} fit, as in \texttt{nCTEQ15}~\cite{Kovarik:2015cma}, so that agreement of the NLO predictions with these data is expected and the comparisons below are not an independent test of the nuclear PDFs.
All NLO predictions presented in this section have been cross-checked against the internal calculations of the nCTEQ26 fitting framework, finding agreement at the percent level.

\subsubsection{E772 nuclear-target ratios}
\label{sec:FTDY-E772}

\begin{figure}[tbp]
    \centering
    \includegraphics[width=0.49\textwidth]{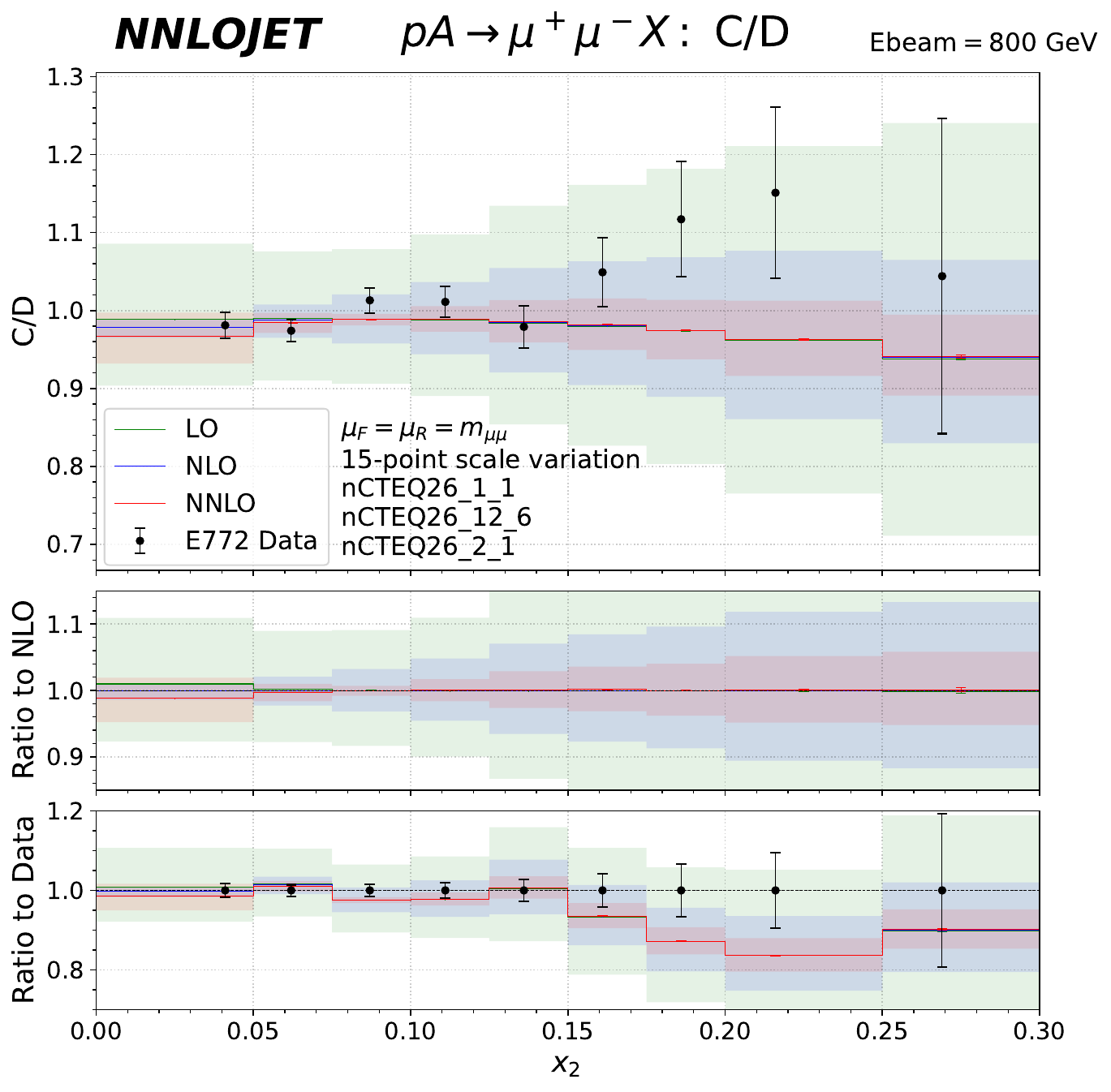}
    \includegraphics[width=0.49\textwidth]{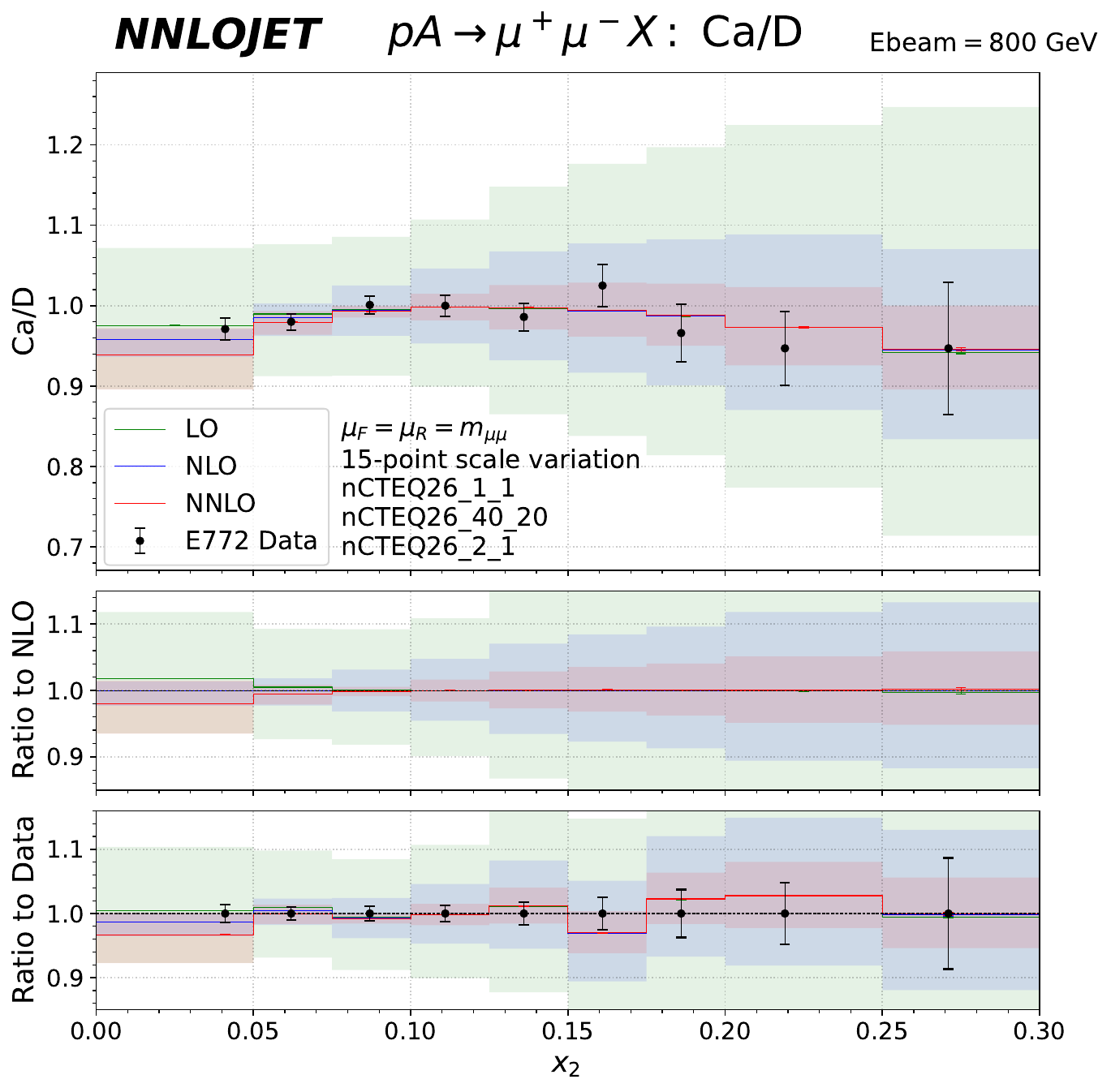}
    \includegraphics[width=0.49\textwidth]{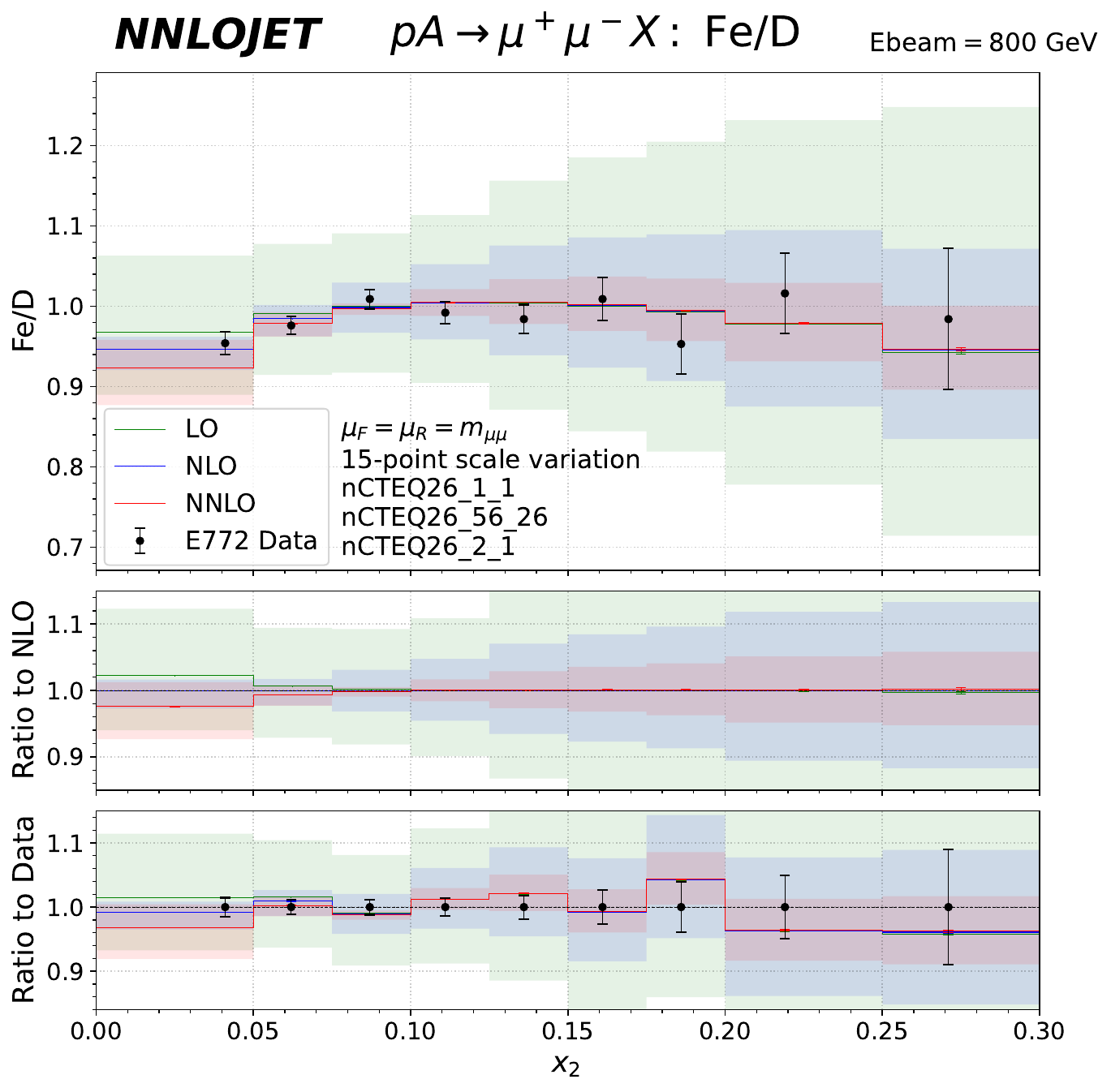}
    \includegraphics[width=0.49\textwidth]{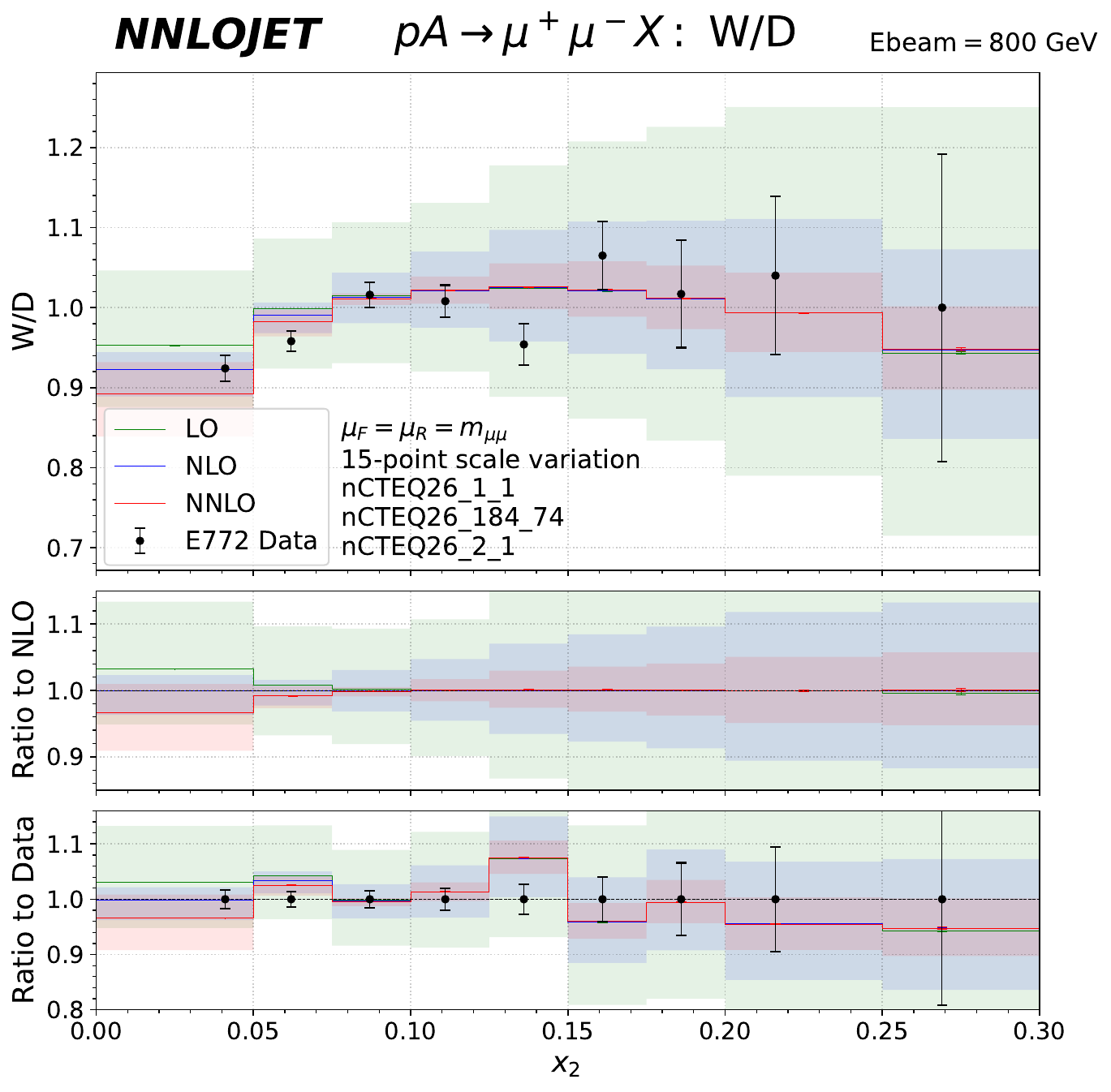}
\caption{Per-nucleon DY cross-section ratios $\mathrm{C}/\mathrm{D}$, $\mathrm{Ca}/\mathrm{D}$, $\mathrm{Fe}/\mathrm{D}$ and $\mathrm{W}/\mathrm{D}$ as functions of the target momentum fraction $x_2$, defined in eq.~\eqref{eq:li-reconstructed-x}, compared with the Fermilab \href{https://doi.org/10.17182/hepdata.19997.v1}{E772 measurements}~\cite{Alde:1990im} at a proton beam energy of $800~\GeV$.
The predictions are evaluated up to NNLO QCD using the \texttt{nCTEQ26} PDF sets at the central scale $\mu_F=\mu_R=m_{\mu\mu}$.
The three PDF labels in each plot correspond, from top to bottom, to the projectile proton, the nuclear target in the numerator and the reference nucleus in the denominator, respectively.
The coloured bands show the partially correlated 15-point scale-variation uncertainty of eq.~\eqref{eq:scale15}, and the error bars on the predictions indicate numerical integration uncertainties.
The data error bars denote statistical uncertainties; systematic uncertainties are not shown.
The middle and lower panels show the predictions normalised to the central NLO result and to the corresponding E772 data, respectively.
Upper row: carbon (left) and calcium (right); lower row: iron (left) and tungsten (right), each relative to deuterium.}
\label{fig:FTDY_E772}
\end{figure}

Figure~\ref{fig:FTDY_E772} shows the per-nucleon cross-section ratios $\mathrm{C}/\mathrm{D}$, $\mathrm{Ca}/\mathrm{D}$, $\mathrm{Fe}/\mathrm{D}$ and $\mathrm{W}/\mathrm{D}$ as functions of reconstructed target momentum fraction $x_2$ up to NNLO QCD, compared with the E772 measurement~\cite{Alde:1990im}, with the middle and lower panels displaying predictions normalised to the central NLO prediction and to the data, respectively. 
The coloured bands denote the partially correlated 15-point scale variation, and the theory and data error bars indicate numerical integration and statistical uncertainties, respectively.
At NNLO, all four predicted ratios lie below unity in the lowest bin, $x_2<0.05$, by an amount that grows with the mass number, from $3\%$ for carbon to $11\%$ for tungsten, return to within $3\%$ of unity for $0.075<x_2<0.2$, exceeding it only for iron and tungsten, then decrease again to $0.94$--$0.95$ at the outermost end of the measured $x_2$ regions.
The perturbative corrections cancel almost completely in the ratios: for $x_2>0.075$ the NLO corrections lie within $\pm0.5\%$ and the NNLO corrections within $\pm0.2\%$ for all targets.
At smaller $x_2$, both corrections reduce the ratios: in the lowest bin, the NLO corrections range from $-1.0\%$ for carbon to $-3.1\%$ for tungsten, and the NNLO corrections from $-1.2\%$ to $-3.3\%$. 
The NNLO correction is therefore comparable to the NLO correction in this region where the shadowing is strongest, although the NNLO central prediction remains inside the NLO scale envelope throughout the measured regions.
Across the four targets, the quoted scale uncertainty increases with $x_2$ at every order for $x_2\geq0.075$, and decreases from $9$--$33\%$ at LO to $1.4$--$13\%$ at NLO and $0.8$--$6\%$ at NNLO. 
In the lowest bin, however, the NNLO uncertainty of $3.6$--$6.0\%$ exceeds the NLO uncertainty of $1.4$--$3.7\%$, whereas for $x_2\geq0.075$ the successive scale bands are nested and the NNLO uncertainty grows from $0.8\%$ to $5.7\%$ towards the upper end of the measured range.

The NNLO predictions follow the broad behaviour of the data, but target-dependent differences remain, as visible from the lower panels of figure~\ref{fig:FTDY_E772}. 
Differences between the central NNLO predictions and the data are quoted in units of the statistical uncertainty $\sigma_{\rm stat}$ of the data, shown by the error bars in the figures; the systematic uncertainties, less than $2\%$ for E772~\cite{Alde:1990im} and a $1\%$ normalisation uncertainty for E866~\cite{NuSea:1999egr}, are not included.
The central predictions lie within the experimental statistical error bars in 22 of the 36 bins and within $2\sigma_{\rm stat}$ in 33.
The three larger deviations are deficits of $2.3\sigma_{\rm stat}$ and $2.2\sigma_{\rm stat}$ for calcium and iron in $x_2<0.05$, respectively, and an excess of $2.8\sigma_{\rm stat}$ for tungsten in $0.125<x_2<0.15$.
Although the carbon prediction lies below the data for $x_2\geq0.15$, the differences remain below $2\sigma_{\rm stat}$ owing to the growing experimental uncertainties.
These ratios are primarily sensitive to the nuclear modification of the target antiquark distributions relative to deuterium~\cite{Alde:1990im,Klasen:2023uqj}, and the suppression at small $x_2$ is characteristic of shadowing~\cite{NuSea:1999egr}.
The shallow maxima of the predicted ratios and their decline at larger $x_2$ are consistent with anti-shadowing and the onset of the EMC effect. 
However, they cannot be established as nuclear effects from these comparisons alone: as for the collider observables of section~\ref{sec:AP}, a quantitative interpretation requires the proton- and nuclear-PDF uncertainties, which are not included in the depicted scale bands, as well as the experimental systematic uncertainties and their correlations.

\subsubsection{E866 nuclear-target ratios}
\label{sec:FTDY-E866}

\begin{figure}[tbp]
    \centering
    \includegraphics[width=0.49\textwidth]{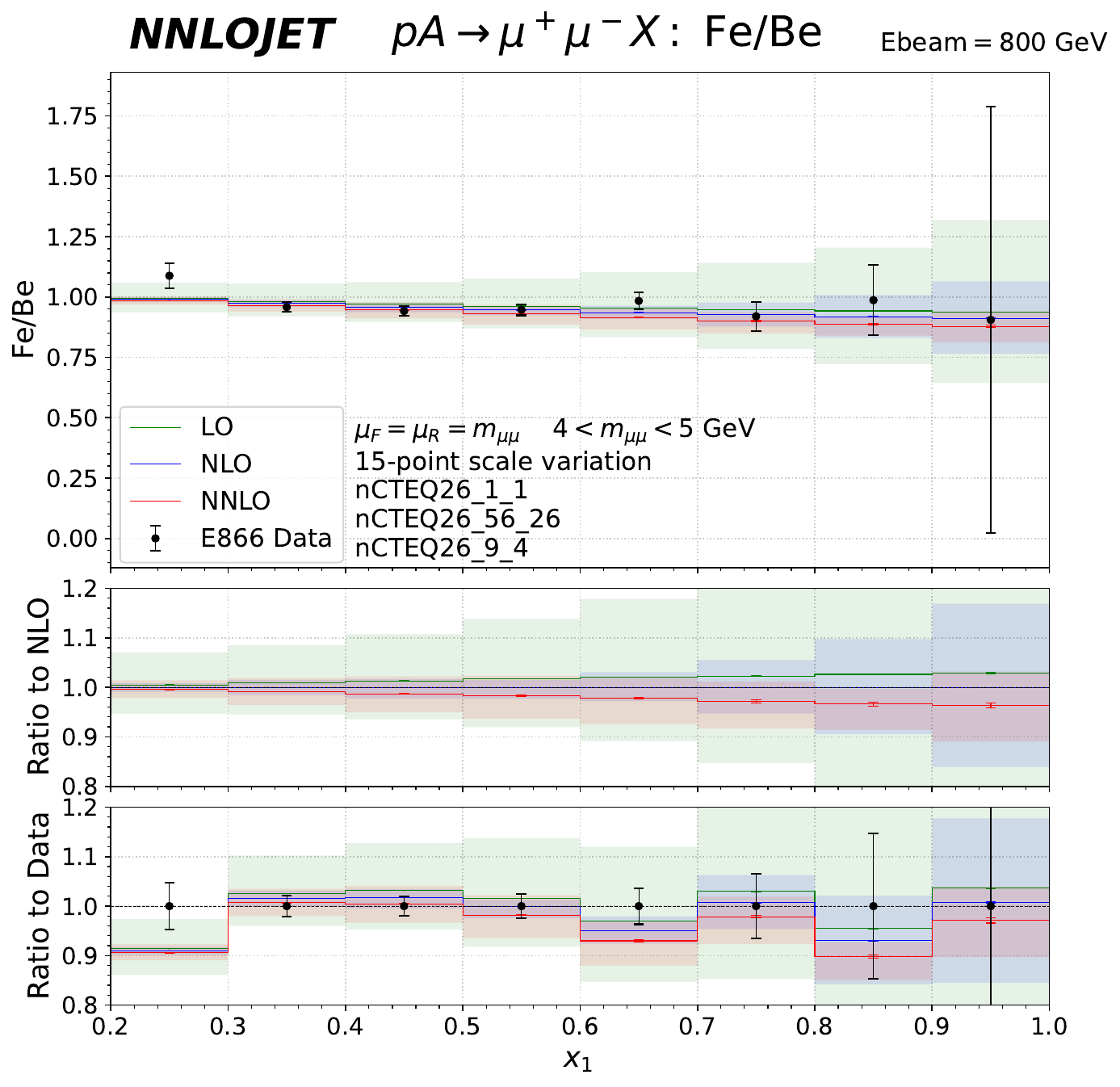}
    \includegraphics[width=0.49\textwidth]{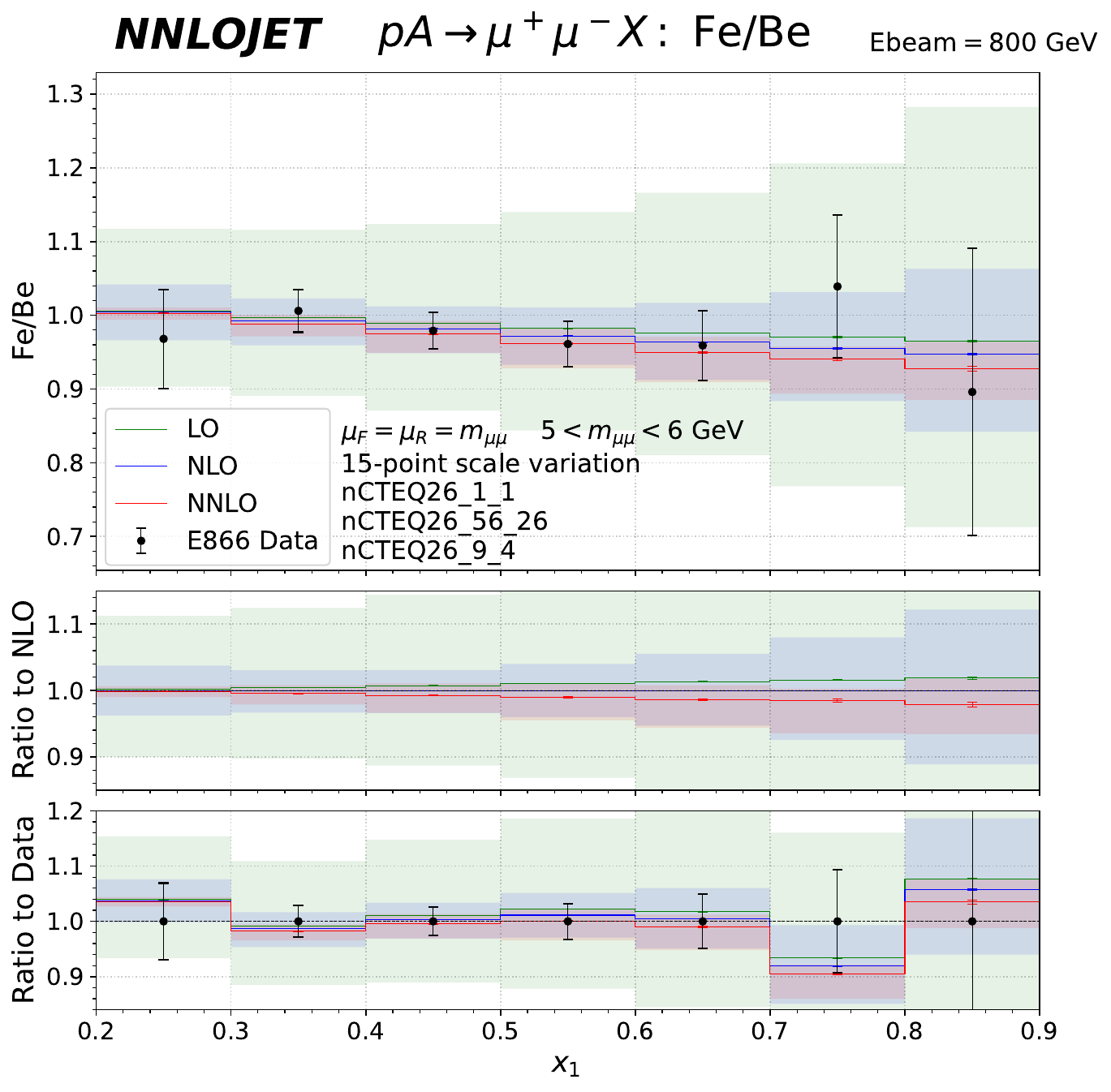}
    \includegraphics[width=0.49\textwidth]{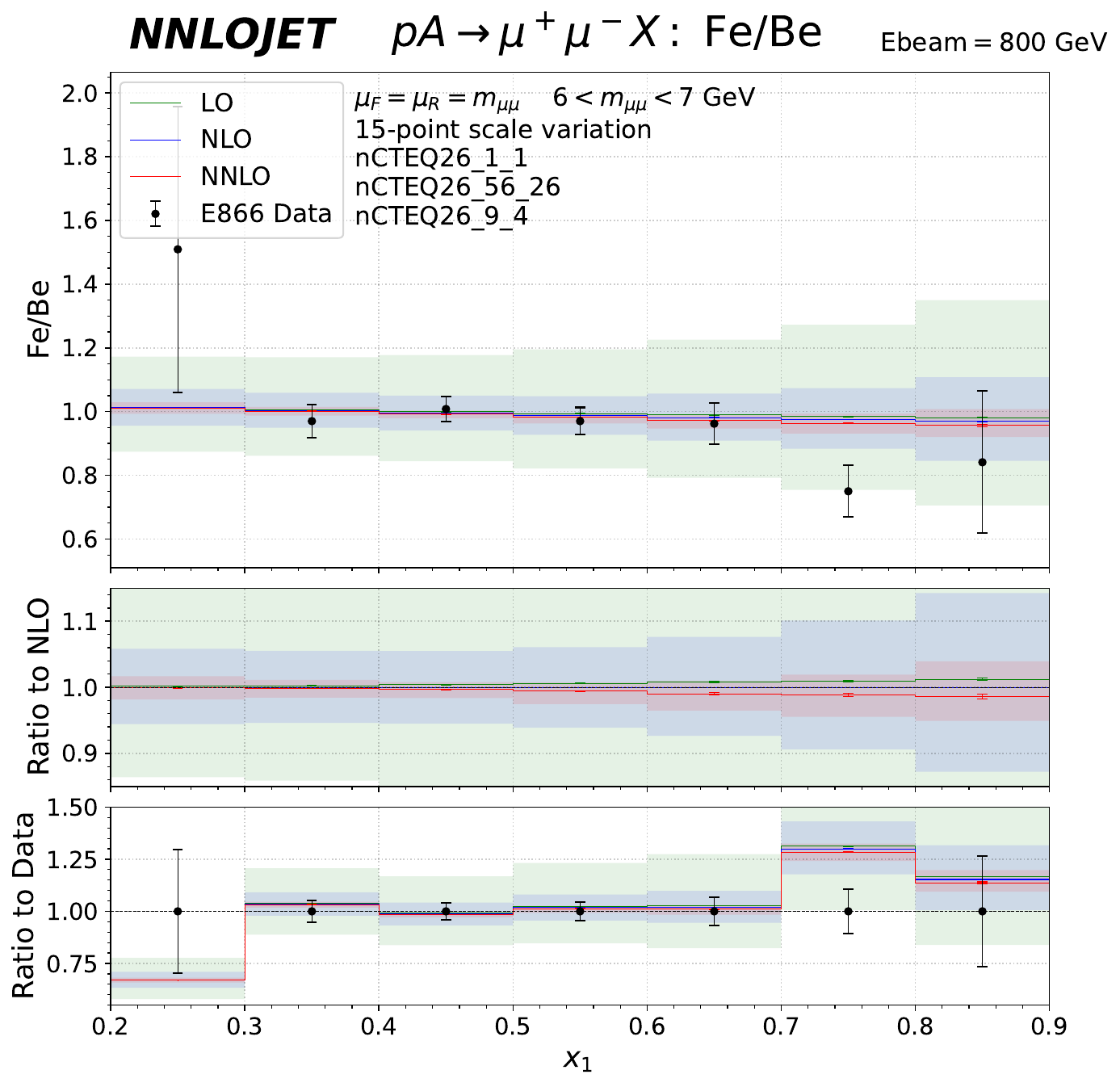}
    \includegraphics[width=0.49\textwidth]{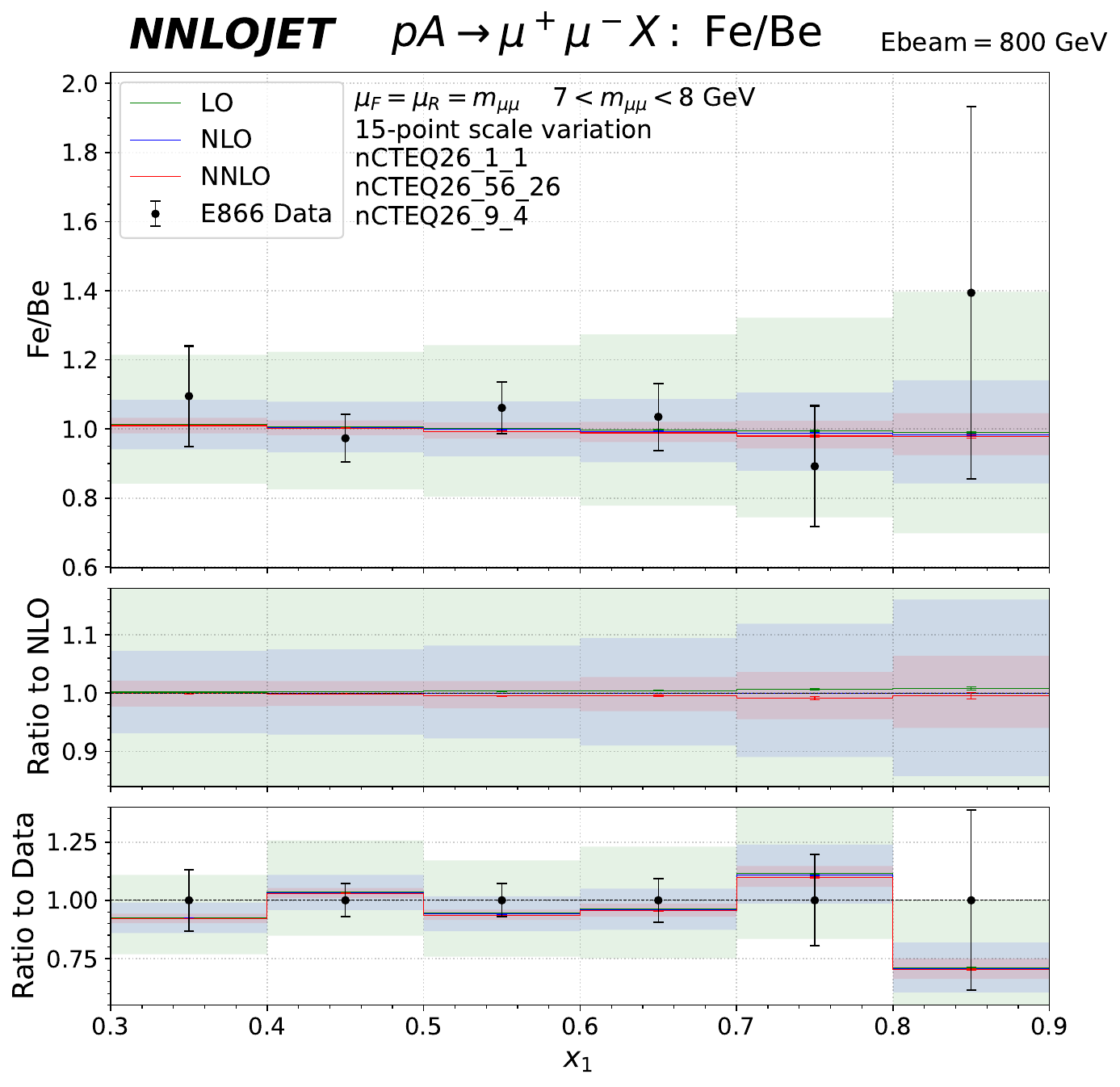}
\caption{Same as figure~\ref{fig:FTDY_E772}, but for the per-nucleon cross-section ratio $\mathrm{Fe}/\mathrm{Be}$ as a function of the beam momentum fraction $x_1$, defined in eq.~\eqref{eq:li-reconstructed-x}, compared with the Fermilab \href{https://doi.org/10.17182/hepdata.19445.v1}{E866 measurements}~\cite{NuSea:1999egr} at a proton beam energy of $800~\GeV$.
The panels correspond to the dimuon invariant-mass windows $4<m_{\mu\mu}<5~\GeV$ (upper left), $5<m_{\mu\mu}<6~\GeV$ (upper right), $6<m_{\mu\mu}<7~\GeV$ (lower left) and $7<m_{\mu\mu}<8~\GeV$ (lower right).}
\label{fig:FTDY_E866_FEoBE}
\end{figure}

\begin{figure}[tbp]
    \centering
    \includegraphics[width=0.49\textwidth]{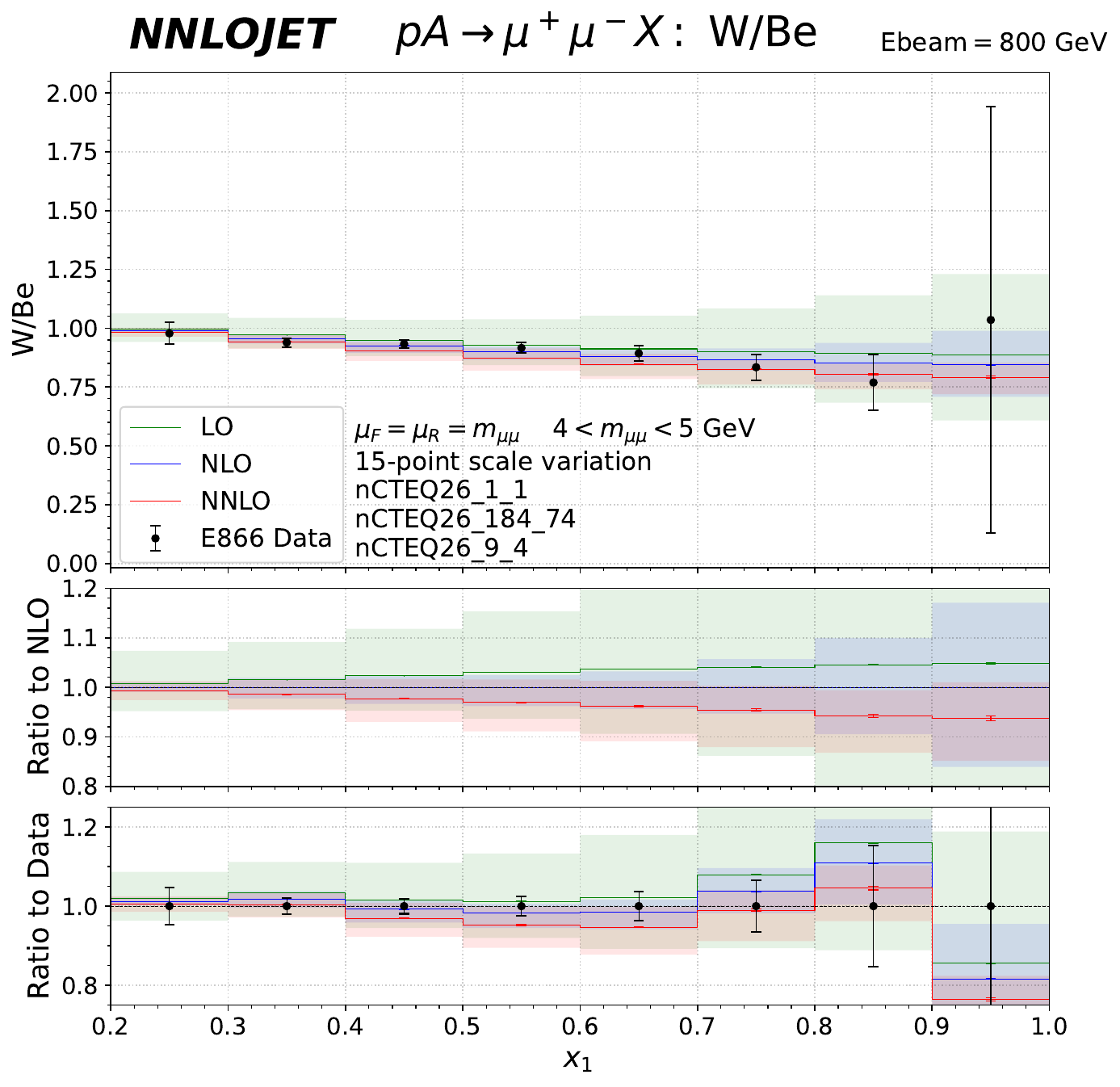}
    \includegraphics[width=0.49\textwidth]{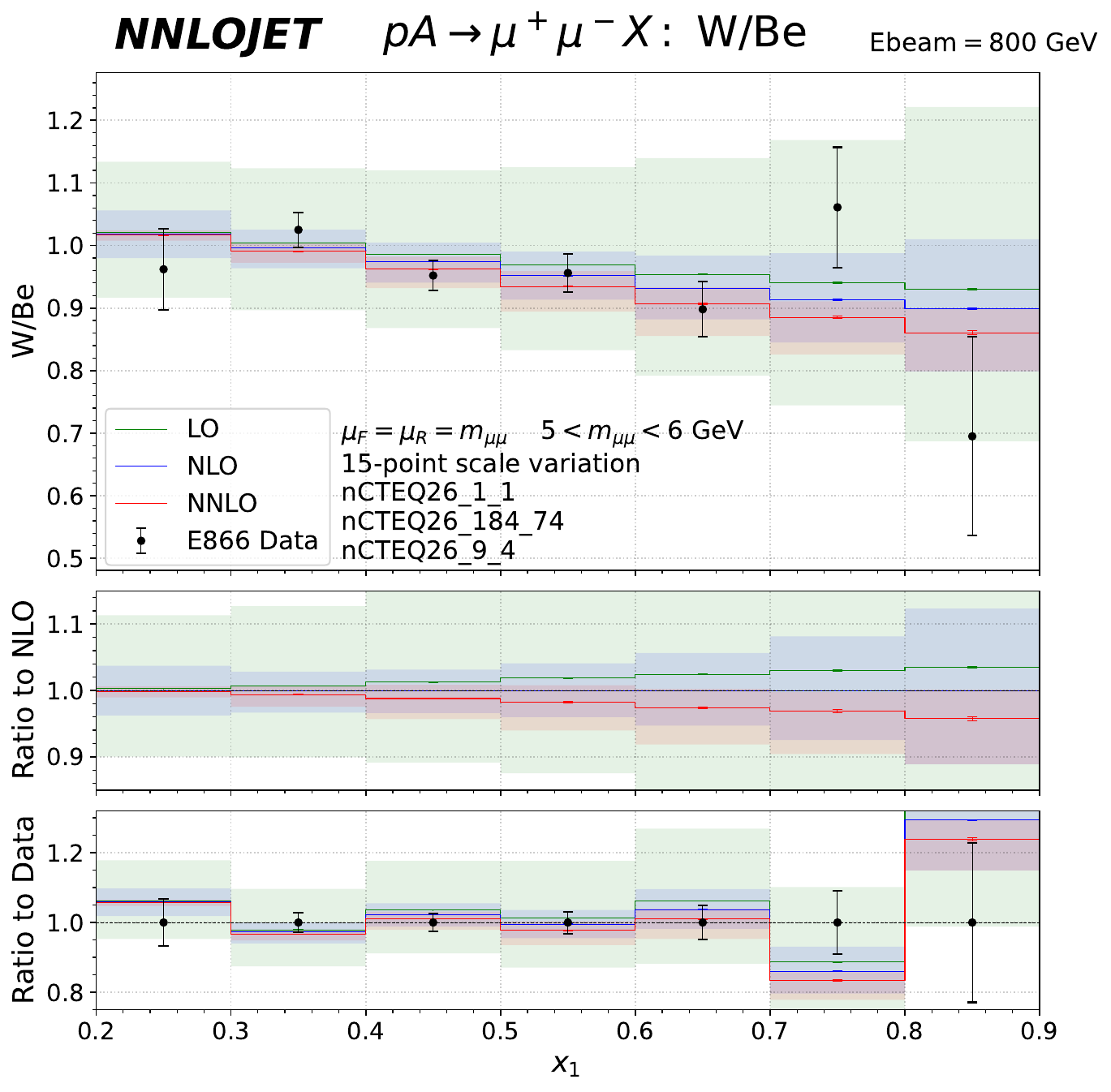}
    \includegraphics[width=0.49\textwidth]{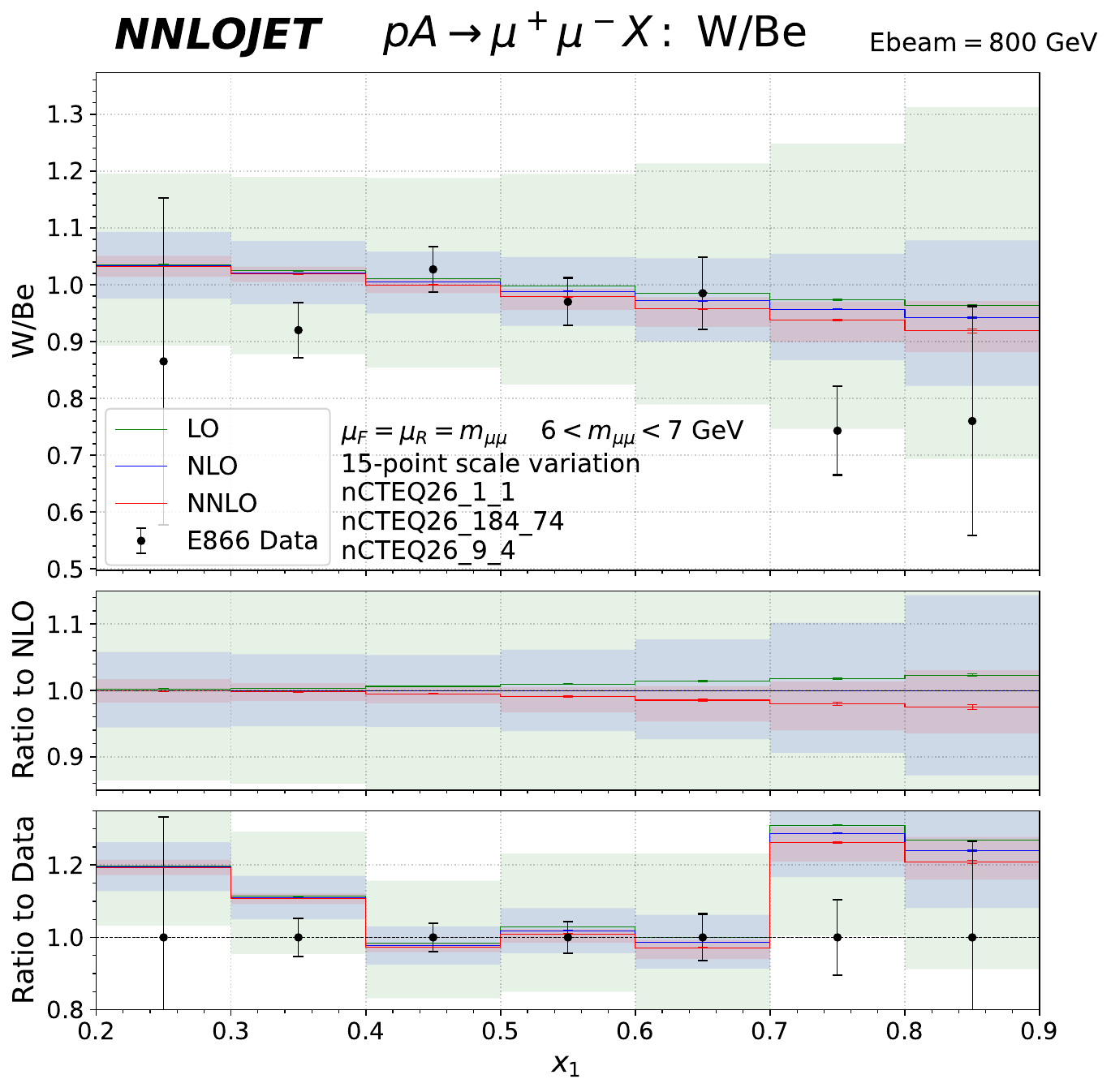}
    \includegraphics[width=0.49\textwidth]{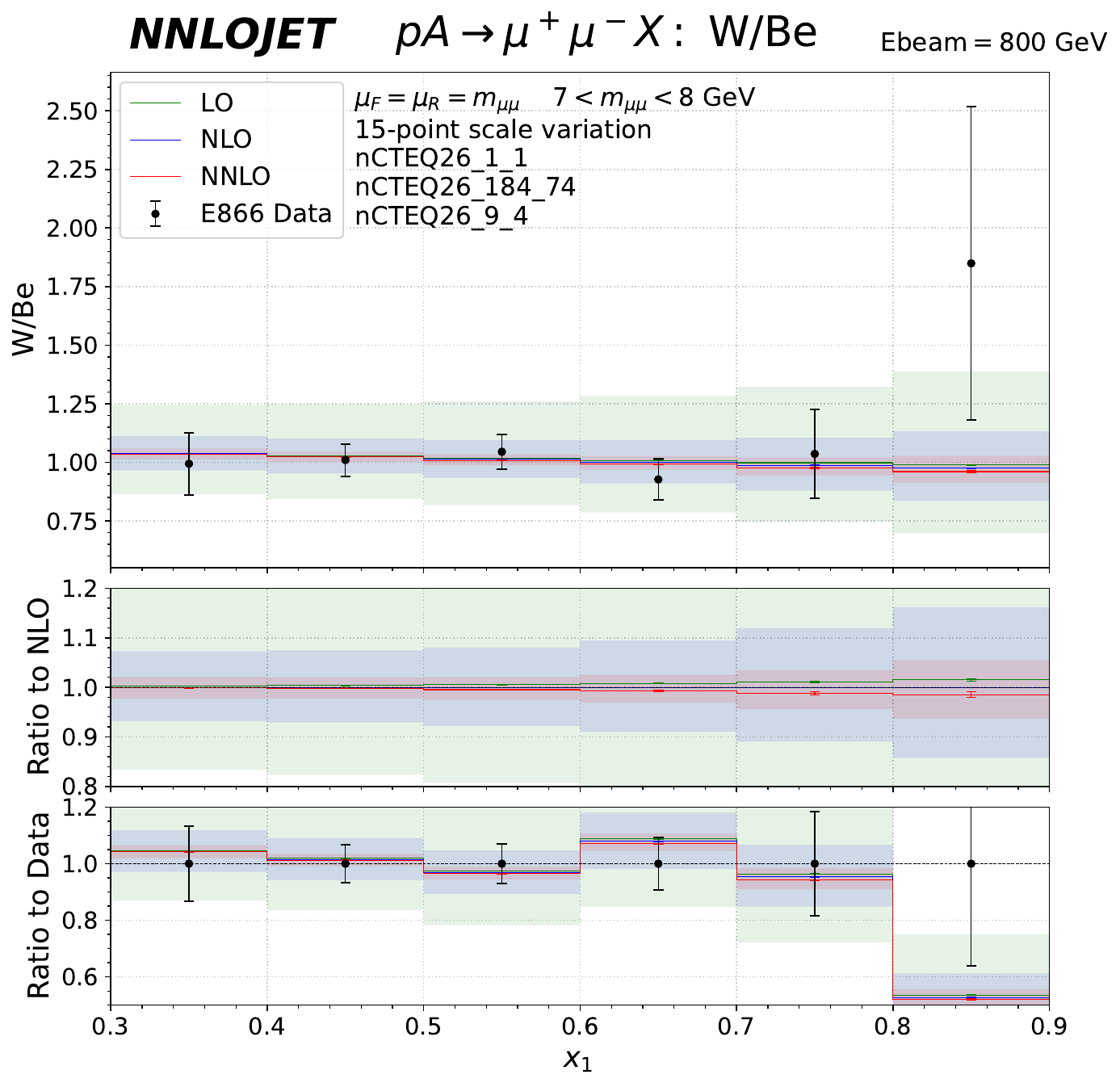}
\caption{Same as figure~\ref{fig:FTDY_E866_FEoBE}, but for the per-nucleon cross-section ratio $\mathrm{W}/\mathrm{Be}$.}
\label{fig:FTDY_E866_WoBE}
\end{figure}

Figures~\ref{fig:FTDY_E866_FEoBE} and~\ref{fig:FTDY_E866_WoBE} show the per-nucleon cross-section ratios $\mathrm{Fe}/\mathrm{Be}$ and $\mathrm{W}/\mathrm{Be}$ as functions of beam momentum fractions $x_1$ in the four dimuon-mass windows between $4$ and $8~\GeV$ up to NNLO QCD, compared with the E866 measurements~\cite{NuSea:1999egr}, with the panels, scale bands and error bars defined as in figure~\ref{fig:FTDY_E772}.  %
The NNLO predicted ratios decrease with $x_1$ in every mass window, with a steeper decline for tungsten than for iron, and a more pronounced one in the low-mass region than in the high-mass region: $\mathrm{Fe}/\mathrm{Be}$ falls from $0.99$ to $0.88$ and $\mathrm{W}/\mathrm{Be}$ from $0.98$ to $0.79$ across $0.2<x_1<1.0$ in the $4$--$5~\GeV$ window, whereas in the $7$--$8~\GeV$ window they decrease only from $1.01$ to $0.98$ and from $1.04$ to $0.96$, respectively, over $0.3<x_1<0.9$.
This trend is consistent with stronger nuclear shadowing (suppression) in iron and tungsten than in beryllium at the smaller target momentum fractions, since increasing $x_1$ within a given mass window generally probes smaller target momentum fractions, as indicated by the Born-level relation $x_2=m_{\mu\mu}^2/(s_{\rm NN}x_1)$.
Such an interpretation is in line with the E866 analysis, which found that shadowing at small $x_2$ could account for most of the nuclear dependence of its mass-integrated $x_1$ ratios~\cite{NuSea:1999egr}.

The cancellation of higher-order corrections between the heavier-target and beryllium cross sections is less effective than in the E772 ratios, particularly for tungsten: both the NLO and NNLO corrections reduce the ratios throughout, with the largest reductions at low mass and large $x_1$, that is, at the smallest $x_2$, where the E772 ratios also receive their largest corrections.
In the $4$--$5~\GeV$ window, the NLO corrections range from $-0.5\%$ to $-2.8\%$ for $\mathrm{Fe}/\mathrm{Be}$ and from $-0.7\%$ to $-4.6\%$ for $\mathrm{W}/\mathrm{Be}$, and the NNLO corrections, of comparable or larger size, from $-0.5\%$ to $-3.6\%$ and from $-0.7\%$ to $-6.3\%$, whereas in the $7$--$8~\GeV$ window both corrections remain below $1.5\%$ in magnitude.
The NNLO corrections are thus not uniformly smaller than the NLO corrections, although the NNLO central prediction lies within the NLO scale band throughout the measured regions.
With the partially correlated 15-point prescription of eq.~\eqref{eq:scale15}, the scale uncertainty across both targets and all mass windows spans $6.6$--$41\%$ at LO and $0.9$--$17\%$ at NLO, decreasing overall to $0.9$--$7.6\%$ for iron and $0.9$--$9.1\%$ for tungsten at NNLO.
The NNLO uncertainties are generally largest towards high $x_1$, reaching $7.6\%$ and $9.1\%$ in the last bin of the lowest mass window.
This reduction is not uniform: in the lowest mass window, $4$--$5~\GeV$, the NNLO uncertainty exceeds its NLO counterpart over $0.2<x_1<0.8$ for both targets, whereas in the two highest mass windows $6<m_{\mu\mu}<8~\GeV$ the successive scale bands are nested order by order throughout the measured regions.

The relative statistical uncertainties of the E866 ratios are $1.9$--$2.4\%$ for $0.3<x_1<0.6$ in the $4$--$5~\GeV$ window, but reach $22$--$98\%$ in the highest-$x_1$ bin of each mass window.
The NNLO predictions broadly describe the measured ratios: with the experimental statistical uncertainties alone, 42 of the 56 central predictions lie within one standard deviation of the data and 52 within two.
For $0.3<x_1<0.7$, the predictions agree with the data within $2\sigsta$, except for $\mathrm{W}/\mathrm{Be}$ in $0.3<x_1<0.4$ at $6$--$7~\GeV$, where the prediction exceeds the data by $11\%$, corresponding to $2.1\sigma_{\rm stat}$.
The remaining three deviations exceeding $2\sigma_{\rm stat}$ occur for $\mathrm{Fe}/\mathrm{Be}$ in $0.2<x_1<0.3$ at $4$--$5~\GeV$, where the prediction lies $9\%$ below the data, and for both ratios in $0.7<x_1<0.8$ at $6$--$7~\GeV$, where the iron and tungsten predictions exceed the data by $2.6\sigsta$ and $2.5\sigsta$, respectively.
In the $4$--$5~\GeV$ window, where the statistical uncertainties are smallest, the NNLO $\mathrm{W}/\mathrm{Be}$ prediction lies $3.1$--$5.3\%$ below the data for $0.4<x_1<0.7$, corresponding to deviations of $1.4$--$2.0\sigma_{\rm stat}$.
The NLO central prediction lies within the statistical error bars in these ranges, and the NNLO corrections lower the prediction by a further $2.3$--$3.9\%$.
The measured central value lies within the NNLO scale band for $0.4<x_1<0.5$, whereas in the two bins covering $0.5<x_1<0.7$ it lies just above the upper edge of the band; the experimental error bars overlap the band in all three bins.
Since these ratios are included in the fit of the nuclear PDFs used here, the agreement at NLO is expected. %
In the three $\mathrm{W}/\mathrm{Be}$ bins discussed above, the NNLO corrections, although smaller than the NNLO scale uncertainty, are comparable to the statistical uncertainties of the data ($1.9$--$3.7\%$), so that the cancellation of higher-order corrections cannot be assumed when these ratios are included in NNLO nuclear-PDF analyses.

In summary, we have presented the first NNLO QCD predictions for the E772 and E866 fixed-target DY nuclear-target ratios, using the state-of-the-art nCTEQ26 nuclear PDFs. These calculations are performed within what is, to our knowledge, the first publicly available, fully differential NNLO framework capable of generating interpolation grids for fixed-target DY production involving arbitrary nuclear targets. Our results are available in the form of interpolation grids for convolution with any nPDF set. 
The predictions broadly describe the measured distributions. While higher-order QCD corrections largely cancel in the ratios over much of the measured kinematic range, this cancellation becomes less effective at small target momentum fractions, particularly for heavier nuclei. In these regions, the NNLO corrections can reach several percent and be comparable to or larger than the NLO corrections, as well as the experimental uncertainties in the most precise bins. These results demonstrate the relevance of NNLO corrections for future nuclear-PDF analyses incorporating fixed-target DY data.

\FloatBarrier

\section{Impact of NNLO Corrections on Nuclear PDF Fits}
\label{sec:grid-applications}

The perturbative accuracy of the QCD calculations entering global nPDF analyses is an important component of the theoretical framework underlying these fits.
Most modern nPDF analyses are nevertheless still performed at NLO QCD accuracy.
This is partly motivated by the fact that nuclear data are, in general, less precise than the corresponding proton data, such that the additional theoretical accuracy offered by NNLO corrections has traditionally been expected to have a more limited phenomenological impact. 
There are, however, already several NNLO nPDF analyses based on more restricted data selections~\cite{Khanpour:2020zyu,Helenius:2021tof}, and the TUJU21~\cite{Helenius:2021tof} analysis in particular found indications that some of the recent $p$Pb collision data from LHC may be sufficiently precise to benefit from NNLO theoretical predictions. 
In this work, we revisit this question in the context of the \texttt{nCTEQ26} nPDF analysis and with dedicated theoretical predictions $p$Pb DY production at NLO and NNLO QCD obtained in this study. 

Moving from NLO to NNLO requires an improvement of both ingredients entering a prediction: the DGLAP evolution of the PDFs and the hard-scattering coefficient functions, or matrix-element contribution. 
Consequently, for the purpose of PDF extractions, the impact of the higher perturbative order cannot be assessed by evaluating the NLO and NNLO hard cross sections with the same PDFs, extracted at the same perturbative accuracy. 
The ratio of NNLO to NLO predictions evaluated with identical PDFs isolates only the higher-order correction to the partonic calculation, which we refer to here as the NNLO matrix-element (ME) $K$-factor. 
A consistent comparison must instead also account for the change induced by NNLO DGLAP evolution, using PDFs evolved at the corresponding perturbative order in the numerator and denominator. 
This distinction is particularly relevant for DY production: while the NNLO ME $K$-factors are often modest, the combined effect of the higher-order hard cross section and the modified PDF evolution can lead to a noticeably larger shift in the full $K$-factor.

To assess the potential importance of these effects for the quality of the data description for a selection of datasets, we therefore compare NLO predictions based on the \texttt{nCTEQ26} PDFs with predictions in which the same fitted input parametrisation is evolved using NNLO DGLAP evolution and combined with NNLO hard-scattering calculations. 
The resulting shift should be interpreted with some care. 
Since the underlying \texttt{nCTEQ26} parametrisation was obtained in an NLO fit, the difference between the NLO and NNLO predictions may somewhat overestimate the change that would remain after a genuine NNLO global refit: in such a fit, part of the effect induced by the higher-order evolution and coefficient functions could be compensated by a readjustment of the fitted PDF parameters. 
For a global analysis of this scope, however, the size of this compensation cannot be established without actually performing the NNLO fit. 
Further details on the distinction between the ME and full $K$-factors are given in Appendix~\ref{app:MEvsFullKfacs}.
Nevertheless, this comparison remains useful as a diagnostic of the sensitivity of the fitted datasets to higher-order QCD corrections. 
It allows us to identify observables for which the perturbative shift is comparable to, or larger than, the relevant experimental uncertainties, and thereby to assess where an upgrade of the global analysis to NNLO may provide a tangible improvement in theoretical accuracy.

To quantify the potential impact of higher-order predictions on nuclear PDF fits, we investigate the $\chi^2$ breakdown of a selection of the pPb DY data by comparing NLO and NNLO predictions obtained with \texttt{nCTEQ26}.
As discussed above, the NNLO predictions are obtained by evolving the NLO-fitted nPDFs at NNLO while keeping the underlying fit parameters fixed. The analysis is performed within the \texttt{nCTEQ++} framework, using the nPDF fit and the corresponding $\chi^2$ definition introduced in Ref.~\cite{nCTEQ:2026ewr}.
The LHC Drell--Yan data considered here are already included in the original fit at NLO. For the present analysis, however, the NLO theory calculations used in that fit are replaced by the \texttt{fastNLO} grids generated in this work, which are employed consistently for both the NLO and NNLO evaluations. Consequently, the NLO $\chi^2$ values reported here are not expected to exactly coincide with those of Ref.~\cite{nCTEQ:2026ewr}. In both cases, the $\chi^2$ values are evaluated using the best-fit parameters of the existing \texttt{nCTEQ26} fit, without performing a new fit.

Normalisation uncertainties of the datasets included in the fit are treated using the $T$-method~\cite{Ball:2009qv,osti_1891663}. This means that the theory-values in the $\chi^2$ function are scaled by a nuisance parameter, which can then be minimised analytically. 
To analyse the theory-data agreement and the impact of the NNLO corrections on a point-by-point basis, the $\chi^2$ function is decomposed into an uncorrelated contribution associated with each data point and a correlated penalty contribution due to normalisation adjustment. 
After minimisation with respect to the nuisance parameters, their optimal values are incorporated into the shifted theory predictions $T_i$:
\begin{equation}
  \hat T_i = r_\mathrm{norm} T_i + \sum_\alpha \bar \sigma_{i\alpha} \bar r_\alpha,
\end{equation}
where $\bar\sigma_{i\alpha}$ is the correlated systematic uncertainty of the data point $i$ from the $\alpha$-th source of uncertainty. Here, $r_\mathrm{norm}$ and $\bar{r}_\alpha$ are the respective minimal values of the nuisance parameters, which are re-optimised in each calculation of each NLO and NNLO theory prediction corresponding to a given dataset. 
Theoretical predictions shown in the remainder of this section correspond to the shifted theory predictions $\hat T_i$.

We now turn to the discussion of our results.
In table~\ref{tab:normalization_penalties} we present the corresponding penalty terms for the datasets, which share a common normalisation in the fit. The NNLO predictions require systematically smaller normalisation shifts than the corresponding NLO predictions. In particular, the combined normalisation-penalty contribution decreases from $12.08$ to $3.06$.
\begin{table}[tbp]
    \centering
    \renewcommand{\arraystretch}{1.3}
    \begin{tabular}{ccccc}
        \toprule[1.2pt]
        Dataset 
        & $\chi^{2}_{\mathrm{pen,NLO}}$
        & $r_{\mathrm{norm,NLO}}$
        & $\chi^{2}_{\mathrm{pen,NNLO}}$
        & $r_{\mathrm{norm,NNLO}}$ \\
        \midrule
        \midrule
        $W/Z$ prod. (CMS, Run~1) & 1.40 & 1.04 & 0.00 & 1.00 \\
       $Z$ prod. (ATLAS, Run~1) & 5.93 & 1.06 & 2.51 & 1.04 \\
       $Z$ prod. (CMS, Run~2, fixed scale)& 4.75 & 1.08 & 0.55 & 1.03 \\
        $Z$ prod. (CMS, Run~2, dynamic scale)& 5.24 & 1.08 & 0.59 & 1.03 \\
        \bottomrule[1.2pt]
    \end{tabular}
    \caption{Nuisance-parameter values and corresponding penalty contributions at the global minimum of the $\chi^2$ function for the existing \texttt{nCTEQ26} fit. All values are rounded to two decimal places.}
    \label{tab:normalization_penalties}
\end{table}

\begin{figure}[tbp]
    \centering
    \includegraphics[width=0.49\textwidth]{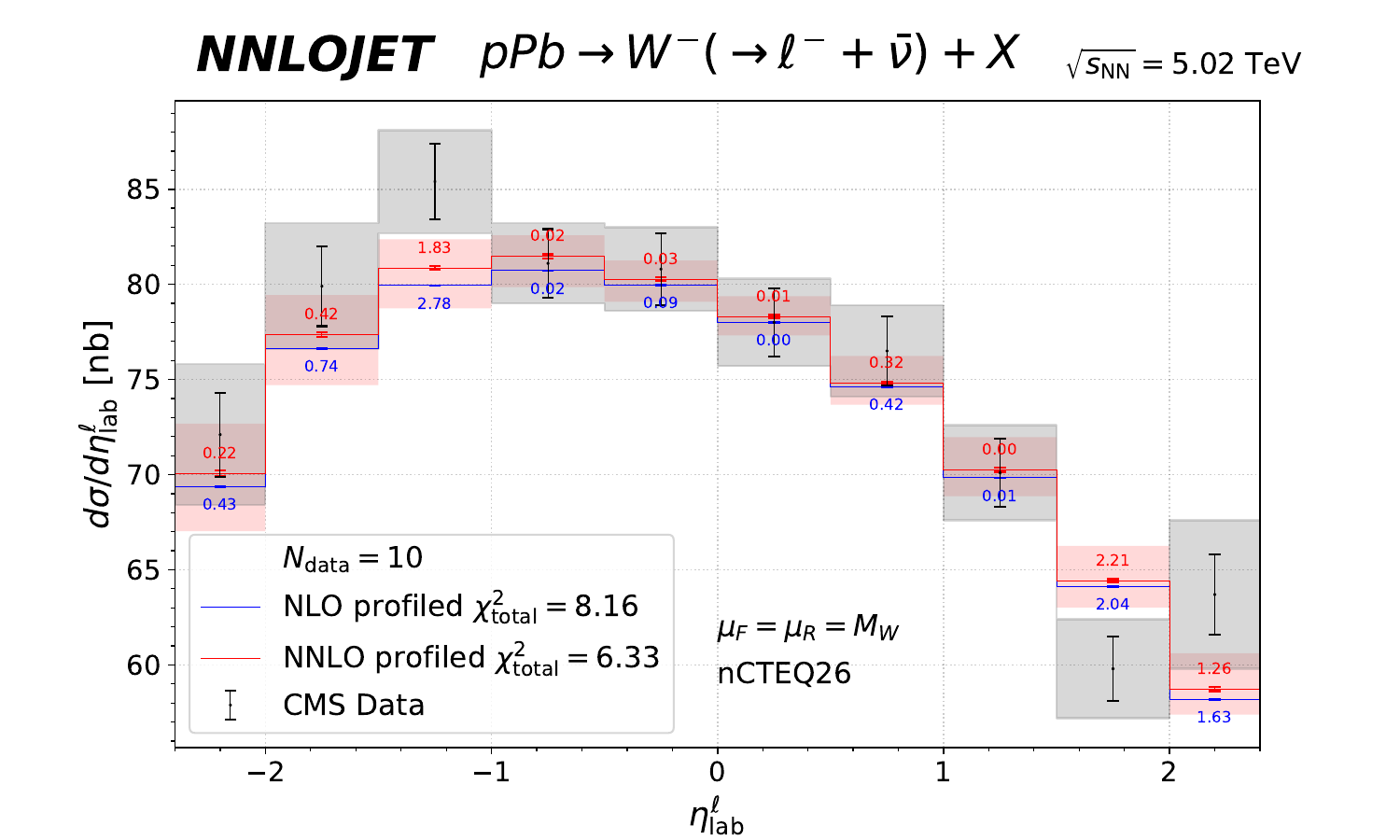}
    \includegraphics[width=0.49\textwidth]{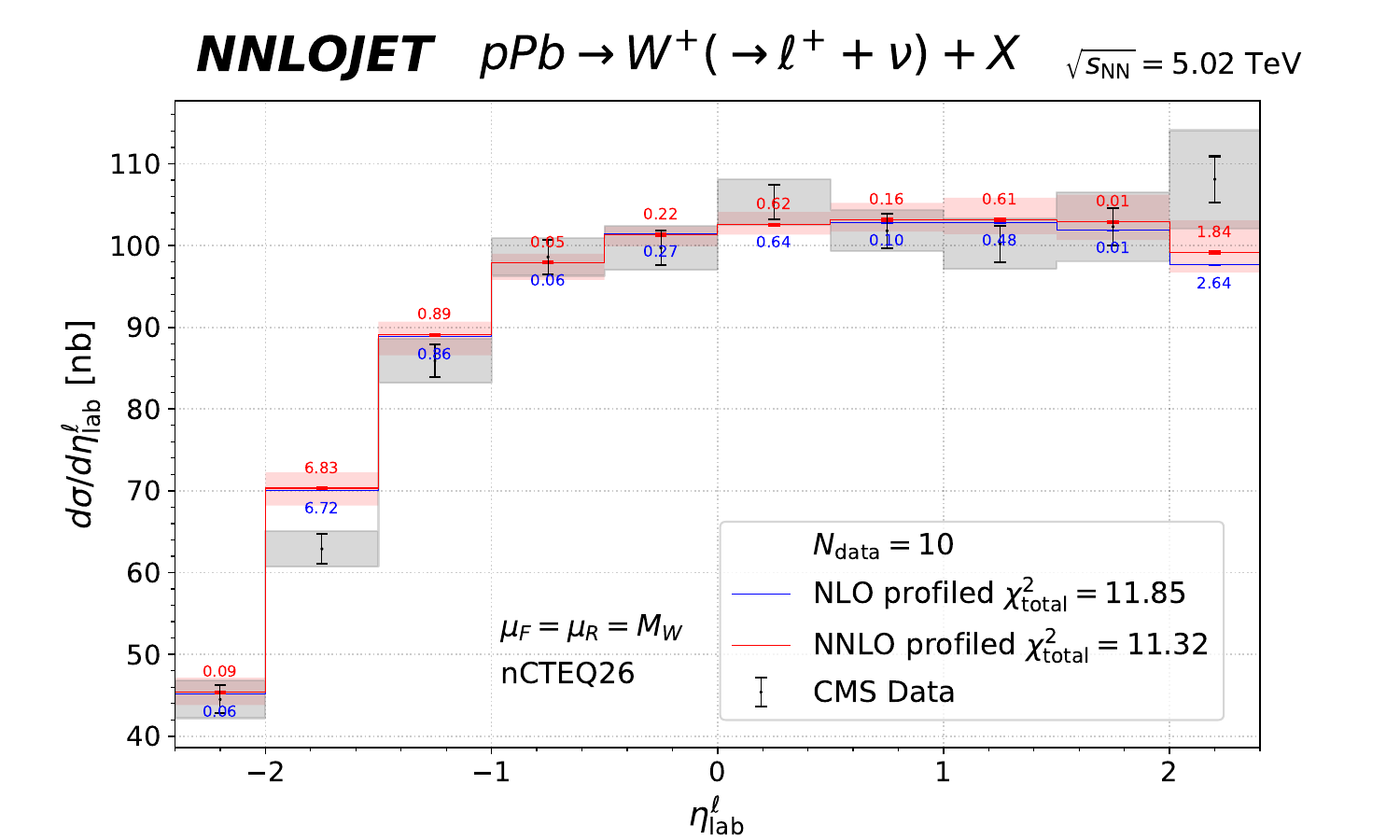}
    \includegraphics[width=0.49\textwidth]{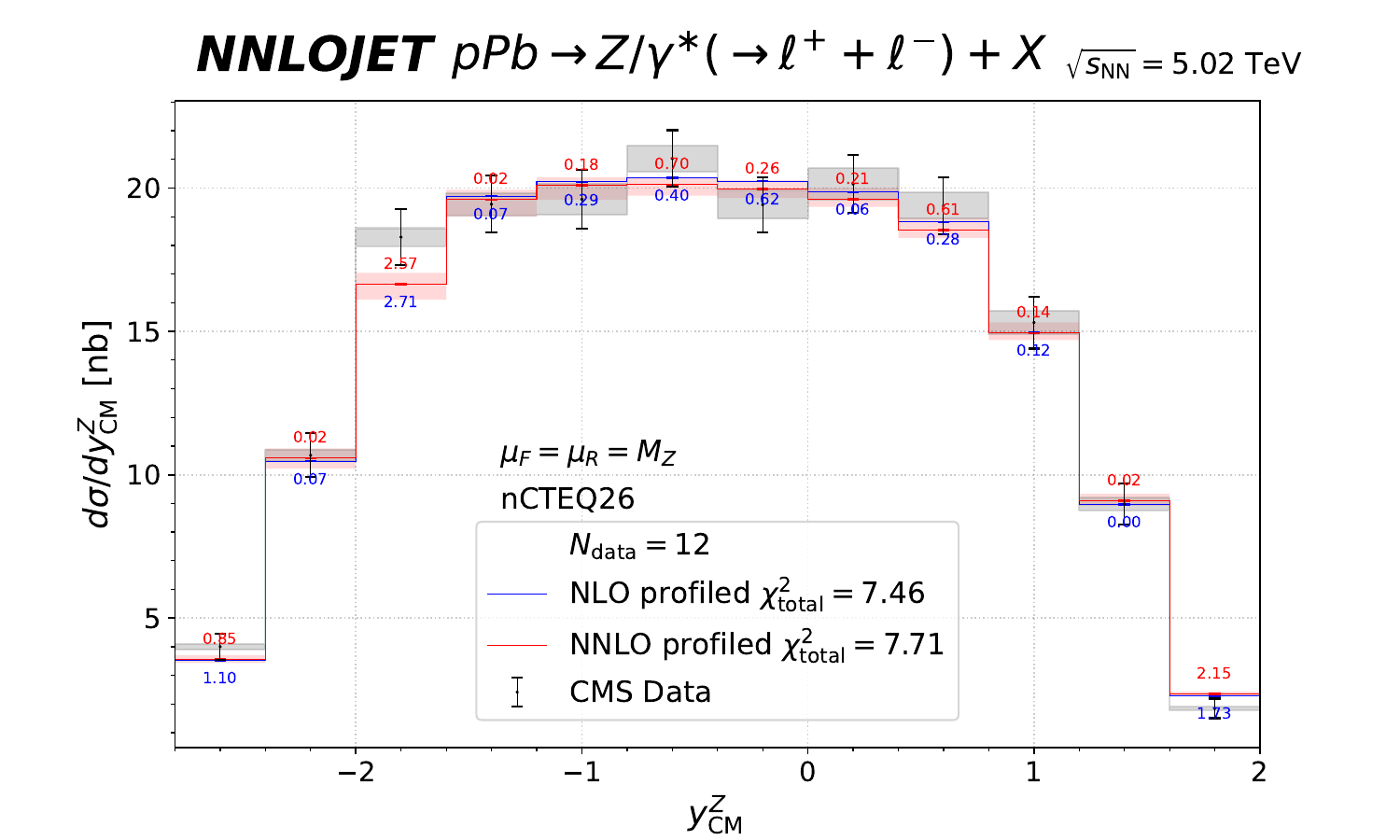}
     \includegraphics[width=0.49\textwidth]{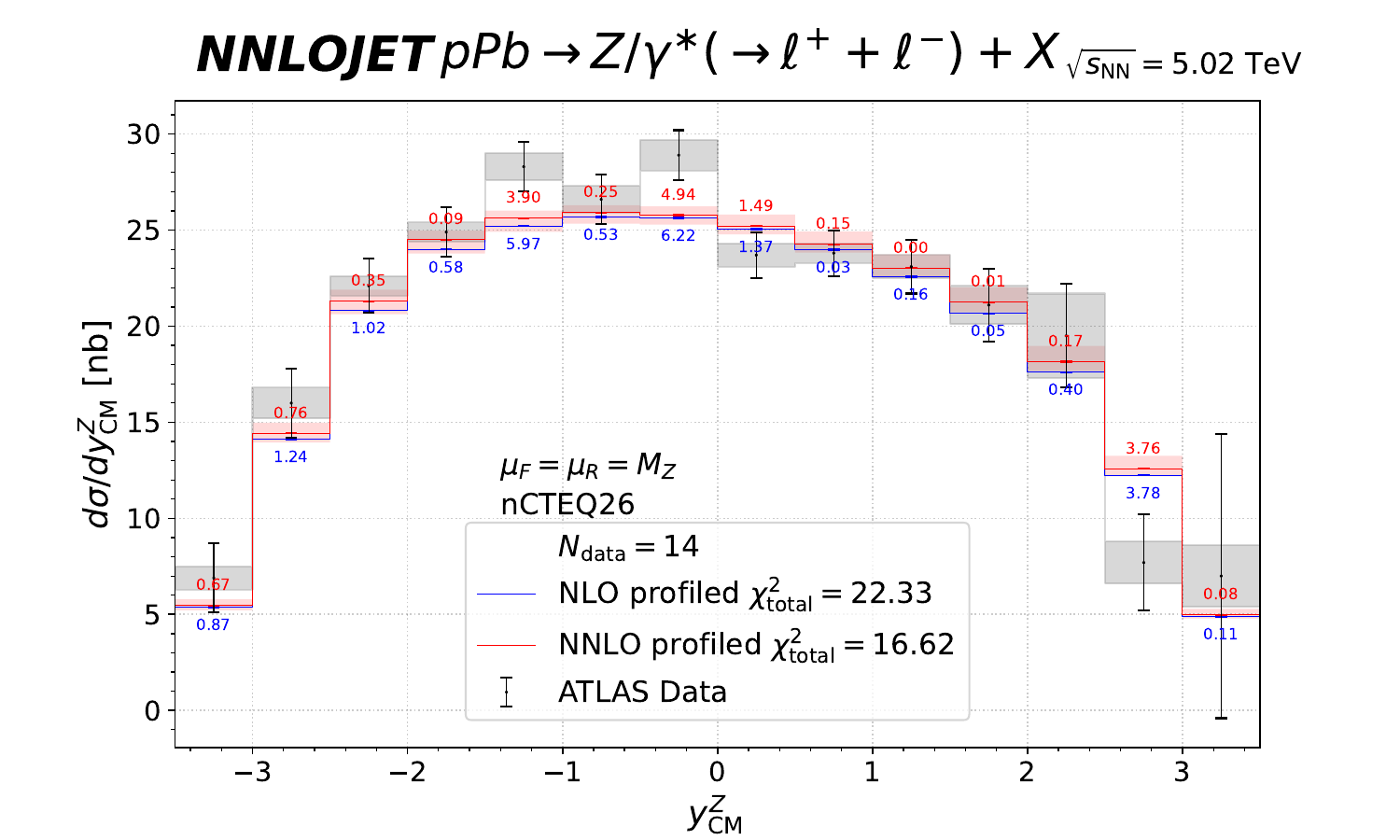}
    \caption{Point-by-point $\chi^2$ contributions for differential cross sections measured in $p\mathrm{Pb}$ collisions of Run~1 at
$\sqrt{s_{\mathrm{NN}}}=5.02~\mathrm{TeV}$, comparing NLO and NNLO
predictions with \texttt{nCTEQ26} PDFs. The NNLO predictions use PDFs obtained
from an NLO QCD fit and evolved at NNLO. The shaded bands indicate
the PDF uncertainties of the NNLO predictions.
Upper row: CMS $W^\pm \to \ell^\pm\nu$ production.
Lower row: $Z \to \ell^+\ell^-$ production measured by CMS (left)
and ATLAS (right).}
  \label{fig:chi2_run1}
\end{figure}

Figure~\ref{fig:chi2_run1} presents the theory predictions at (N)NLO with per-point $\chi^2$ breakdown for the CMS run~1 data discussed in section~\ref{sec:AP}. For visibility, the PDF uncertainties are only shown for the NNLO prediction, as the corresponding uncertainties for the NLO prediction are expected to be very similar.
These uncertainties, corresponding to 90\% C.L., were obtained using using nCTEQ26 error PDFs evolved by NNLO DGLAP evolution with the normalisation correction re-profiled for each member.
An improved description of the data is observed for the CMS $W^-$ and ATLAS $Z$ datasets, for which the total $\chi^2$ decreases by approximately $22\%$ and $25\%$, respectively, upon including NNLO corrections. For the CMS $W^+$ and $Z$ datasets, the NLO predictions already provide a satisfactory description of the data, such that the inclusion of NNLO corrections leads to no or only minor improvements.
\begin{figure}[tbp]
    \centering
    \includegraphics[width=0.49\textwidth]{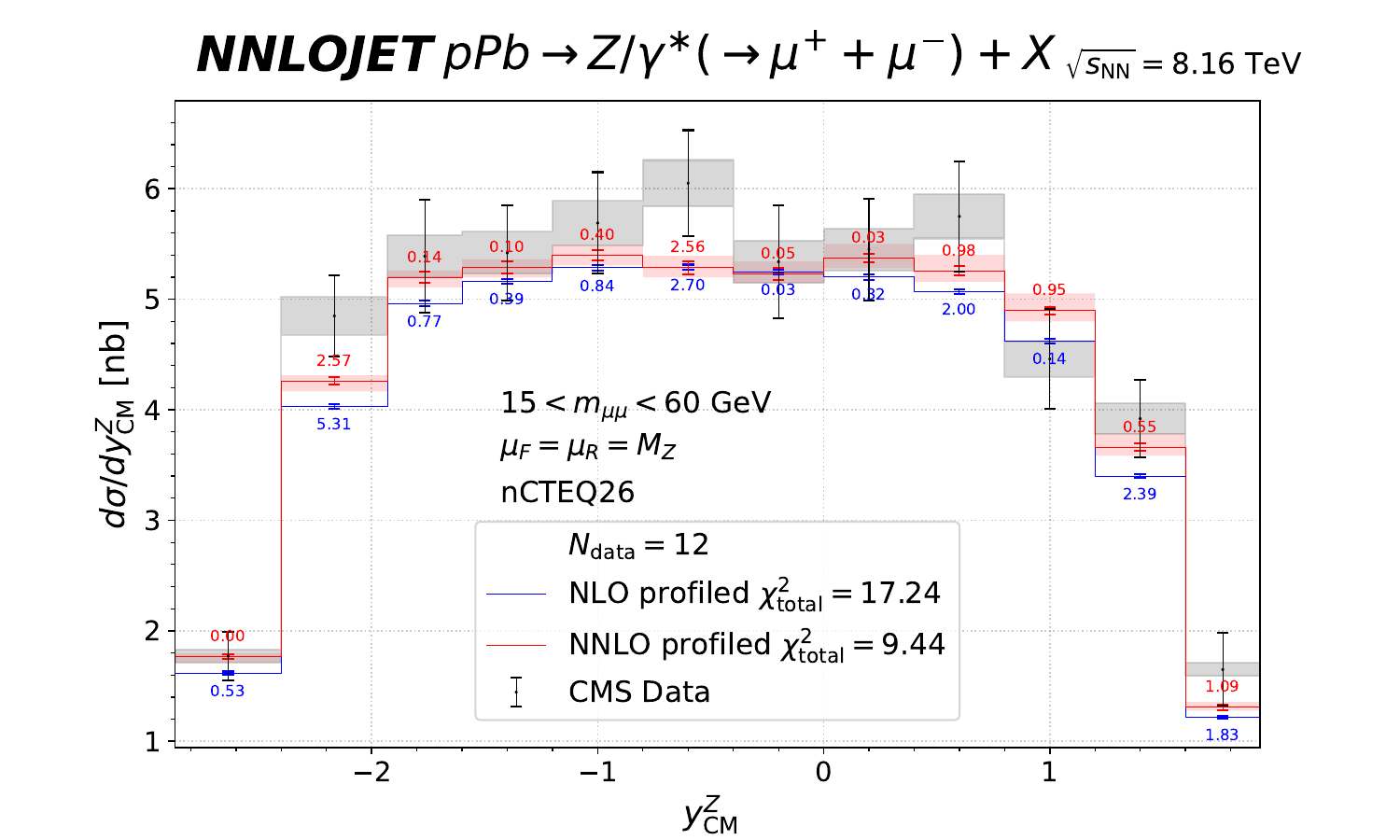}
    \includegraphics[width=0.49\textwidth]{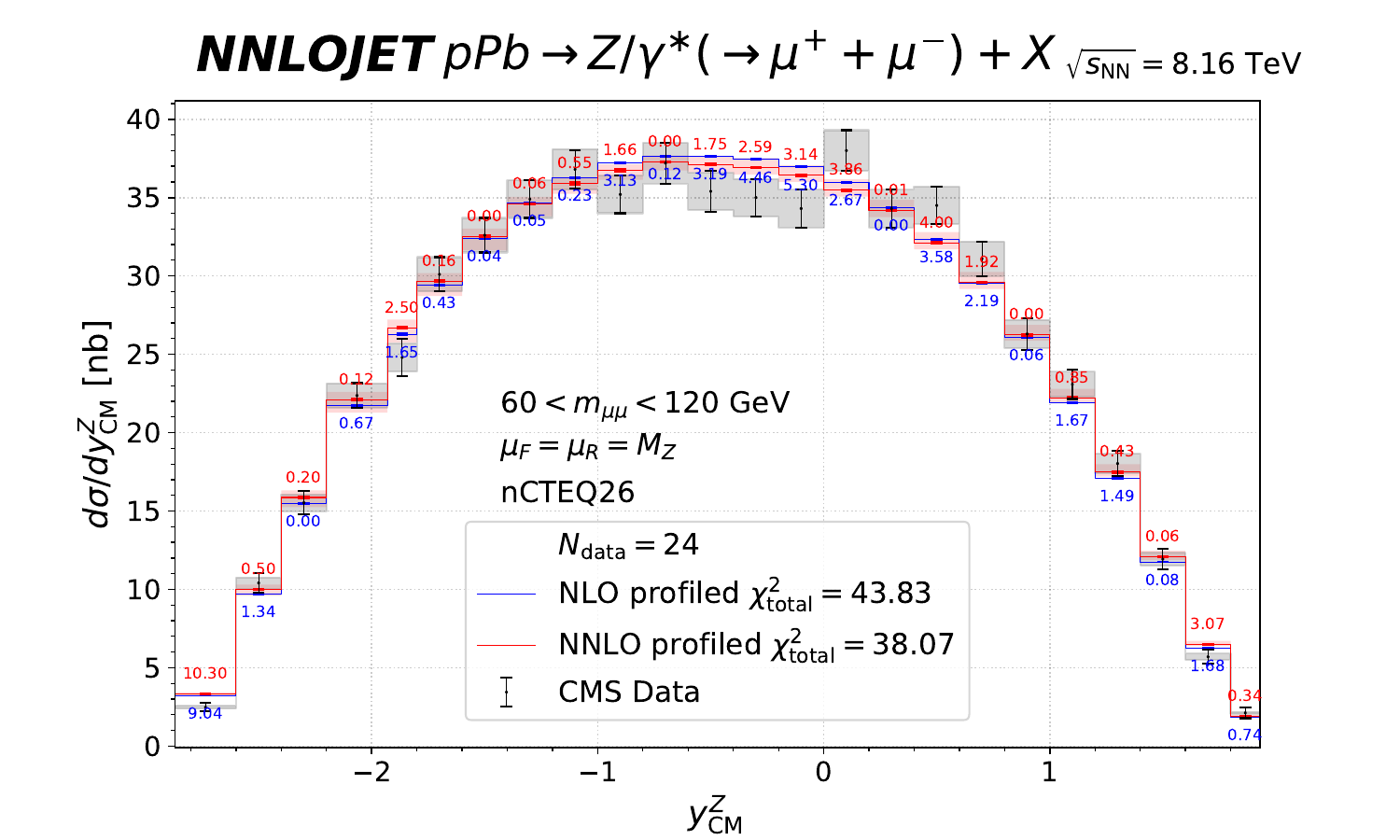}
    \includegraphics[width=0.49\textwidth]{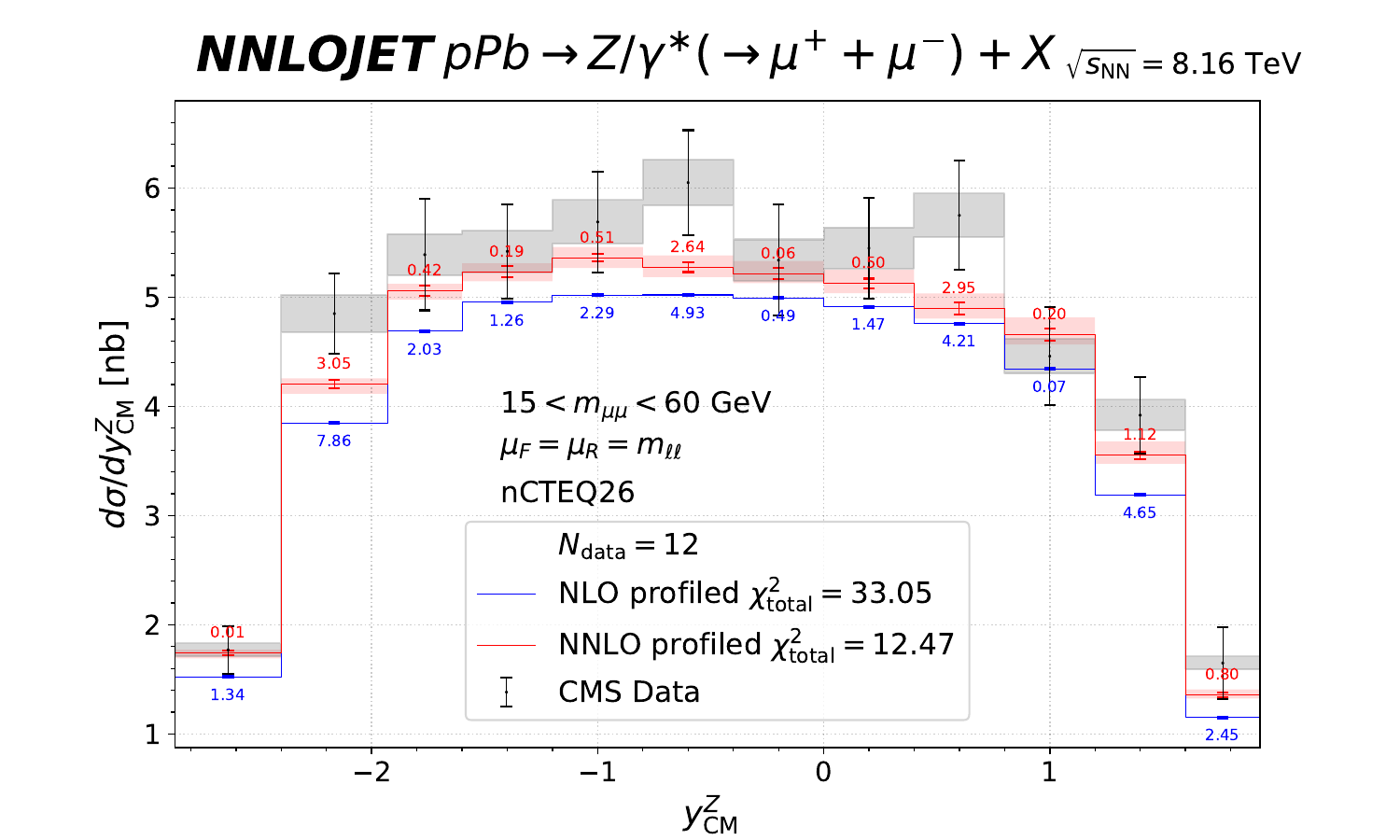}
    \includegraphics[width=0.49\textwidth]{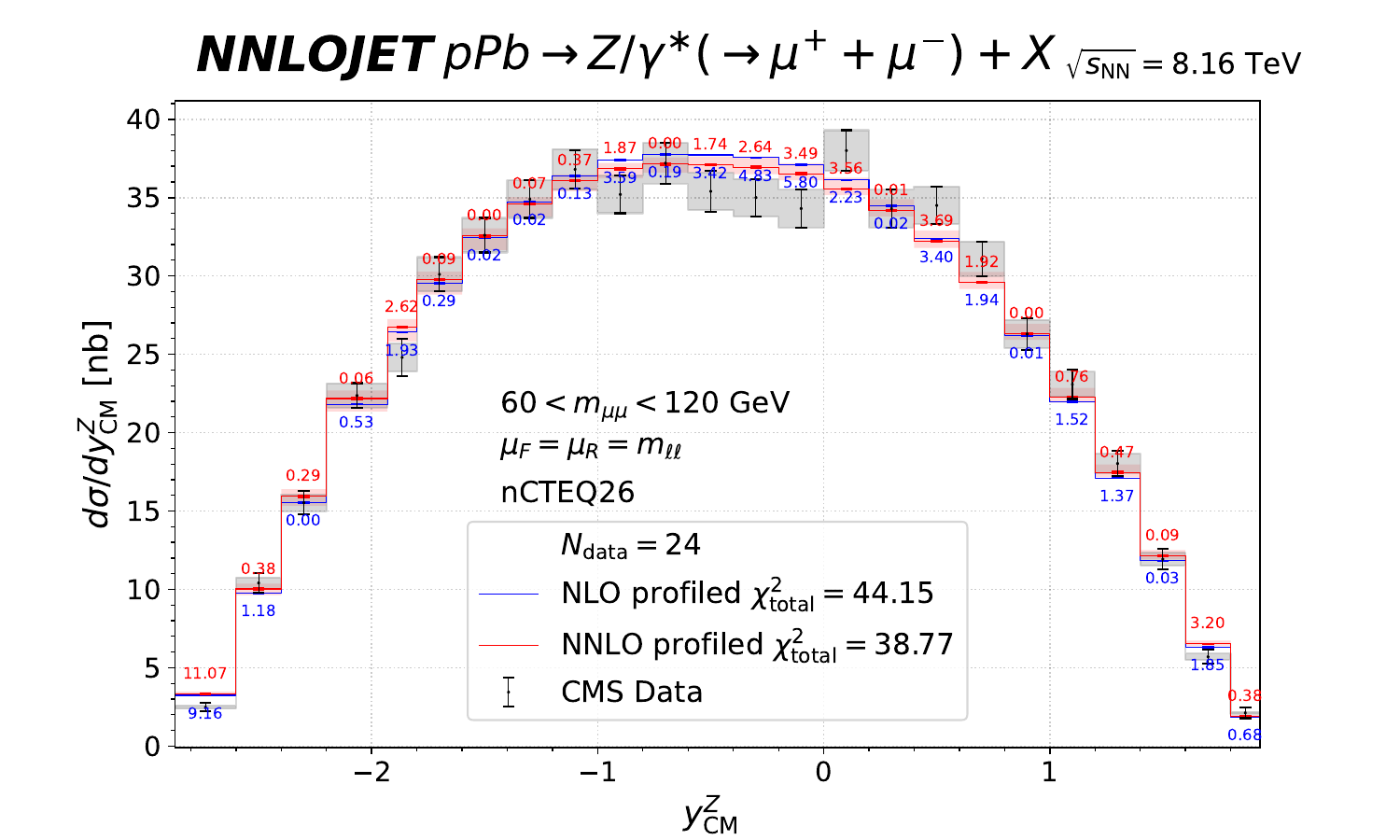}
    \caption{Point-by-point $\chi^2$ contributions for differential cross sections measured by CMS in $p\mathrm{Pb}$ collisions during Run~2 at $\sqrt{s_{\mathrm{NN}}}=8.16~\mathrm{TeV}$, comparing NLO and NNLO predictions with \texttt{nCTEQ26} PDFs. The NNLO predictions use PDFs obtained from an NLO QCD fit and evolved
at NNLO. The shaded bands indicate the PDF uncertainties of the NNLO
predictions. Upper row: fixed scales $\mu_R=\mu_F=M_Z$.
Lower row: dynamical scales $\mu_R=\mu_F=m_{\ell\ell}$.
The left and right columns show dimuon production in the mass ranges
$15<m_{\mu\mu}<60~\mathrm{GeV}$ and
$60<m_{\mu\mu}<120~\mathrm{GeV}$, respectively.}
    \label{fig:chi2_run2_Z}
\end{figure}
A different behaviour is visible for the low mass region of the $Z$ data of CMS run~2, as shown in the upper left panel of figure~\ref{fig:chi2_run2_Z}. Here, the NLO predictions yields a comparatively large $\chi^2$ contribution with $\chi^2/N_\mathrm{data} = 1.43$, while the inclusion of NNLO corrections significantly improves the agreement with the data, reducing $\chi^2/N_\mathrm{data}$ by almost a factor of two to a value of $0.79$.
Also in the high-mass region, the extension towards NNLO accuracy leads to an improvement, with  $\chi^2/N_\mathrm{data}$ decreasing from $1.83$ at NLO to $1.59$ at NNLO. In the central rapidity region, a few residual deviations between theory and data remain. While NNLO corrections improve the overall description, these remaining differences are not expected to be fully accounted for by higher-order corrections alone.

We can further compare the $\chi^2$ decomposition for this measurement using a fixed scale choice, $\mu = M_Z$, with that obtained using a dynamic scale, $\mu = m_{ll}$ (see lower panel of figure~\ref{fig:chi2_run2_Z}). 
Both scale choices show an improvement in the description of the data when going from NLO to NNLO, with a more pronounced improvement for the dynamic scale. However, the overall agreement with the data remains worse for the dynamic scale choice. This is not unexpected, since the nCTEQ26 PDFs were fitted using the fixed scale, and changing the scale without refitting the PDFs need not preserve the quality of the description. The substantial differences between the two scale choices demonstrate the sensitivity of these data to the scale prescription. We argue that the dynamic scale, which follows the dilepton invariant mass, is more appropriate for the low-mass DY region and should therefore be adopted consistently in future nuclear-PDF analyses. This may help reduce internal tensions between datasets probing different kinematic regions, which could otherwise be absorbed into the fitted PDFs.

\section{Conclusion}
\label{sec:Conclusion}

We have presented NNLO QCD predictions for DY production in proton--nucleus collisions, covering the ATLAS and CMS measurements of $W^{\pm}$ and $Z/\gamma^{*}$ production in $p\mathrm{Pb}$ collisions at $\sNN=5.02$ and $8.16~\TeV$ and the per-nucleon cross-section ratios measured by the fixed-target experiments E772 and E866/NuSea.
The NNLO calculations of both types of collisions are obtained within a common \texttt{NNLOJET} implementation, which accounts for asymmetric beam energies, frame conventions and experimental selections, and provides predictions for the ATLAS and CMS $W^{\pm}$ and $Z/\gamma^{*}$ measurements and the E772 and E866 per-nucleon nuclear-target ratios directly in the momentum-fraction variables of the published measurements.
The accompanying interpolation grids, allowing repeated evaluations with independently specified proton and nuclear PDFs and scale choices, generated by \texttt{fastNLO} through the \texttt{APPLfast} interface, are filled during these perturbative fixed-order calculations and are released alongside this paper.

The NNLO corrections to the $W^{\pm}$ and $Z$-mass distributions remain within about $3\%$, accompanied by a substantial reduction in scale dependence. 
The CMS low-mass window, $15<m_{\mu\mu}<60~\GeV$, behaves differently: the NLO corrections nearly double the LO cross section, and the NNLO corrections reach $15\%$ near the acceptance boundaries, reduce but do not remove the deficit relative to the data.
This enhanced sensitivity to QCD radiation arises in part from the fiducial cuts, which exclude masses below $30~\GeV$ at Born level.
Higher-order corrections largely cancel in the $W^{\pm}$ forward--backward ratios and over much of the E772 range, but not at the smallest target momentum fractions: the NNLO correction to the E866 $\mathrm{W}/\mathrm{Be}$ ratio reaches $-6.3\%$ at low mass and large beam momentum fraction.
These results show that small corrections to some ratios do not justify neglecting the NNLO contributions throughout the datasets.
Residual differences between data and theory cannot be attributed uniquely to nuclear modifications without accounting for the proton- and nuclear-PDF uncertainties and the experimental correlations.
    
Finally, we have investigated the impact of NNLO corrections on the description of LHC $p$Pb data using the nCTEQ26 nuclear PDFs. We confirm the findings of the TUJU21 analysis that some nuclear datasets are sufficiently precise to benefit from NNLO theoretical predictions. While the inclusion of NNLO corrections leads to only modest improvements for several datasets, substantial reductions in $\chi^2$ are observed for CMS $W^-$ and ATLAS $Z$ production at 5.02 TeV, and particularly for CMS low-mass Drell--Yan production at 8.16 TeV. The latter also exhibits a pronounced sensitivity to the scale choice, highlighting the importance of adopting a consistent, physically motivated scale prescription in future nPDF analyses. Although these results demonstrate the potential benefits of NNLO accuracy, a definitive assessment requires a global NNLO nPDF fit, including NNLO DGLAP evolution and a refitting of the PDF parameters.

\acknowledgments
T.\,J., M.\,K. and J.\,N. are grateful to their nCTEQ colleagues, in particular Katrin Greve, Aleksander Kusina and Jan Wissmann for valuable discussions.
X.\,C. gratefully acknowledges the valuable discussions and insights provided by Sven Moch. 
X.\,C. and A.\,H. are grateful to their \texttt{APPLfast} colleagues, especially their efforts to develop the interface to \texttt{NNLOJET}.
Work at the University of Münster has been partially funded by the European Union’s Horizon Europe research and innovation programme under Grant Agreement No. 101292522 (HADRON-2030). 
Work at the Shandong University has been funded by the
National Natural Science Foundation of China under grants No.12321005 and No.12475085. 

\appendix
\section{Nuclear PDF closure test}
\label{app:closure}

\begin{figure}[tbp]
    \centering
    \includegraphics[width=0.49\textwidth]{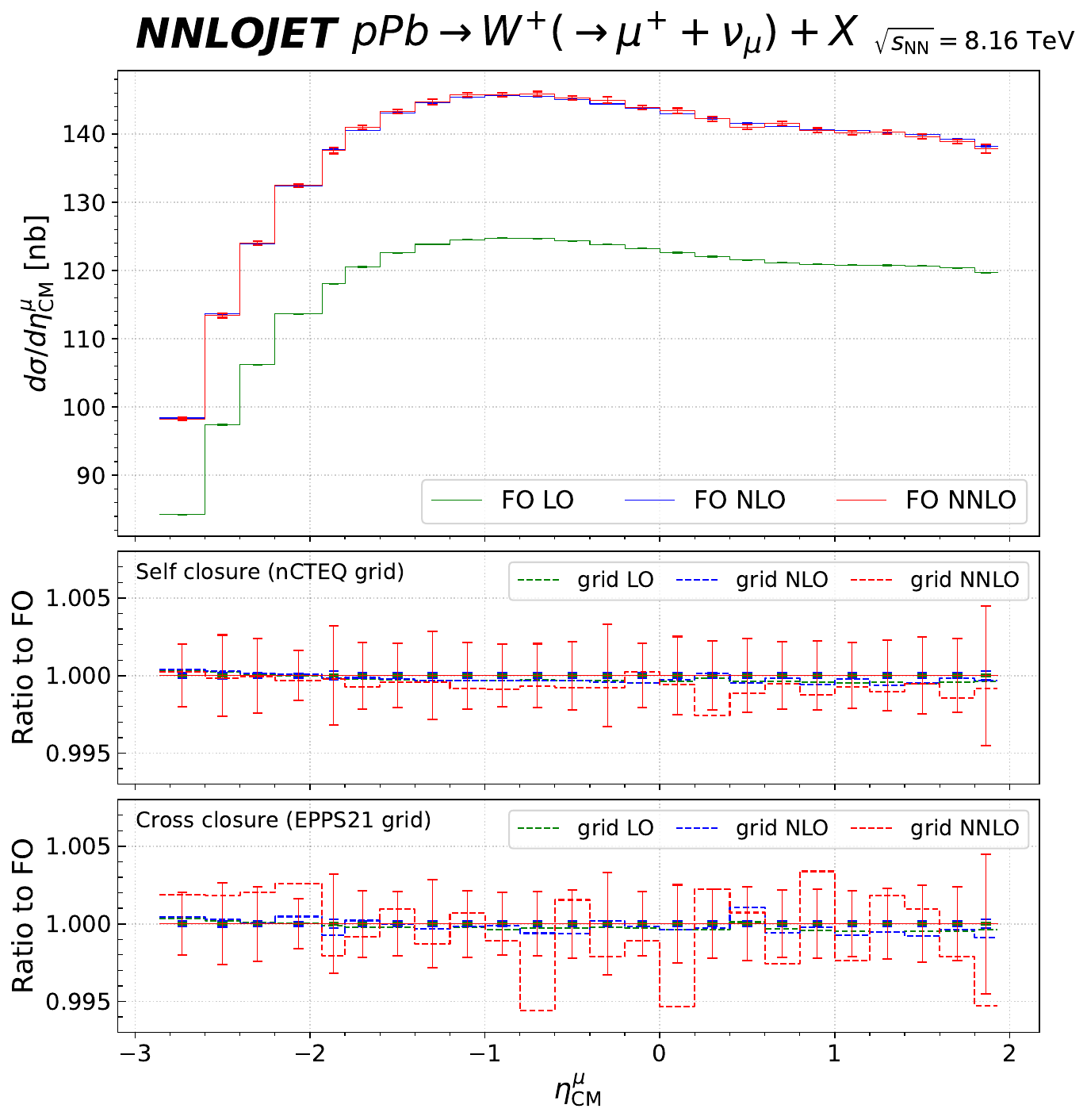} 
    \includegraphics[width=0.49\textwidth]{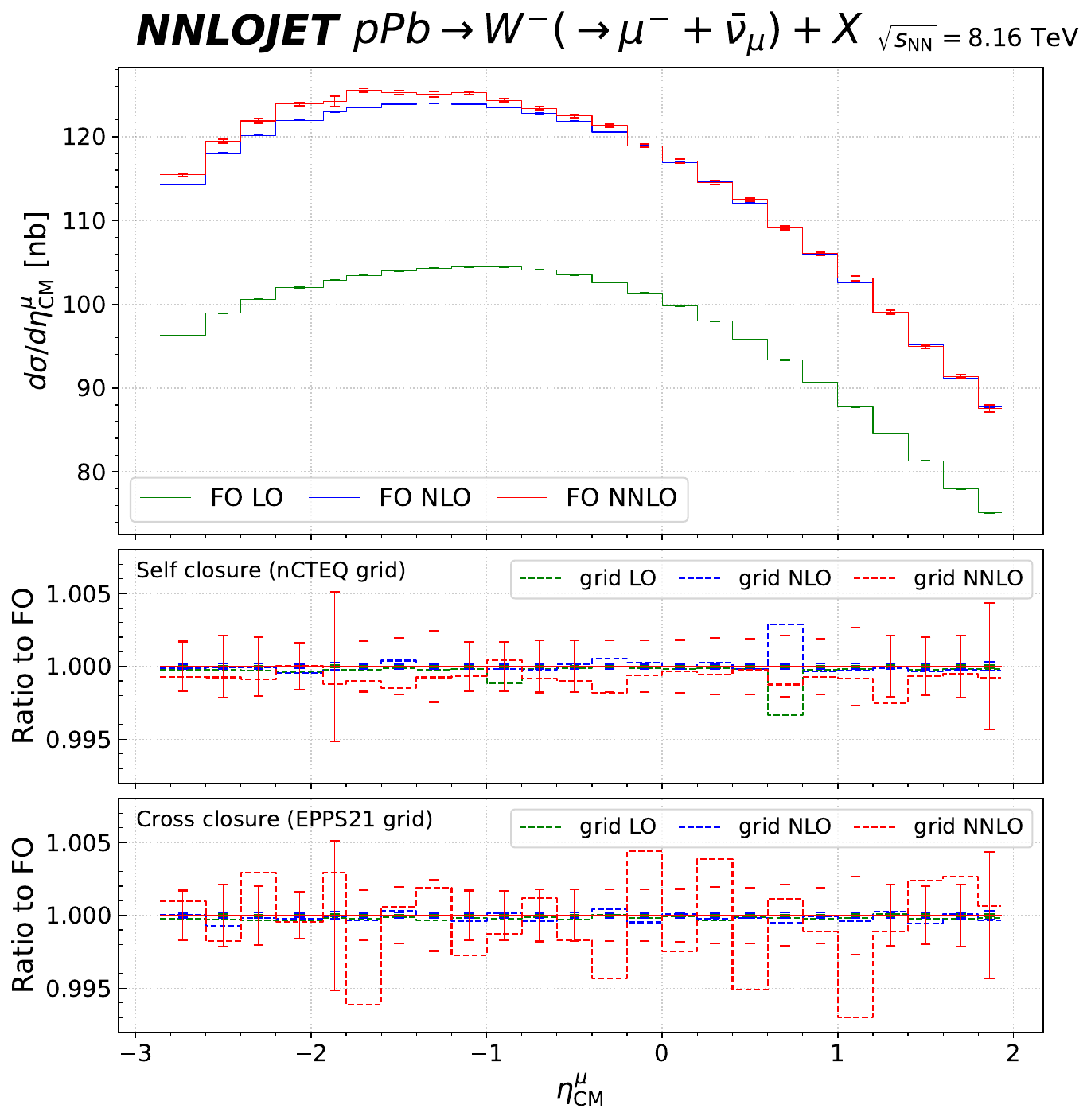} 
    \includegraphics[width=0.49\textwidth]{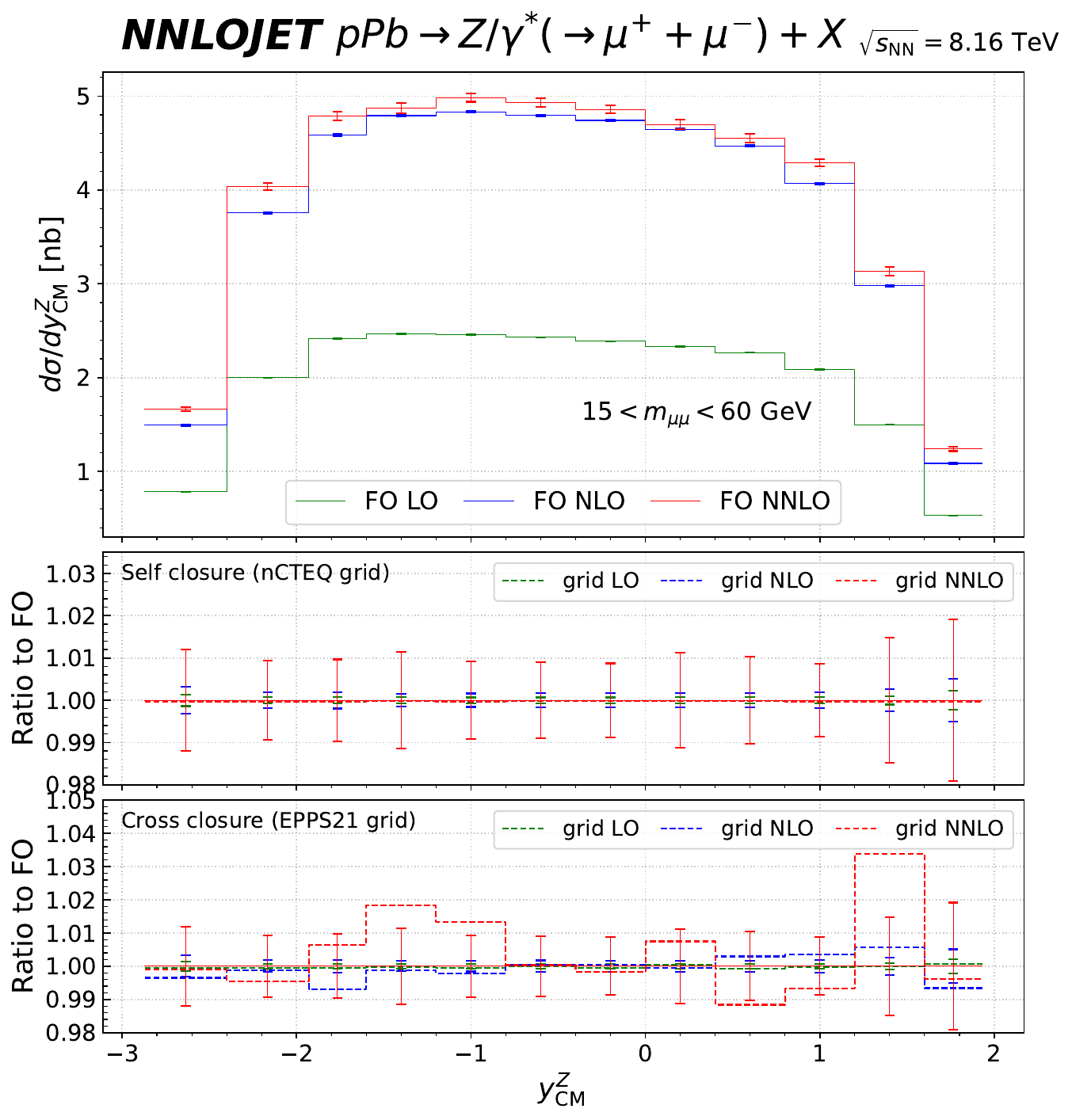} 
    \includegraphics[width=0.49\textwidth]{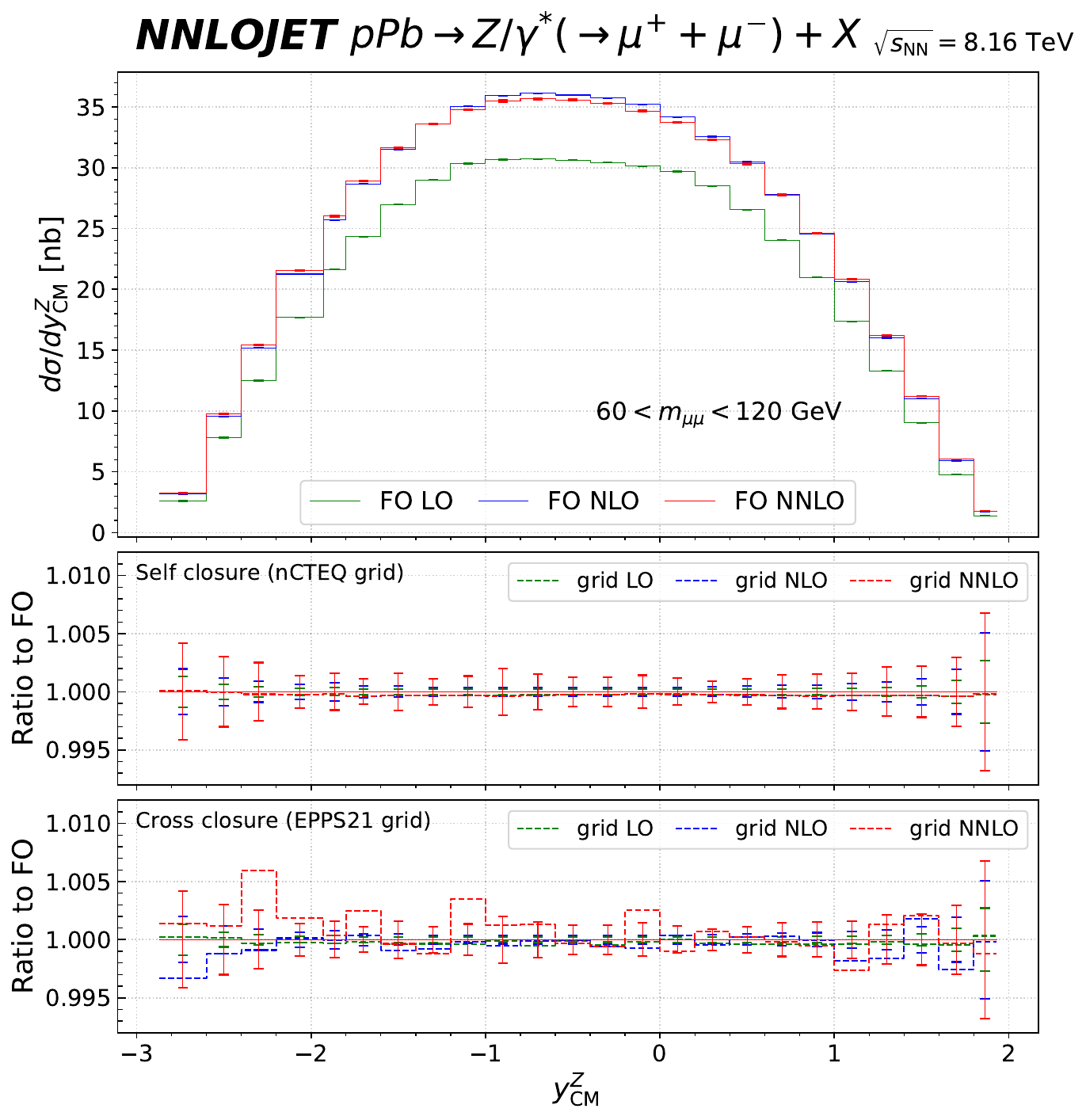} 
\caption{Closure test of the interpolation grids for the CMS Run~2 observables at $\sNN=8.16~\TeV$: the $W^{+}$ (upper left) and $W^{-}$ (upper right) charged-lepton pseudorapidity distributions and the dilepton rapidity distributions in the mass windows $15<m_{\mu\mu}<60~\GeV$ (lower left) and $60<m_{\mu\mu}<120~\GeV$ (lower right).
The upper panels show the direct fixed-order (FO) \texttt{NNLOJET} predictions at LO, NLO and NNLO with the \texttt{nCTEQ26} lead set and the \texttt{CT18NNLO\_as\_0118} proton PDF.
The middle and lower panels show the ratio of the grid evaluation to the FO result for grids filled with \texttt{nCTEQ26} (self closure) and with \texttt{EPPS21nlo\_CT18Anlo\_Pb208} (cross closure), respectively, evaluated with the same PDFs as the FO predictions; the error bars, centred at unity in the ratio panels, indicate the numerical integration uncertainties of the FO calculation.}
\label{fig:PDF_closure}
\end{figure}

In this appendix, we demonstrate the reliability and robustness of our interpolation grids through a nuclear-PDF closure test.
The interpolation grids used in section~\ref{sec:grid-applications} are filled during the \texttt{NNLOJET} calculations of section~\ref{sec:AP}, which employ the \texttt{CT18NNLO\_as\_0118} proton and \texttt{EPPS21nlo\_CT18Anlo\_Pb208} nuclear PDFs.
Their subsequent evaluation with other nuclear PDFs relies on the accuracy of the interpolation in the parton momentum fractions and scales, which we verify with a closure test across PDF sets: the grids for the CMS Run~2 observables at $\sNN=8.16~\TeV$ are evaluated a posteriori with the \texttt{nCTEQ26} lead set \texttt{nCTEQ26\_208\_82}~\cite{nCTEQ:2026ewr}, retaining the \texttt{CT18NNLO\_as\_0118} proton PDF, and compared with a dedicated fixed-order \texttt{NNLOJET} calculation performed directly with these PDFs, under identical fiducial selections and dynamical scales.
In addition, grids filled with the \texttt{nCTEQ26} set itself are compared with this direct calculation (self closure).
Figure~\ref{fig:PDF_closure} shows these comparisons for the $W^{+}$ and $W^{-}$ charged-lepton pseudorapidity distributions and for the dilepton rapidity distributions in the two mass windows, $15<m_{\mu\mu}<60~\GeV$ and $60<m_{\mu\mu}<120~\GeV$, at $\sNN=8.16~\TeV$, with the middle (self closure) and lower (cross closure) panels displaying the ratio of the grid evaluation to the fixed-order result at each perturbative order.

In the self closure, the grid and direct predictions differ by less than $0.05\%$ in both mass windows at all orders.
For $W^{+}$ and $W^{-}$ production, the differences remain below $0.07\%$ at LO and NLO, except in two isolated $W^{-}$ bins, $-1.0<\eta^{\mu}_{\rm CM}<-0.8$ and $0.6<\eta^{\mu}_{\rm CM}<0.8$, where they reach $0.12\%$ and $0.33\%$, respectively.
At NNLO, all differences lie within one standard deviation of MC error ($\sigma_{\rm MC}$), %
except in three $W^{\pm}$ bins, $0.2<\eta^{\mu}_{\rm CM}<0.4$ for $W^{+}$ and $-0.4<\eta^{\mu}_{\rm CM}<-0.2$ and $1.2<\eta^{\mu}_{\rm CM}<1.4$ for $W^{-}$, where they reach up to 1.2 standard deviations. %
In the cross closure, the LO relative differences between the grid and direct predictions remain below $0.06\%$ for $W^{+}$ production, $0.04\%$ for $W^{-}$ production, $0.05\%$ in the $Z$-mass region and $0.08\%$ in the low-mass window.
At NLO, the magnitudes of the differences, averaged over all bins, are $2.3\ \sigma_{\rm MC}$ and $1.5\ \sigma_{\rm MC}$ for $W^{+}$ and $W^{-}$ production, $1.1\ \sigma_{\rm MC}$ and $1.3\ \sigma_{\rm MC}$ in the $Z$-mass region and the low-mass window, with relative differences below $0.11\%$, $0.07\%$, $0.4\%$ and $0.7\%$, respectively.
At NNLO, the corresponding averaged differences are $0.9\ \sigma_{\rm MC}$, $1.2\ \sigma_{\rm MC}$, $0.9\ \sigma_{\rm MC}$ and $0.8\ \sigma_{\rm MC}$, respectively.
The error bars show the numerical integration uncertainties of the direct calculations alone, denoted by $\sigma_{\rm MC}$ and used to normalise the differences quoted above; those of the grid evaluation, which for the cross closure are inherited from the calculations of section~\ref{sec:AP}, are not displayed.
The observed differences therefore do not by themselves isolate the interpolation error from numerical integration fluctuations, and a full statistical assessment of the closure would also require the uncertainties of the grid evaluation.
Nevertheless, these comparisons provide a solid numerical cross-check of the interpolation grids under a change of nuclear-PDF input for the applications presented in section~\ref{sec:grid-applications}.

\section{\texorpdfstring{$K$}{K}-factors}
\label{app:MEvsFullKfacs}
\begin{figure}[tbp]
    \centering
    \includegraphics[width=0.49\textwidth]{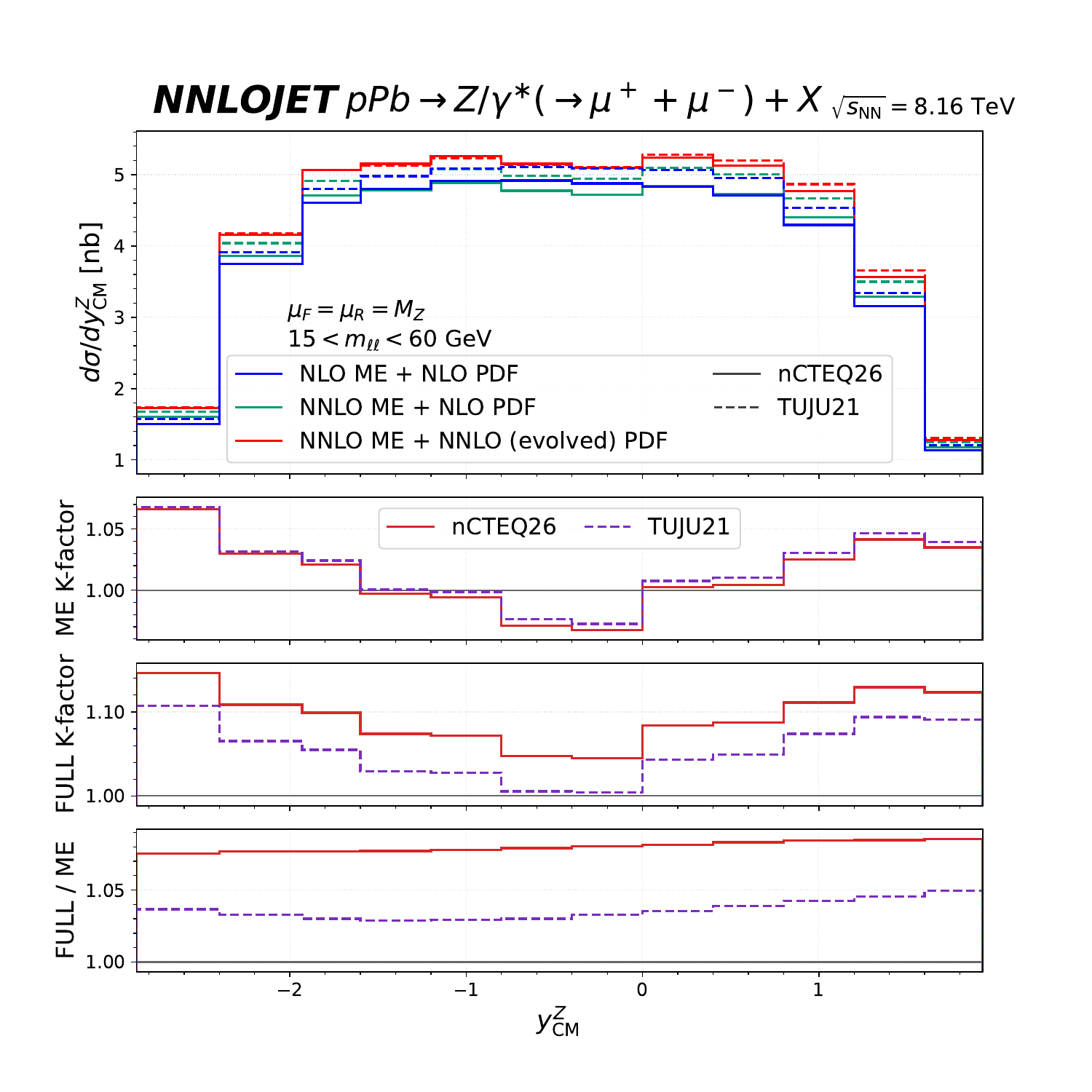}
    \includegraphics[width=0.49\textwidth]{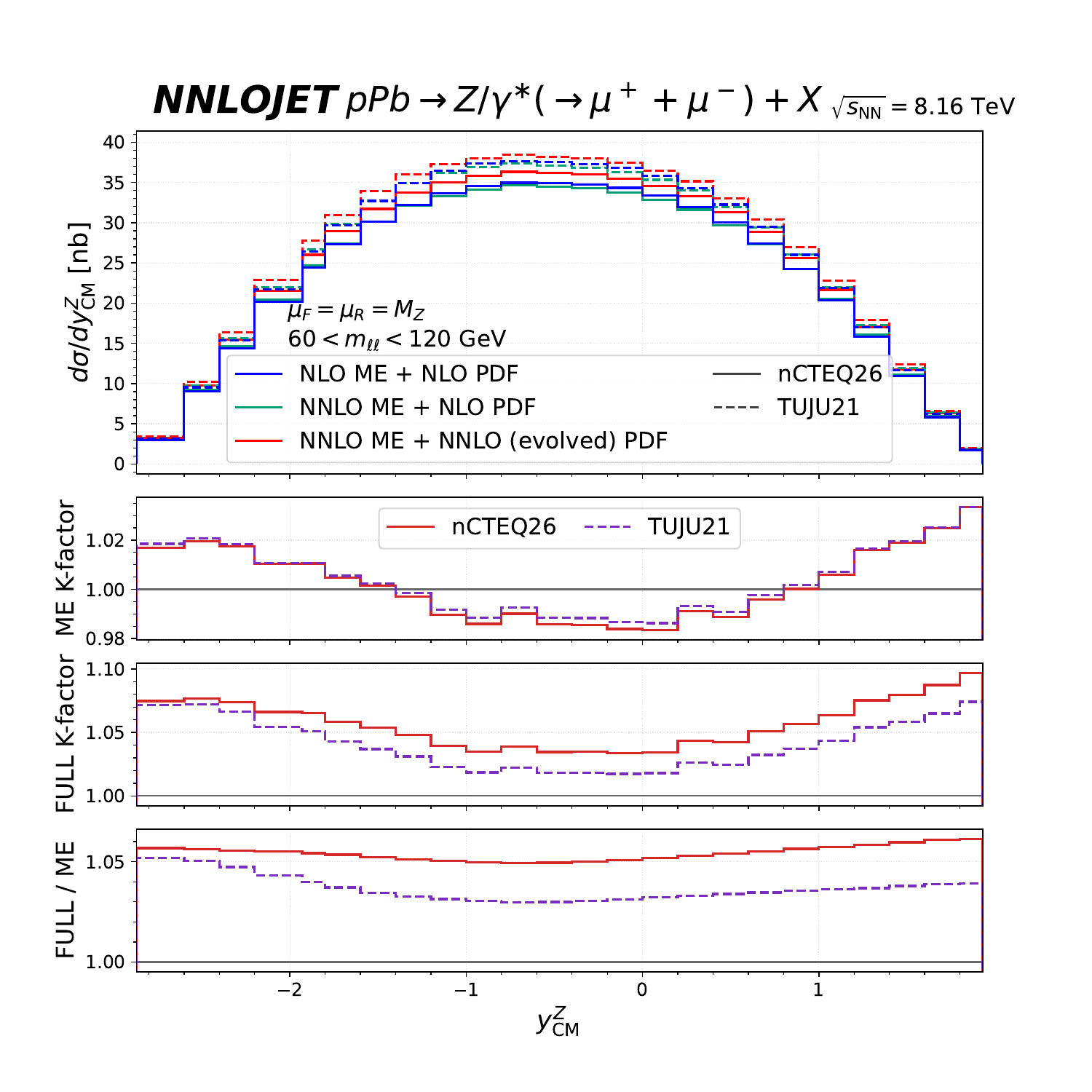}
    \caption{Predictions for dimuon production in CMS Run~2 pPb collisions at $\sqrt{s_{\mathrm{NN}}}=8.16~\mathrm{TeV}$ in the mass ranges $15<m_{\mu\mu}<60~\mathrm{GeV}$ (left) and
$60<m_{\mu\mu}<120~\mathrm{GeV}$ (right), comparing NLO and NNLO QCD
calculations using \texttt{nCTEQ26} and TUJU21 PDFs.
The NNLO-evolved \texttt{nCTEQ26} PDFs are obtained from an NLO QCD fit
and subsequently evolved using NNLO evolution.
The ratio panels show, from top to bottom, the effect of NNLO
matrix elements with NLO PDFs (ME $K$-factor), the combined effect
of NNLO matrix elements and NNLO PDFs (FULL $K$-factor), and the
ratio of the FULL to the ME $K$-factor.
All $K$-factors are normalized to the corresponding NLO prediction
with NLO PDFs.}
\label{fig:ME_VS_FULL_K}
\end{figure}

We assess the impact of higher-order perturbative calculations on PDF extraction by isolating the effects arising from higher-order matrix-element calculations through the definition of a matrix-element (ME) $K$-factor
\begin{equation}
   K_\mathrm{ME} = \frac{\sigma^\mathrm{NNLO}[\mathrm{PDF_\mathrm{NLO}}]}{\sigma^\mathrm{NLO}[\mathrm{PDF_\mathrm{NLO}}]}.
\end{equation}
This is compared to a full $K$-factor,
\begin{equation}
\label{eq:full_K}
    K_\mathrm{FULL} = \frac{\sigma^\mathrm{NNLO}[\mathrm{PDF_\mathrm{NNLO}}]}{\sigma^\mathrm{NLO}[\mathrm{PDF_\mathrm{NLO}}]},
\end{equation}
which additionally incorporates the effects induced by the NNLO DGLAP evolution of the PDFs. 

In figure~\ref{fig:ME_VS_FULL_K} we compare the two $K$-factors and their ratio for the differential cross sections for neutral-current DY production measured by CMS in Run~2 in the low- and high-mass regions, using the \texttt{nCTEQ26} and TUJU21 PDF sets.
For the case of \texttt{nCTEQ26}, the NNLO PDFs in eq.~\ref{eq:full_K} are obtained by evolving the underlying NLO-fitted PDFs with NNLO DGLAP evolution, while keeping the fitted parameters unchanged. In contrast, the TUJU21 NNLO PDFs are extracted from a dedicated NNLO global fit.
As can be seen, the ME $K$-factors differ only at the sub-percent level, which is expected since both NLO PDF sets are obtained from a global fit. 
Since the \texttt{nCTEQ26} NNLO PDF is not based on a fit, the deviations to TUJU21 in the full $K$-factor are larger, reaching around $2\%$ in the low-mass and around $1\%$ in the high-mass region.
The full $K$-factor is particularly pronounced in the forward (backward) region, reaching values of more than $1.1$ for \texttt{nCTEQ26}.
Even if the higher-order matrix-element corrections are relatively small, the combined effect of higher-order hard-scattering calculations and the corresponding PDF evolution can lead to sizeable changes in the predicted cross sections. This indicates that these observables can point to kinematic regions where an NNLO PDF upgrade may have a notable impact on the theoretical prediction.
By considering the ratio of the ME and full $K$-factors, the \texttt{nCTEQ26} result highlights the relative importance of the higher-order correction associated with the NNLO DGLAP evolution, reaching around $7-8\%$ in the low-mass and $5\%$ in the high-mass region.
For TUJU21, the ratio instead reflects the combined impact of the change in the perturbative order of the PDF fit and the corresponding hard-scattering calculation.

\bibliographystyle{JHEP}
\bibliography{references}

\end{document}